\documentclass[letterpaper,11pt]{article}

\usepackage[arxiv]{optional}

\usepackage[margin=1in]{geometry}
\usepackage{setspace}
\usepackage{graphicx}
\usepackage{epsfig}

\usepackage{microtype} %
\usepackage{array} %
\usepackage{booktabs} %

\usepackage{amssymb}
\usepackage{amsmath}
\usepackage{amsthm}
\usepackage{mathtools} %
\usepackage{stmaryrd} %
\usepackage{stackrel} %
\usepackage{latexsym} %

\usepackage{tikz}
\usetikzlibrary{arrows.meta}
\usetikzlibrary{decorations.pathreplacing}
\usetikzlibrary{cd}
\usepackage[all]{xy} %

\usepackage{algorithm}
\usepackage[noend]{algpseudocode}
\algnewcommand{\LineComment}[1]{\Statex\hspace{\algorithmicindent}\(\triangleright\) #1}
\algnewcommand\algorithmicforeach{\textbf{for each}}
\algdef{S}[FOR]{ForEach}[1]{\algorithmicforeach\ #1\ \algorithmicdo}
\usepackage{etoolbox} %

\usepackage[labelformat=parens]{subfig}

\usepackage[hidelinks]{hyperref} %
\usepackage{cleveref} %
\usepackage{makeidx} %
\usepackage{url} %
\usepackage{color}
\usepackage{xcolor}
\usepackage{xspace} %
\usepackage[normalem]{ulem} %
\usepackage{soul} %
\usepackage{ifpdf} %

\usepackage{newtxtext}
\usepackage[scr=boondoxo,scrscaled=0.9]{mathalfa}
\DeclareMathAlphabet{\mathsfit}{T1}{\sfdefault}{\mddefault}{\sldefault}
\SetMathAlphabet{\mathsfit}{bold}{T1}{\sfdefault}{\bfdefault}{\sldefault}

\makeatletter
\expandafter\patchcmd\csname\string\algorithmic\endcsname{\itemsep\z@}{\itemsep=0.25ex}{}{}
\makeatother

\newcounter{usesmallsep}
\ifnum\the\value{usesmallsep}=1

    \newlength{\myitemsep}
    \newlength{\mytopsep}
    \usepackage{enumitem}
    \setlist[itemize]{leftmargin=\parindent,parsep=\parskip,
      listparindent=\parindent,itemsep=\myitemsep,topsep=\myitemsep}
    \setlist[enumerate]{leftmargin=\parindent,parsep=\parskip,
      listparindent=\parindent,itemsep=\myitemsep,,topsep=\myitemsep}
    \setlist[description]{font=\bfseries,leftmargin=\parindent,parsep=\parskip,
      listparindent=\parindent,itemsep=\myitemsep,topsep=\myitemsep}
    
    \newlength{\mypartitlesep}
    \usepackage[explicit]{titlesec}
    \titlespacing{\paragraph}{0pt}{\mypartitlesep}{\mypartitlesep}
    
    \newlength{\mythmsep}
    \newtheoremstyle{mythmstyle}
      {\mythmsep} %
      {\mythmsep} %
      {\itshape} %
      {} %
      {\bfseries} %
      {.} %
      {.5em} %
      {} %
    
    \newtheoremstyle{mydefstyle}
      {\mythmsep} %
      {\mythmsep} %
      {} %
      {} %
      {\bfseries} %
      {.} %
      {.5em} %
      {} %
    
    \theoremstyle{mythmstyle}
        \newtheorem{theorem}{Theorem}
        \newtheorem{proposition}[theorem]{Proposition}

        \newtheorem*{fact*}{Fact}
    \theoremstyle{mydefstyle}
        \newtheorem{definition}{Definition}
        \newtheorem{problem}{Problem}
        \newtheorem{assumption}{Assumption}
        \newtheorem{remark}{Remark}
        \newtheorem{algr}[algorithm]{Algorithm}

    \newenvironment{proof}
        {\vspace{-0.9em}\begin{proof}}
        {\end{proof}\vspace{-0.4em}}

\else

    \theoremstyle{plain}
        \newtheorem{theorem}{Theorem}
        \newtheorem{proposition}{Proposition}
        \newtheorem{observation}{Observation}
        \newtheorem{algr}[algorithm]{Algorithm}
    \theoremstyle{definition}
        \newtheorem{definition}{Definition}

        \newtheorem{remark}{Remark}
        
    \usepackage{enumitem}
    \setlist[itemize]{leftmargin=\parindent}
    \setlist[enumerate]{leftmargin=\parindent}
    \setlist[description]{font=\bfseries,leftmargin=\parindent}

\fi

\newcommand{\Hm}{\mathsf{H}}

\newcommand{\homolog}{\sim}

\newcommand{\Real}{\mathbb{R}}

\newcommand{\fsimp}[2]{\sigma_{#2}}
\newcommand{\morph}[2]{\psi_{#2}}
\newcommand{\fmorph}[2]{\varphi_{#2}}

\renewcommand{\ker}{\mathrm{ker}}
\newcommand{\img}{\mathrm{img}}
\newcommand{\interior}{\mathsf{Int}}
\newcommand{\cof}{\mathsf{cof}}

\newcommand{\Pers}{\mathsf{PD}}

\newcommand{\bd}{\mathsf{bd}}

\renewcommand{\bar}[1]{\overline{#1}}

\newcommand{\inv}{^{-1}}

\newcommand{\lbarrowspace}{\;}

\let\leftrightarrowsp\lrarrowsp

\newcommand{\incto}{\hookrightarrow}
\newcommand{\inctosp}[1]{\xhookrightarrow{\lbarrowspace#1\lbarrowspace}}
\newcommand{\bakincto}{\hookleftarrow}
\newcommand{\bakinctosp}[1]{\xhookleftarrow{\lbarrowspace#1\lbarrowspace}}

\newcommand{\given}{\,|\,}
\newcommand{\Set}[1]{\{#1\}}
\newcommand{\bigSet}[1]{\big\{#1\big\}}
\newcommand{\bigCard}[1]{\big|#1\big|}
\newcommand{\sCommaT}{}
\newcommand{\vsimpset}[2][\Dim]{{\mathbb{K}}^#1_{(#2)}}

\let\wdhat\widehat
\let\wdtild\widetilde

\let\emptyset\varnothing
\let\intersect\cap
\let\intsec\intersect
\let\union\cup
\let\bigunion\bigcup

\newcommand{\Acal}{\mathcal{A}}
\newcommand{\Bcal}{\mathcal{B}}

\newcommand{\Dcal}{\mathcal{D}}

\newcommand{\Fcal}{\mathcal{F}}
\newcommand{\Ical}{\mathcal{I}}

\newcommand{\Mcal}{\mathcal{M}}

\newcommand{\Scal}{\mathcal{S}}

\newcommand{\Xcal}{\mathcal{X}}

\newcommand{\Zbb}{\mathbb{Z}}

\newcommand{\Fsfit}{\mathsfit{F}}

\newcommand{\aG}{\alpha}
\newcommand{\bG}{\beta}
\newcommand{\dG}{\delta}
\newcommand{\DG}{\Delta}

\newcommand{\iG}{\iota}

\newcommand{\lG}{\lambda}
\newcommand{\LG}{\Lambda}

\newcommand{\sG}{\sigma}
\newcommand{\SG}{\Sigma}
\newcommand{\tG}{\tau}
\newcommand{\thG}{\theta}
\newcommand{\thGvar}{\theta}
\newcommand{\zG}{\zeta}

\newcommand{\Dim}{p}
\newcommand{\diml}{q}
\newcommand{\birth}{b}
\newcommand{\death}{d}
\newcommand{\fbirth}{\beta}
\newcommand{\fdeath}{\delta}
\newcommand{\filtcnt}{r}
\newcommand{\lfiltcnt}{\ell}
\newcommand{\dgmcnt}{m}

\newcommand{\critval}{\alpha}
\newcommand{\pcritval}{{\alpha}^\Dim}
\newcommand{\critv}{v}
\newcommand{\pcritv}{v^\Dim}
\newcommand{\critvcnt}{n}
\newcommand{\pcritvcnt}{m}
\newcommand{\regval}{s}
\newcommand{\gsrc}{\mathscr{s}}
\newcommand{\gsink}{\mathscr{t}}
\newcommand{\simpset}{\DG}
\newcommand{\cplx}{\mathbb{K}}
\newcommand{\subsp}{\mathbb{X}}
\newcommand{\lcplx}{X}
\newcommand{\lvldgm}{\mathcal{L}}
\newcommand{\clvldgm}{\mathcal{L}^\mathsf{c}}
\newcommand{\lvlfilt}{\mathcal{F}}
\newcommand{\lfilt}{\Xcal}
\newcommand{\subfilt}{\Fsfit}
\newcommand{\pcritind}{\lG}
\newcommand{\dummyV}{\phi}
\newcommand{\comp}{\mathcal{C}}

\newcommand{\pcycind}{i}
\newcommand{\dgraphsec}{G}
\newcommand{\cyc}{z}

\makeatletter
\newcommand{\algmargin}{\the\ALG@thistlm}
\makeatother
\newlength{\whilewidth}
\algdef{SE}[parWHILE]{parWhile}{EndparWhile}[1]
  {\parbox[t]{\dimexpr\linewidth-\algmargin}{%
     \hangindent\whilewidth\strut\algorithmicwhile\ #1\ \algorithmicdo\strut}}{\algorithmicend\ \algorithmicwhile}%
\algdef{SE}[parIF]{parIf}{EndparIf}[1]
  {\parbox[t]{\dimexpr\linewidth-\algmargin}{%
     \hangindent\whilewidth\strut\algorithmicif\ #1\ \algorithmicthen\strut}}{\algorithmicend\ \algorithmicif}%
\algnewcommand{\parState}[1]{\State%
  \parbox[t]{\dimexpr\linewidth-\algmargin}{\strut #1\strut}}

\begin{document}

\newlength{\mydisplaymathsep}
\setlength{\mydisplaymathsep}{0.7em}
\setlength{\abovedisplayskip}{\mydisplaymathsep}
\setlength{\belowdisplayskip}{\mydisplaymathsep}
\setlength{\abovedisplayshortskip}{\mydisplaymathsep}
\setlength{\belowdisplayshortskip}{\mydisplaymathsep}

\title{Computing Optimal Persistent Cycles for Levelset Zigzag on Manifold-like Complexes}

\author{Tamal K. Dey\thanks{Department of Computer Science, Purdue University. 
\texttt{tamaldey@purdue.edu}}
\and Tao Hou\thanks{Department of Computer Science, University of Oregon. 
\texttt{taohou@uoregon.edu}}
\and Anirudh Pulavarthy\thanks{School of Computing, DePaul University. \texttt{apulavar@depaul.edu}}
}

\date{}

\maketitle
\thispagestyle{empty}

\newenvironment{myproof}{\begin{proof}}{\end{proof}}

\begin{abstract}
In standard persistent homology, a persistent cycle
born and dying with a persistence interval (bar)
associates the 
bar with a 
concrete topological representative,
which provides
means to effectively navigate back from the barcode to the topological space.
Among the possibly many, optimal persistent cycles bring forth further
information due to having guaranteed quality. However, 
topological features
usually go through variations in the lifecycle of a bar which
a single persistent cycle may not capture.
Hence, for persistent homology induced from PL functions,
we propose 
{\it levelset persistent cycles}
consisting of a sequence of cycles 
that depict the evolution of homological features from birth to death.
Our definition is based on levelset zigzag persistence 
which involves four types of persistence intervals as opposed to the two
types in standard persistence. For each of the four types,
we present a polynomial-time algorithm computing 
an optimal sequence of levelset persistent $p$-cycles
for the so-called {\it weak $(p+1)$-pseudomanifolds}.
Given that optimal cycle problems for homology are NP-hard in general,
our results are useful in practice because 
weak pseudomanifolds do appear in applications. Our algorithms draw upon an idea
of relating optimal cycles to min-cuts in a graph that was 
exploited earlier for standard persistent cycles. Notice that levelset zigzag
poses non-trivial challenges for the approach because a sequence of
optimal cycles instead of a single one needs to be computed
in this case.
We show some empirical evidence that optimal cycles produced by our implemented software have nice quality.
\end{abstract}

\newpage
\setcounter{page}{1}

\section{Introduction}

\begin{figure}
  \centering
  \includegraphics[width=\linewidth]{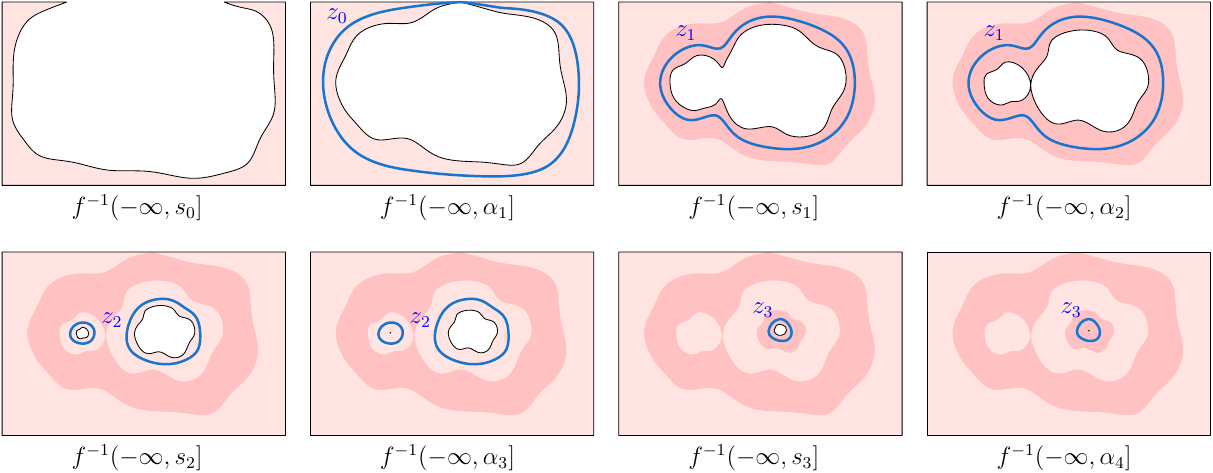}
  
  \opt{DCG}{\vspace{10pt}}
  \caption{Evolution of a homological feature across different critical points.}
  \label{fig:motiv}
\end{figure}

Given a filtered topological space,
persistent homology~\cite{edelsbrunner2000topological} 
produces a stable~\cite{cohen2007stability} topological signature called {\it barcode} (or {\it persistence diagram})
which has proven useful in many applications.
Though being widely adopted,
a persistence interval in a barcode only indicates that
a certain topological feature gets born and dies with the interval
but does not provide a canonical and concrete representative of the feature.
In view of this,
{\it persistent cycles}~\cite{de2011dualities,dey2019persistent,obayashi2018volume} were proposed
as concrete representatives
for standard (i.e., non-zigzag) persistent homology,
which also enables one to navigate back to the topological space
from a barcode.
Among the many,
optimal persistent cycles (or ones with a quality measure)~\cite{dey2019persistent,dey2020computing,obayashi2018volume,wu2017optimal}
are of special interest  for applications in different domains~\cite{wu2017optimal,DBLP:journals/tvcg/Iuricich22,strommen2023topological} due to having guaranteed quality.
However,
one drawback of standard persistent cycles
is that
only a single cycle born at the start
is used,
while homological features may vary continuously inside an interval.
For example, in \Cref{fig:motiv}, 
let the growing space be the {\it sub-levelset filtration}
of a real-valued function $f$, in which $\critval_1,\ldots,\critval_4$
are consecutive critical values and $\regval_0,\ldots,\regval_3$
are regular values in between.
If we consider the changes of homology after each critical point,
then a non-trivial 1-cycle $z_0$ is first born in $f\inv(-\infty,\critval_1]$
and splits into two in $f\inv(-\infty,\regval_2]$.
The two separate cycles eventually shrink and die independently,
generating a (standard) persistence interval $[\critval_1,\critval_4)$.
Using standard persistent cycles~\cite{dey2020computing,obayashi2018volume}, only $z_0$ would be picked
as a representative for $[\critval_1,\critval_4)$, 
which fails to depict the subsequent behaviors.

In this paper, 
we propose alternative persistent cycles capturing the dynamic behavior
shown in \Cref{fig:motiv}.
We focus on a special but important type of persistent homology --
those generated by piecewise linear (PL) functions~\cite{edelsbrunner2010computational}.
We also base our definition on an extension of standard persistence
called the {\it levelset zigzag} persistence~\cite{carlsson2009zigzag-realvalue},
which tracks the survival of homological features
at and in between the critical points.
Given a persistence interval from levelset zigzag,
we define a {\it sequence} of cycles
called {\it levelset persistent cycles}
so that
there is a cycle 
between each consecutive critical points
within the interval.
For example, in \Cref{fig:motiv}, 
$[\critval_1,\critval_4)$ is also a persistence interval (i.e., a {\it closed-open} interval~\cite{carlsson2009zigzag-realvalue})
in the levelset zigzag of $f$.
The cycles $z_0,z_1,z_2,z_3$ forming a sequence of levelset persistent {1-cycles}
for $[\critval_1,\critval_4)$
capture all the variations across the critical points.
Section~\ref{sec:prob-stat} details the definition.

Levelset zigzag on a PL function relates to
the standard sub-levelset version in the following way: 
{\it finite} intervals from
the sub-levelset  version
on the original function and its negation 
produce {\it closed-open} and {\it open-closed} intervals 
in levelset zigzag,
while levelset zigzag additionally provides
{\it closed-closed} and {\it open-open} intervals~\cite{carlsson2009zigzag-realvalue}.
Thus,
levelset persistent cycles are oriented toward richer types of intervals
(see also  extended persistence~\cite{cohen2009extending}).

Computationally, optimal cycle problems for homology
in both persistence and 
non-persis\-tence settings are
NP-hard in general~\cite{chambers2009minimum,chen2011hardness,dey2019persistent,dey2020computing}.
Other than the optimal homology basis algorithms
in dimension one~\cite{busaryev2012annotating,dey2018efficient,dey2010approximating},
to our knowledge,
all polynomial-time algorithms for such problems
aim at manifolds or manifold-like 
complexes~\cite{borradaile2017minimum,chambers2009minimum,chen2011hardness,dey2020computing,erickson2005greedy}.
In particular, the existing algorithms
for \emph{general dimensions}~\cite{chen2011hardness,dey2020computing} 
exploit the dual graph structure of given complexes
and reduce the optimal cycle problem in codimension one
to a minimum cut problem.
In this paper,
we find a way of applying this technique
to computing an \emph{optimal sequence} of levelset persistent cycles -- 
one that has the minimum {\it sum} of weight.
Our approach which also works for general dimensions
differs from previous ones
to account for the fact that 
a sequence of optimal cycles instead of a single one need to be computed.
We assume the input to be a generalization of 
$(\Dim+1)$-manifold 
called {\it weak $(\Dim+1)$-pseudomanifold}~\cite{dey2020computing}:

\opt{DCG}{\medskip}
\begin{definition}
A weak $(\Dim+1)$-pseudomanifold
is a simplicial complex
in which each $\Dim$-simplex has
no more than two $(\Dim+1)$-cofaces.
\end{definition}
\opt{DCG}{\medskip}

Given an arbitrary PL function on a weak $(\Dim+1)$-pseudomanifold ($\Dim\geq 1$),
we show that 
an optimal sequence of levelset persistent $\Dim$-cycles
can be computed in polynomial time
for {\it any} type of levelset zigzag intervals of dimension $\Dim$.
This is in contrast to the standard persistence setting, 
where 
computing optimal persistent $\Dim$-cycles for 
one type of intervals (the {\it infinite} intervals)
is NP-hard even for weak $(\Dim+1)$-pseudo\-manifolds~\cite{dey2020computing}.
Notice that among the four mentioned types of intervals
in levelset zigzag,
closed-open and open-closed intervals are symmetric so that
everything concerning open-closed intervals can be derived directly
from the closed-open case.
Hence, for these two types of intervals,
we address everything \emph{only for the closed-open case}.

We propose three algorithms for the three types of intervals by utilizing 
minimum \sCommaT{}cuts on the dual graphs.
Specifically, levelset persistent $\Dim$-cycles for an open-open interval have 
direct correspondence to \sCommaT{}cuts on a dual graph,
and so the optimal ones can be computed directly from the minimum \sCommaT{}cut.
For the remaining cases, the crux is to deal with the so-called ``monkey saddles''
and the computation spans two phases.
The first phase computes minimum $\Dim$-cycles
in certain components of the complex;
then, using minimum cuts, the second phase determines the optimal combination
of the components by introducing some {\it augmenting} edges.
All three algorithms run in $O(n^2)$ time dominated by the complexity
of the minimum cut computation, for which we use
Orlin's max-flow algorithm~\cite{orlin2013max}.
Section~\ref{sec:compute} details the computation.

We note that there have been
recent progresses made on computing representatives for 
zigzag persistence~\cite{DBLP:conf/soda/Dey0M25}. 
However, the work~\cite{DBLP:conf/soda/Dey0M25} only concerns computing an arbitrary 
representative for a zigzag interval.
The optimal representative problem for zigzag persistence
appears to be more  complicated due to its nature (e.g., a sequence of optimal cycles
need to be defined and computed).
To our knowledge, our work is the first to address the problem in the zigzag setting.

We also implemented our proposed algorithms
(available online at: \url{https://github.com/taohou01/LvlsetPersCyc})
and performed experiments on triangular meshes.
The computed optimal cycles show nice quality while
capturing the variations of the topological features inside a persistence interval.
See \Cref{sec:exp} for details.

\section{Preliminaries}\label{sec:prelim}

\subsection{Simplicial homology}
We only briefly review simplicial homology here;
see~\cite{edelsbrunner2010computational}
for a detailed treatment.
Let $K$ be a simplicial complex.
Since coefficients for homology are in $\Zbb_2$ in this paper,
a {\it $\Dim$-chain} $c$ of $K$ is a {\it set} of $\Dim$-simplices of $K$
and can also be expressed as the formal sum $\sum_{\sG\in c}\sG$;
these two forms of $\Dim$-chains 
are used interchangeably.
The sum of two $\Dim$-chains
is the symmetric difference of sets
and is denoted as both ``$+$'' and ``$-$''
because plus and minus are the same in $\Zbb_2$.
A {\it $\Dim$-cycle} is a $\Dim$-chain 
in which any $(\Dim-1)$-face
adjoins even number of $\Dim$-simplices;
a {\it $\Dim$-boundary} is a $\Dim$-cycle
being the boundary of a $(\Dim+1)$-chain.
Two $\Dim$-cycles $\zG$, $\zG'$ are {\it homologous},
denoted $\zG\homolog\zG'$,
if their sum is a $\Dim$-boundary.
The set of all $\Dim$-cycles homologous to a fixed $\Dim$-cycle $\zG\subseteq K$
forms a {\it homology class} $[\zG]$,
and all these homology classes form 
the $\Dim$-th {\it homology group} $\Hm_\Dim(K)$ of $K$.
Note that $\Hm_\Dim(K)$ is a vector space over $\Zbb_2$.

\subsection{Zigzag modules, barcodes, and filtrations}\label{sec:prelim-zigzag}
A {\it zigzag module}~\cite{carlsson2010zigzag} (or {\it module} for short)
is a sequence of vector spaces 
\[\Mcal: V_0 
\leftrightarrow
V_1 
\leftrightarrow
\cdots
\leftrightarrow
V_\dgmcnt\]
in which
each $V_i\leftrightarrow V_{i+1}$ is a linear map and is either forward, i.e., $V_i\to V_{i+1}$,
or backward, i.e., $V_i\leftarrow V_{i+1}$.
In this paper, vector spaces are taken over $\Zbb_2$.
A module
$\Scal:W_0 
\leftrightarrow
W_1 
\leftrightarrow
\cdots 
\leftrightarrow
W_\dgmcnt$
is called a {\it submodule} of $\Mcal$ if each $W_i$ is a subspace of $V_i$ and
each map $W_i\leftrightarrow W_{i+1}$ 
is the restriction of $V_i\leftrightarrow V_{i+1}$.
For an interval $[b,d]\subseteq[0,\dgmcnt]$,
$\Scal$ is called an {\it interval submodule} of $\Mcal$ over $[b,d]$
if $W_i$
is one-dimensional for $i\in[\birth,\death]$
and is trivial for $i\not\in[\birth,\death]$,
and $W_i\leftrightarrow W_{i+1}$  is an isomorphism for $i\in[\birth,\death-1]$.
By the Krull-Schmidt principle and Gabriel's theorem~\cite{carlsson2010zigzag},
$\Mcal$ admits 
an {\it interval decomposition},
$\Mcal=\bigoplus_{k\in\LG}\Ical^{[\birth_k,\death_k]}$,
in which each $\Ical^{[\birth_k,\death_k]}$
is an interval submodule of $\Mcal$ over $[\birth_k,\death_k]$.
We call the (multi-)set of intervals
\[\Set{[\birth_k,\death_k]\given k\in\LG}\]
the {\it zigzag barcode} (or {\it barcode} for short) of $\Mcal$,
and denote it as $\Pers(\Mcal)$.
Each interval in a zigzag barcode is called a {\it persistence interval}.

A {\it zigzag filtration} (or {\it filtration} for short)
is a sequence of simplicial complexes or general topological spaces
\[\Xcal: X_0 \leftrightarrow X_1 \leftrightarrow 
\cdots \leftrightarrow X_\dgmcnt\]
in which each
$X_i\leftrightarrow X_{i+1}$ is either a forward inclusion $X_i\incto X_{i+1}$
or a backward inclusion $X_i\bakincto X_{i+1}$.
If not mentioned otherwise, a zigzag filtration is always assumed to be 
a sequence of simplicial complexes.
Applying the $\Dim$-th homology functor 
with $\Zbb_2$ coefficients,
we have the {\it $\Dim$-th zigzag module} of $\Xcal$:
\[\Hm_\Dim(\Xcal): 
\Hm_\Dim(X_0) 
\leftrightarrow
\Hm_\Dim(X_1) 
\leftrightarrow
\cdots 
\leftrightarrow
\Hm_\Dim(X_\dgmcnt) \]
in which
each $\Hm_\Dim(X_i)\leftrightarrow \Hm_\Dim(X_{i+1})$
is the linear map induced by inclusion.
The barcode of $\Hm_\Dim(\Xcal)$ is also called the {\it$\Dim$-th zigzag barcode}
of $\Xcal$ and is alternatively denoted as $\Pers_\Dim(\Xcal):=\Pers(\Hm_\Dim(\Xcal))$,
where each interval in $\Pers_\Dim(\Xcal)$ is called a {\it $\Dim$-th persistence interval}.
For an interval $[\birth,\death]\in\Pers_\Dim(\Xcal)$, we also conveniently denote
the interval as $[X_\birth,X_\death]\in\Pers_\Dim(\Xcal)$, i.e.,
by its starting and ending spaces.
This is helpful when a filtration is not naturally indexed
by consecutive integers, as seen in Section~\ref{sec:prob-stat}.
In this case, an element $X_i\in[X_\birth,X_\death]$ is just a space in $\Xcal$
with $\birth\leq i\leq\death$.

A special type of filtration
called {\it simplex-wise} filtration
is frequently used in this paper,
in which 
each forward (resp. backward) inclusion 
is an addition (resp. deletion) of a single simplex. 
Any $\Dim$-th zigzag module induced by a simplex-wise filtration
has the property of being {\it elementary},
meaning that all linear maps in the module are of the three forms:
(\text{i}) an isomorphism; 
(\text{ii}) an injection with rank 1 cokernel; 
(\text{iii}) a surjection with rank 1 kernel.
This property
is useful for
the definitions and computations presented later.

\subsection{Graph \sCommaT{}cuts}
For a  graph $G=(V(G),E(G))$  with  a weight function $w:E(G)\to[0,\infty]$,
let $\gsrc$ be a set of {\it sources} 
and $\gsink$ be a set of {\it sinks} 
which are two disjoint non-empty subsets of $V(G)$.
A {\it cut} $(S,T)$ of the tuple $(G,\gsrc,\gsink)$ consists of two sets
such that $S\intsec T=\emptyset$, $S\union T=V(G)$,
$\gsrc\subseteq S$, and $\gsink\subseteq T$.
Define $E(S,T)$ as the set of
all edges of $G$
connecting a vertex in $S$ and a vertex in $T$,
in which each edge is said to {\it cross} the cut $(S,T)$.
The weight of the cut is defined as 
$w(S,T)=\sum_{e\in E(S,T)}w(e)$. 
The {\it minimum \sCommaT{}cut} of $(G,\gsrc,\gsink)$
is a \sCommaT{}cut with the minimum weight.

\subsection{Dual graphs for manifolds}
A manifold-like complex (e.g., a weak pseudomanifold)
often has an undirected dual graph structure,
which is utilized extensively in this paper. 
Let the complex be $(\Dim+1)$-dimensional.
Then, each $(\Dim+1)$-simplex is dual to a vertex
and each $\Dim$-simplex is dual to an edge
in the dual graph.
For a $\Dim$-simplex
with two $(\Dim+1)$-cofaces $\tG_1$ and $\tG_2$,
its dual edge connects the vertex
dual to $\tG_1$ and the vertex dual to $\tG_2$.
For a $\Dim$-simplex of other cases, 
its dual edge is problem-specific 
and is explained in the corresponding paragraphs.

\section{Problem statement}
\label{sec:prob-stat}

In this section, we develop the definitions for levelset persistent cycles
and the optimal ones. 
Levelset persistent cycles are sometimes simply called {\it persistent cycles}
for brevity,
which should not cause  confusions.
We begin the section by defining levelset zigzag persistence in Section~\ref{sec:p-levelzz},
where we present an alternative version of the classical one proposed 
by Carlsson et al.~\cite{carlsson2009zigzag-realvalue}.
Adopting this alternative version enables us to focus on
critical values (and the changes incurred) in a specific dimension.
Section~\ref{sec:p-levelzz} also defines a simplex-wise levelset filtration, 
which provides an elementary view of levelset zigzag and
is helpful to our subsequent definition and computation.

Section~\ref{sec:pers-cyc-dfn} details the definition of levelset persistent cycles.
The cycles in the middle of the sequence 
are the same for all types of intervals,
while the cycles for the endpoints
differ according to the types of ends.

Finally, in Section~\ref{sec:lvlset-compat}, we address an issue left over 
from Section~\ref{sec:p-levelzz}, which is the validity of
the discrete levelset filtration.
The validity is found to be relying
on the triangulation representing 
the underlying shape.
We also argue that 
the triangulation has to be fine enough in order
to obtain accurate depictions of persistence intervals by levelset
persistent cycles.

\subsection{$\Dim$-th levelset zigzag persistence}
\label{sec:p-levelzz}

Throughout the section, 
let $\Dim\geq 1$,
$K$ be a finite simplicial complex with underlying space $X=|K|$,
and $f:X\to\Real$ be a PL function~\cite{edelsbrunner2010computational}
derived by interpolating values on vertices.
We consider PL functions that are {\it generic}, 
i.e., having distinct values on the vertices.
Notice that the function values can be slightly perturbed to satisfy this
if they are not initially.
An open interval $I\subseteq\Real$ is called {\it regular} if
there exist a topological space $Y$ and a homeomorphism 
\[\Phi:Y\times I\to f\inv(I)\]
such that $f\circ\Phi$ is the projection onto $I$
and $\Phi$ extends to a continuous function $\bar{\Phi}:Y\times \bar{I}\to f\inv\big(\bar{I}\big)$
with $\bar{I}$ being the closure of $I$~\cite{carlsson2009zigzag-realvalue}.
It is known that $f$ is of {\it Morse type}~\cite{carlsson2009zigzag-realvalue},
meaning that
each levelset $f\inv(s)$ has finitely generated homology,
and there are finitely many {\it critical values}
\[\critval_0=-\infty<\critval_1<\cdots<\critval_\critvcnt<\critval_{\critvcnt+1}=\infty\]
such that each interval $(\critval_i,\critval_{i+1})$
is regular.
Notice that critical values of $f$ can only be function values
of $K$'s vertices.

\begin{figure}
  \centering
  \opt{arxiv}{\includegraphics[width=0.65\linewidth]{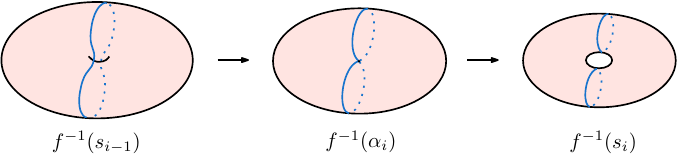}}
  \opt{DCG}{\includegraphics[width=0.8\linewidth]{fig/squeeze}\bigskip}
  \caption{A critical value 
  $\critval_i$
  across which the 0th and 2nd homology stays the same;
  $f$ is defined on a 3D domain
  and $\regval_{i-1}$, $\regval_{i}$ are two regular values 
  with $\regval_{i-1}<\critval_i<\regval_{i}$.
  The levelset $f\inv(\regval_{i-1})$ is a 2-sphere where two antipodal points
  are getting close and eventually pinch in $f\inv(\critval_{i})$.
  Crossing the critical value, $f\inv(\regval_{i})$ becomes a torus.}
  \label{fig:squeeze}
\end{figure}

As mentioned, levelset persistent cycles 
for a $\Dim$-th interval 
should capture the changes of $\Dim$-th homology
across different critical values.
However, some critical values may cause no change to the 
$\Dim$-th homology. 
\Cref{fig:squeeze} illustrates such a critical value 
around which
only the 1st homology changes and the 0th and 2nd homology stays the same.
Thus, to capture
the most essential variation, the persistent $\Dim$-cycles
should stay the same across such critical values. 
The following definition
characterizes those critical values that we are interested in:

\opt{DCG}{\medskip}
\begin{definition}[$\Dim$-th homologically critical value]
\label{dfn:homo-critp}
A~critical value $\critval_i\neq-\infty,\infty$ of $f$ 
is called {\it$\Dim$-th homologically critical}
(or {\it$\Dim$-th critical} for short)
if one of the two linear maps induced by inclusion is not an isomorphism: 
\begin{align*}
&\;\Hm_\Dim\big(f\inv{(\critval_{i-1},\critval_i)}\big)\to
\Hm_\Dim\big(f\inv{(\critval_{i-1},\critval_{i+1})}\big),
\\
&\;\Hm_\Dim\big(f\inv{(\critval_{i-1},\critval_{i+1})}\big)
\leftarrow\Hm_\Dim\big(f\inv{(\critval_i,\critval_{i+1})}\big).
\end{align*}
For convenience, we also let $-\infty,\infty$ be $\Dim$-th critical.
Moreover,
a vertex $v$ of $K$ is {\it $\Dim$-th critical}
if $f(v)$ is a 
$\Dim$-th critical.
\end{definition}
\opt{DCG}{\medskip}
\begin{remark}
By inspecting the (classical) levelset barcode~\cite{carlsson2009zigzag-realvalue} of $f$
(see also Section~\ref{sec:levelzz}),
it can be easily determined whether a critical value is $\Dim$-th critical.
\end{remark}
\opt{DCG}{\medskip}

Throughout this section,
let
\[\pcritval_0=-\infty<\pcritval_{1}<\cdots<\pcritval_{\pcritvcnt}<\pcritval_{\pcritvcnt+1}=\infty\]
denote all the 
$\Dim$-th homologically critical values of $f$,
and $\pcritv_{1},\ldots,\pcritv_{\pcritvcnt}$ denote the corresponding 
$\Dim$-th critical vertices.

\begin{figure}
  \centering
  \opt{arxiv}{\includegraphics[width=0.9\linewidth]{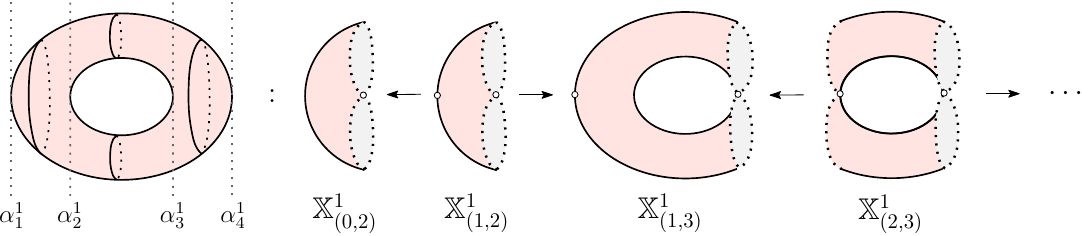}}
  \opt{DCG}{\includegraphics[width=\linewidth]{fig/torus}\bigskip}
  \caption{A torus with the height function $f$ taken over the horizontal line.
  The 1st levelset barcode is
  $\bigSet{\big(\critval^1_1,\critval^1_4\big),\big[\critval^1_2,\critval^1_3\big]}$. 
  We list the first half of $\clvldgm_1(f)$
  but excluding $\subsp^1_{(0,1)}=\emptyset$;
  the remaining half is symmetric.
  An empty dot indicates the point is not included in the space.}
  \label{fig:torus}
\end{figure}

\opt{DCG}{\medskip}
\begin{definition}[$\Dim$-th levelset zigzag]
\label{dfn:p-lvlpers}
Denote $f\inv(\pcritval_i,\pcritval_j)$ as $\subsp^\Dim_{(i,j)}$
for any $i<j$.
The continuous version of {\it $\Dim$-th levelset filtration} 
of $f$, denoted $\clvldgm_\Dim(f)$, 
is defined as
\[
\clvldgm_\Dim(f): 
\subsp^\Dim_{(0,1)} \incto 
\subsp^\Dim_{({0},{2})}\bakincto \subsp^\Dim_{(1,2)} \incto 
\subsp^\Dim_{(1,3)} \bakincto 
\cdots
\incto \subsp^\Dim_{({\pcritvcnt-1},{\pcritvcnt+1})}
\bakincto \subsp^\Dim_{({\pcritvcnt},{\pcritvcnt+1})}.
\]
The barcode $\Pers_\Dim(\clvldgm_\Dim(f))$ is called the
$\Dim$-th {\it levelset barcode} of $f$,
in which each interval 
is called
a $\Dim$-th {\it levelset persistence interval} of $f$.
\end{definition}
\opt{DCG}{\medskip}
\begin{remark}
Notice that we generally do not consider the barcode 
$\Pers_q(\clvldgm_\Dim(f))$ where $q\neq \Dim$
for a $\Dim$-th levelset filtration $\clvldgm_\Dim(f)$.
\end{remark}
\opt{DCG}{\medskip}
\begin{remark}
See \Cref{fig:torus} for an example of $\clvldgm_1(f)$
and its 1st levelset barcode.
\end{remark}
\opt{DCG}{\medskip}

We postpone the justification of Definition~\ref{dfn:p-lvlpers}
to Section~\ref{sec:equiv-p-levelzz}, 
where we prove that the $\Dim$-th levelset barcode in Definition~\ref{dfn:p-lvlpers}
is equivalent to the classical one
defined in~\cite{carlsson2009zigzag-realvalue}.
In $\clvldgm_\Dim(f)$, 
$\subsp^\Dim_{({i},{i+1})}$ is called 
a {\it $\Dim$-th regular subspace}, and
a homological feature in $\Hm_\Dim(\subsp^\Dim_{({i},{i+1})})$
is considered to be \emph{alive} in the entire real-value interval $\big(\pcritval_{i},\pcritval_{{i+1}}\big)$;
$\subsp^\Dim_{({i-1},{i+1})}$ is called 
a {\it $\Dim$-th critical subspace},
and a homological feature in $\Hm_\Dim(\subsp^\Dim_{({i-1},{i+1})})$
is considered to be alive at the critical value $\pcritval_{i}$.
Intervals in $\Pers_\Dim(\clvldgm_\Dim(f))$ can then be mapped to
real-value intervals in which the homological features persist,
and are classified into four types based on the open and closeness
of the ends; see Table~\ref{tab:p-intv-map}.
From now on, levelset persistence intervals 
can be of the two forms shown in Table~\ref{tab:p-intv-map},
which we consider as interchangeable.

\renewcommand{\arraystretch}{1.6}
\begin{table}[ht]
\normalsize
\centering
\begin{tabular}{ l l c l }
 \toprule
 \textbf{closed-open:}   & $\big[\subsp^\Dim_{({\birth-1},{\birth+1})},
 \subsp^\Dim_{({\death-1},\death)}\big]$ & 
 $\Leftrightarrow$ & $\big[\pcritval_{\birth},\pcritval_{\death}\big)$ \\

 \textbf{open-closed:}   & $\big[\subsp^\Dim_{(\birth,{\birth+1})},
 \subsp^\Dim_{({\death-1},{\death+1})}\big]$ & 
 $\Leftrightarrow$ & $\big(\pcritval_{\birth},\pcritval_{\death}\big]$ \\

 \textbf{closed-closed:} & $\big[\subsp^\Dim_{({\birth-1},{\birth+1})},
 \subsp^\Dim_{({\death-1},{\death+1})}\big]$ & 
 $\Leftrightarrow$ & $\big[\pcritval_{\birth},\pcritval_{\death}\big]$ \\

 \textbf{open-open:}     & $\big[\subsp^\Dim_{(\birth,{\birth+1})},
 \subsp^\Dim_{({\death-1},\death)}\big]$ & 
 $\Leftrightarrow$ & $\big(\pcritval_{\birth},\pcritval_{\death}\big)$ \\[4pt]
  \bottomrule
\end{tabular}
\opt{DCG}{\medskip}
\caption{Four types of intervals in $\Pers_\Dim(\clvldgm_\Dim(f))$ and their mapping to real-value intervals.}
\label{tab:p-intv-map}
\end{table}

\paragraph{Discrete version.}
Since the optimal persistent cycles can only be 
computed on the discrete domain $K$, 
we provide a discrete version of our construction.
First, let the subcomplex $\cplx^\Dim_{(i,j)}$ of $K$ denote
the discrete version of $\subsp^\Dim_{(i,j)}$:
\begin{equation}\label{eqn:dgm-cplx-dfn}
\cplx^\Dim_{(i,j)}:=\bigSet{\sG\in K\given 
\forall\,v\in \sG, f(v)\in \big(\pcritval_i,\pcritval_j\big)}.
\end{equation}
We also define $\cplx^\Dim_{[i,j)}$ and $\cplx^\Dim_{(i,j]}$ similarly,
in which $f(v)$ in Equation~(\ref{eqn:dgm-cplx-dfn}) belongs to 
$\big[\pcritval_i,\pcritval_j\big)$ and $\big(\pcritval_i,\pcritval_j\big]$
respectively.
Then, the {\it discrete version}
of $\clvldgm_\Dim(f)$, denoted $\lvldgm_\Dim(f)$, is defined as
\begin{equation*}
\label{eqn:lvldgm}
\lvldgm_\Dim(f): 
\cplx^\Dim_{(0,1)} \incto 
\cplx^\Dim_{({0},{2})}\bakincto \cplx^\Dim_{(1,2)} \incto 
\cplx^\Dim_{(1,3)} \bakincto 
\cdots
\incto \cplx^\Dim_{({\pcritvcnt-1},{\pcritvcnt+1})}
\bakincto \cplx^\Dim_{({\pcritvcnt},{\pcritvcnt+1})}.
\end{equation*}
In $\lvldgm_\Dim(f)$,
$\cplx^\Dim_{({i},{i+1})}$ is called 
a {\it $\Dim$-th regular complex} and
$\cplx^\Dim_{({i-1},{i+1})}$ is called 
a {\it $\Dim$-th critical complex}.
At this moment,
we assume that $\subsp^\Dim_{(i,j)}$ deformation retracts
to $\cplx^\Dim_{(i,j)}$ whenever $i<j$, and hence
$\clvldgm_\Dim(f)$ 
and $\lvldgm_\Dim(f)$ 
are equivalent.
We discuss this assumption in detail in Section~\ref{sec:disc-p-dgm}.

\paragraph{Simplex-wise levelset filtration.}

For defining and computing levelset persistent cycles,
besides the filtration $\lvldgm_\Dim(f)$,
we also work on a simplex-wise version expanding $\lvldgm_\Dim(f)$.
We do this to harness the property that a simplex-wise filtration 
induces an elementary $\Dim$-th module (see Section~\ref{sec:prelim-zigzag}),
which eliminates ambiguities in definitions and computations.

\opt{DCG}{\medskip}
\begin{definition}[Simplex-wise levelset filtration]
\label{dfn:lvlfilt}
For the PL function $f$,
the $\Dim$-th {\it simplex-wise levelset filtration} of $f$,
denoted $\lvlfilt_\Dim(f)$,
is derived from $\lvldgm_\Dim(f)$ by expanding
each forward (resp. backward) inclusion in $\lvldgm_\Dim(f)$ into
a sequence of additions (resp. deletions) of a single simplex.
We also let the additions and deletions follow the order of the function values:
\begin{itemize}
    \item For the forward inclusion 
    $\cplx^\Dim_{({i},{i+1})} \incto \cplx^\Dim_{({i},{i+2})}$ in $\lvldgm_\Dim(f)$,
let 
$u_1=\pcritv_{i+1},u_2,\ldots,\allowbreak u_k$
be all the vertices 
with function values in $\big[\pcritval_{i+1},\pcritval_{i+2}\big)$
such that $f(u_1)<f(u_2)<\cdots<f(u_k)$.
Then, the lower stars~\cite{edelsbrunner2010computational} 
of $u_1,\ldots,u_{k}$ are added by $\lvlfilt_\Dim(f)$ following the order.

\opt{DCG}{\medskip}
    \item Symmetrically, for the backward inclusion 
$\cplx^\Dim_{({i},{i+2})}\bakincto \cplx^\Dim_{({i+1},{i+2})}$ in $\lvldgm_\Dim(f)$,
let
$ 
u_1,u_2,\ldots,u_k=\pcritv_{i+1}
$
be all the vertices 
with function values in $\big(\pcritval_{i},\pcritval_{i+1}\big]$
such that $f(u_1)<f(u_2)<\cdots<f(u_k)$.
Then, the upper stars of $u_1,\ldots,u_{k}$ are deleted 
by $\lvlfilt_\Dim(f)$ following the order.
\end{itemize}
Note that for each $u_j\in\Set{u_1,\ldots,u_{k}}$, 
we add (resp. delete) simplices
inside the lower (resp. upper) star of $u_j$ in any order maintaining the condition
of a filtration.
\end{definition}
\opt{DCG}{\medskip}

In this paper, 
we  {\it fix} an
$\lvlfilt_\Dim(f)$ 
derived from $\lvldgm_\Dim(f)$. Moreover, $\lvlfilt_\Dim(f)$  is assumed to be of the form 
\[\lvlfilt_\Dim(f): K_0\leftrightarrowsp{\fsimp{}{0}} K_1\leftrightarrowsp{\fsimp{}{1}}\cdots\leftrightarrowsp{\fsimp{}{\filtcnt-1}} K_\filtcnt\]
where each $K_i$, $K_{i+1}$ differ by
a simplex denoted $\fsimp{}{i}$
and each linear map 
in $\Hm_\Dim(\lvlfilt_\Dim(f))$ 
is denoted as
$\fmorph{\Dim}{i}:\Hm_\Dim(K_i)\leftrightarrow\Hm_\Dim(K_{i+1})$.
Notice that each complex in $\lvldgm_\Dim(f)$
equals a $K_j$ in $\lvlfilt_\Dim(f)$,
and specifically, $K_0=\cplx^\Dim_{(0,1)}$, 
$K_\filtcnt=\cplx^\Dim_{({\pcritvcnt},{\pcritvcnt+1})}$.

\paragraph{Simplex-wise intervals.}
By the property of zigzag persistence,
any interval $J$ in $\Pers_\Dim(\lvldgm_\Dim(f))$ 
can be considered to be produced 
by an interval $J'$ in $\Pers_\Dim(\lvlfilt_\Dim(f))$,
and we call $J'$ 
the {\it simplex-wise interval} 
of $J$.
The mapping of intervals of $\Pers_\Dim(\lvlfilt_\Dim(f))$
to those of $\Pers_\Dim(\lvldgm_\Dim(f))$
has the following rule:

\opt{DCG}{\medskip}
\vspace{\topsep}
\noindent\textit{For any 
$[K_\fbirth,K_\fdeath]\in\Pers_\Dim(\lvlfilt_\Dim(f))$,
let 
$\subfilt^{[\fbirth,\fdeath]}:K_\fbirth\leftrightarrow 
K_{\fbirth+1}\leftrightarrow\cdots\leftrightarrow K_\fdeath$ 
be the part of $\lvlfilt_\Dim(f)$ between $K_\fbirth$ and $K_\fdeath$,
and
let $\cplx^\Dim_{(\birth,\birth')}$ and $\cplx^\Dim_{(\death,\death')}$ 
respectively be the first and last complex from $\lvldgm_\Dim(f)$
which appear in $\subfilt^{[\fbirth,\fdeath]}$.
Then, $[K_\fbirth,K_\fdeath]$ produces an interval 
$\big[\cplx^\Dim_{(\birth,\birth')},\cplx^\Dim_{(\death,\death')}\big]$
for $\Pers_\Dim(\lvldgm_\Dim(f))$.
Moreover, if $\subfilt^{[\fbirth,\fdeath]}$ contains no complexes from $\lvldgm_\Dim(f)$,
then $[K_\fbirth,K_\fdeath]$ does not produce any levelset persistence interval
in $\Pers_\Dim(\lvldgm_\Dim(f))$;
such an interval in $\Pers_\Dim(\lvlfilt_\Dim(f))$ is called 
{\rm trivial}.}
\vspace{\topsep}
\opt{DCG}{\medskip}

As can be seen later,
any levelset persistent cycles in this paper are defined on 
{\it both} a levelset persistence interval 
and its simplex-wise interval.
We further notice that persistent cycles for 
trivial intervals in $\Pers_\Dim(\lvlfilt_\Dim(f))$
are exactly the same as standard persistent cycles,
and we refer to~\cite{dey2020computing} for their definition and computation.

\subsection{Definition of levelset persistent cycles}
\label{sec:pers-cyc-dfn}

Representatives for the general zigzag persistence~\cite{maria2014zigzag,DBLP:conf/soda/Dey0M25}
are defined based on the following principle:
for a persistence interval $J$ of a zigzag module $\Mcal$,
its representative should
generate an interval submodule over $J$ so that all such interval submodules
form the interval decomposition of $\Mcal$~\cite{BendichPaul2013Haro};
see also  \Cref{dfn:rep-cyc} in \Cref{sec:conn-to-interv-decomp}.
In this subsection,
we define the levelset persistent cycles
by adapting the general zigzag representatives
following the same principle.
We also explain in detail
the meaning of each aspect of the representative definition in our setting.
We postpone to
\Cref{sec:conn-to-interv-decomp}
the formal justification that
the levelset persistent cycles generate interval submodules
in the interval decompositions for $\Hm_\Dim(\lvldgm_\Dim(f))$ and $\Hm_\Dim(\lvlfilt_\Dim(f))$.

Consider 
a levelset persistence interval in $\Pers_\Dim(\lvldgm_\Dim(f))$
with endpoints $\pcritval_{\birth}$, $\pcritval_{\death}$
produced by a simplex-wise interval
$[K_\fbirth,K_\fdeath]\in\Pers_\Dim(\lvlfilt_\Dim(f))$.
The levelset persistence interval can also be denoted as 
$\big[\cplx^\Dim_{(\birth',{\birth+1})},
\cplx^\Dim_{({\death-1},\death')}\big]$,
where $\birth'=\birth$ or $\birth-1$, and $\death'=\death$ or $\death+1$ 
(see Table~\ref{tab:p-intv-map}).
A sequence of levelset persistent cycles 
should achieve the following for the goal:
\begin{enumerate}
    \item Reflect the changes of homological features across all $\Dim$-th critical values
    between $\pcritval_{\birth}$ and $\pcritval_{\death}$.
    \item Capture the critical events at the birth and death points.
\end{enumerate}

For the first requirement,
we add to the sequence the following $\Dim$-cycles:
\[z_i\subseteq\cplx^\Dim_{(i,i+1)}\text{ for each }\birth\leq i<\death,\]
because $\cplx^\Dim_{(i,i+1)}$ is the complex between 
the two critical values $\pcritval_i$, $\pcritval_{i+1}$.
We do the same for all four types of intervals.
For the second requirement
(capturing critical events at endpoints), 
we have to separately address the 
differently types of ends.
We have the following cases:
\begin{description}
    \item[Open birth:] 
    The starting complex of the levelset persistence interval is $\cplx^\Dim_{(\birth,\birth+1)}$.
    We require the corresponding $\Dim$-cycle 
    $z_\birth$ in $\cplx^\Dim_{(\birth,\birth+1)}$
    to become a boundary when included back into
    $\cplx^\Dim_{(\birth-1,\birth+1)}$,
    so that it represents a new-born class in $\Hm_\Dim(\cplx^\Dim_{(\birth,\birth+1)})$.
    In $\lvlfilt_\Dim(f)$,
    the inclusion 
    $\cplx^\Dim_{(\birth-1,\birth+1)}\bakincto\cplx^\Dim_{(\birth,\birth+1)}$
    is further expanded as follows,
    where the birth happens at $K_{\fbirth-1}\bakincto K_{\fbirth}$:
    \[\cplx^\Dim_{(\birth-1,\birth+1)}\bakincto\cdots\bakincto 
    K_{\fbirth-1}\bakincto K_{\fbirth}\bakincto\cdots\bakincto  
    \cplx^\Dim_{(\birth,\birth+1)}.\]
    We also consider $z_\birth$ as a $\Dim$-cycle in $K_{\fbirth}$
    because $\cplx^\Dim_{(\birth,\birth+1)}\subseteq K_{\fbirth}$.
    Then, in $\lvlfilt_\Dim(f)$, $[z_\birth]\in\Hm_\Dim(K_\fbirth)$ should be the 
    non-zero class in the kernel of 
    $\fmorph{\Dim}{\fbirth-1}:\Hm_\Dim(K_{\fbirth-1})\leftarrow\Hm_\Dim(K_{\fbirth})$
    in order to the capture the birth event.

\opt{DCG}{\medskip}
    \item[Open death:]
    Symmetrically to open birth,
    the corresponding $\Dim$-cycle 
    $z_{\death-1}$ in the ending complex $\cplx^\Dim_{(\death-1,\death)}$
    should become a boundary (i.e., die) entering into
    $\cplx^\Dim_{(\death-1,\death+1)}$.
    The inclusion 
    $\cplx^\Dim_{(\death-1,\death)}\incto\cplx^\Dim_{(\death-1,\death+1)}$    
    is further expanded as follows in the simplex-wise filtration,
    where the death happens at $K_{\fdeath}\incto K_{\fdeath+1}$:
    \[\cplx^\Dim_{(\death-1,\death)}\incto\cdots\incto K_{\fdeath}
    \incto K_{\fdeath+1}
    \incto\cdots\incto\cplx^\Dim_{(\death-1,\death+1)}.\]
    To capture the death event, $[z_{\death-1}]\in\Hm_\Dim(K_\fdeath)$ should be the 
    non-zero class in the kernel of 
    $\fmorph{\Dim}{\fdeath}$,
    where we also consider $z_{\death-1}$ as a $\Dim$-cycle in $K_{\fdeath}$.

\opt{DCG}{\medskip}
    \item[Closed birth:]
    The starting complex of the levelset persistence interval is $\cplx^\Dim_{(\birth-1,\birth+1)}$,
    and the birth event happens when 
    $\cplx^\Dim_{({\birth-1},{\birth})}$ is included into
    $\cplx^\Dim_{({\birth-1},{\birth+1})}$.
    The inclusion 
    is further expanded as follows:
    \[\cplx^\Dim_{(\birth-1,\birth)}\incto\cdots\incto
    K_{\fbirth-1}\inctosp{\fsimp{}{\fbirth-1}} K_{\fbirth}\incto\cdots\incto
    \cplx^\Dim_{(\birth-1,\birth+1)}.\]
    In the simplex-wise filtration, the birth happens at the inclusion 
    $K_{\fbirth-1}\incto K_{\fbirth}$.
    Since no
    $z_i\subseteq\cplx^\Dim_{(i,i+1)}$ for $\birth\leq i<\death$
    can be considered as a $\Dim$-cycle in $K_\fbirth$ (see Proposition~\ref{prop:reg-cmplx-disjoint}),
    we add to the sequence a new-born $\Dim$-cycle
    $z_{\birth-1}$ in $K_\fbirth$
    to capture the birth,
    which is equivalent to saying
    that $z_{\birth-1}$ contains the simplex $\fsimp{}{\fbirth-1}$
    (notice that $\fsimp{}{\fbirth-1}$ is a $\Dim$-simplex; 
    see~\cite{carlsson2009zigzag-realvalue}).

    \opt{DCG}{\medskip}
    \item[Closed death:]
    Symmetrically to closed birth, 
    the death happens when the last complex $\cplx^\Dim_{(\death-1,\death+1)}$
    turns into $\cplx^\Dim_{(\death,\death+1)}$ because of the deletion,
    which is at $K_{\fdeath}\bakincto K_{\fdeath+1}$ in $\lvlfilt_\Dim(f)$:
    \[\cplx^\Dim_{(\death-1,\death+1)}\bakincto\cdots\bakincto K_{\fdeath}
    \bakinctosp{\fsimp{}{\fdeath}} K_{\fdeath+1}
    \bakincto\cdots\bakincto\cplx^\Dim_{(\death,\death+1)}.\]
    Since no $\Dim$-cycles 
    defined above are considered to come from $K_\fdeath$ (Proposition~\ref{prop:reg-cmplx-disjoint}),
    we add to the sequence a $\Dim$-cycle 
    $z_{\death}$ in $K_{\fdeath}\subseteq\cplx^\Dim_{(\death-1,\death+1)}$
    containing $\fsimp{}{\fdeath}$,
    so that it represents a class disappearing in $K_{\fdeath+1}$
    (and hence disappearing in $\cplx^\Dim_{(\death,\death+1)}$).
    Notice that $\fsimp{}{\fdeath}$ is a $\Dim$-simplex~\cite{carlsson2009zigzag-realvalue}.
\end{description}

\begin{proposition}\label{prop:reg-cmplx-disjoint}
If the given levelset persistence interval is closed at birth end,
then $K_\fbirth\subseteq\cplx^\Dim_{(\birth-1,\birth]}$ so that
each $\cplx^\Dim_{(i,i+1)}$ for $\birth\leq i<\death$ is disjoint with $K_\fbirth$.
Similarly, if the persistence interval is closed at death end,
then $K_\fdeath\subseteq\cplx^\Dim_{[\death,\death+1)}$ so that
each $\cplx^\Dim_{(i,i+1)}$ for $\birth\leq i<\death$ is disjoint with $K_\fdeath$.
\end{proposition}
\opt{DCG}{\medskip}
\begin{remark}
Notice that the disjointness of these complexes
also makes computation of the optimal persistent cycles feasible;
see Section~\ref{sec:compute}.
\end{remark}
\opt{DCG}{\medskip}
\begin{myproof}
See Appendix~\ref{sec:pf-prop-reg-cmplx-disjoint}
\end{myproof}

One final thing left for the definition is to relate 
two consecutive $\Dim$-cycles
$z_i$, $z_{i+1}$ in the sequence.
It can be verified that both $z_i$, $z_{i+1}$
reside in $\cplx^\Dim_{(i,i+2)}$,
and hence we require them to be homologous in $\cplx^\Dim_{(i,i+2)}$.
This way, we have
\[[z_i]\mapsto[z_i]=[z_{i+1}]\mapsfrom[z_{i+1}]\]
under the linear maps 
\[\Hm_\Dim\big(\cplx^\Dim_{(i,i+1)}\big)\to\Hm_\Dim\big(\cplx^\Dim_{(i,i+2)}\big)
\leftarrow\Hm_\Dim\big(\cplx^\Dim_{(i+1,i+2)}\big)\]
so that all $\Dim$-cycles in the sequence represent 
corresponding homology classes.

For easy reference,
we formally present the definitions individually for the different types of intervals:

\opt{DCG}{\medskip}
\begin{definition}[Open-open case]
\label{dfn:pers-cyc-oo}
For an open-open 
$\big(\pcritval_{\birth},\pcritval_{\death}\big)\in\Pers_\Dim(\lvldgm_\Dim(f))$
produced by a simplex-wise interval $[K_\fbirth,K_\fdeath]$,
the {\it levelset persistent $\Dim$-cycles} %
is a sequence 
$z_\birth,z_{\birth+1},\ldots,z_{\death-1}$
such that: 
\begin{enumerate}
    \item 
$z_i\subseteq \cplx^\Dim_{(i,i+1)}$ for each $i$;
\item
$[z_\birth]\in\Hm_\Dim(K_\fbirth)$ is the
non-zero class in the kernel of 
$\fmorph{\Dim}{\fbirth-1}:\Hm_\Dim(K_{\fbirth-1})\leftarrow\Hm_\Dim(K_{\fbirth})$;
\item
$[z_{\death-1}]\in\Hm_\Dim(K_{\fdeath})$ is the 
non-zero class in the kernel of 
$\fmorph{\Dim}{\fdeath}:\Hm_\Dim(K_{\fdeath})\to\Hm_\Dim(K_{\fdeath+1})$;
\item
each consecutive $z_i$, $z_{i+1}$ are homologous in $\cplx^\Dim_{(i,i+2)}$.
\end{enumerate}
\end{definition}

\begin{definition}[Closed-open case]
\label{dfn:pers-cyc-co}
For a closed-open 
$\big[\pcritval_{\birth},\pcritval_{\death}\big)\in\Pers_\Dim(\lvldgm_\Dim(f))$
produced by a simplex-wise interval $[K_\fbirth,K_\fdeath]$,
the {\it levelset persistent $\Dim$-cycles} %
is a sequence 
$z_{\birth-1},z_{\birth},\ldots,z_{\death-1}$
such that: 
\begin{enumerate}
    \item 
$\fsimp{}{\fbirth-1}\in z_{\birth-1}\subseteq K_\fbirth$;
\item
$z_i\subseteq \cplx^\Dim_{(i,i+1)}$ for each $i\geq \birth$;
\item
$[z_{\death-1}]\in\Hm_\Dim(K_{\fdeath})$ is the 
non-zero class in the kernel of 
$\fmorph{\Dim}{\fdeath}:\Hm_\Dim(K_{\fdeath})\to\Hm_\Dim(K_{\fdeath+1})$;
\item
each consecutive $z_i$, $z_{i+1}$ are homologous in $\cplx^\Dim_{(i,i+2)}$.
\end{enumerate}
\end{definition}

\begin{definition}[Closed-closed case]
\label{dfn:pers-cyc-cc}
For a closed-closed 
$\big[\pcritval_{\birth},\pcritval_{\death}\big]\in\Pers_\Dim(\lvldgm_\Dim(f)$
produced by a simplex-wise interval $[K_\fbirth,K_\fdeath]$,
the {\it levelset persistent $\Dim$-cycles} %
is a sequence 
$z_{\birth-1},z_{\birth},\ldots,z_{\death}$
such that: 
\begin{enumerate}
    \item 
$\fsimp{}{\fbirth-1}\in z_{\birth-1}\subseteq K_\fbirth$;
\item 
$\fsimp{}{\fdeath}\in z_{\death}\subseteq K_\fdeath$;
\item 
$z_i\subseteq \cplx^\Dim_{(i,i+1)}$ for each $\birth\leq i<\death$;
\item 
each consecutive $z_i$, $z_{i+1}$ are homologous in $\cplx^\Dim_{(i,i+2)}$.
\end{enumerate}
\end{definition}

\Cref{fig:motiv} illustrates a sequence of levelset persistent 1-cycles
for a closed-open interval $[\aG_1,\aG_4)$, where $z_0$ captures
the birth event (created by the corresponding 1st critical vertex\footnote{In
   the discrete setting, $z_0$ is indeed created by an edge incident to the critical vertex.})
and $z_1$, $z_2$, $z_3$
are the ones in the regular complexes.
The cycle $z_3$, which becomes a boundary 
when the last critical vertex is added, captures the death event.
See \Cref{fig:open-open-cyc,fig:closed-ex}
in Section~\ref{sec:compute}
for examples of other types of intervals.
See also \Cref{sec:exp} for optimal levelset persistent 1-cycles
computed on triangular meshes by the software that we implemented.

\paragraph{Optimal levelset persistent cycles.}
To define optimal cycles,
we assign weights to $\Dim$-cycles of $K$ as follows:
let each $\Dim$-simplex $\sG$ of $K$ 
have a non-negative finite weight $w(\sG)$;
then, a $\Dim$-cycle $z$ of $K$ has the weight 
$w(z):=\sum_{\sG\in z}w(\sG)$.

\opt{DCG}{\medskip}
\begin{definition}%
For an interval of $\Pers_\Dim(\lvldgm_\Dim(f))$,
an {\it optimal} sequence of levelset persistent $\Dim$-cycles
is one with the minimum sum of weight.
\end{definition}

\subsection{Validity of discrete levelset filtrations}\label{sec:lvlset-compat}
\label{sec:disc-p-dgm}

\begin{figure}
  \centering
  \opt{arxiv}{\includegraphics[width=0.3\linewidth]{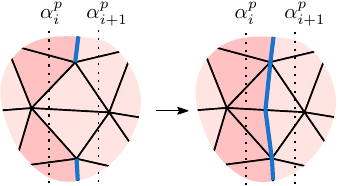}}
  \opt{DCG}{\includegraphics[width=0.4\linewidth]{fig/levelset_compat}}
  
  \opt{DCG}{\medskip}
  \caption{Finer triangulation makes the discrete levelset filtration equivalent with the continuous one.}
  \label{fig:levelset_compat}
\end{figure}

One thing left over from Section~\ref{sec:p-levelzz}
is to justify the validity of the discrete version of $\Dim$-th
levelset filtration.
It turns out that the validity depends on the triangulation of $K$.
For example, let $K$ be the left complex in \Cref{fig:levelset_compat};
then,
$\cplx^\Dim_{(i,i+1)}$ (the blue part) 
is not homotopy equivalent to 
$\subsp^\Dim_{(i,{i+1})}$ (the part between the dashed lines),
and hence $\lvldgm_\Dim(f)$ is
not equivalent to $\clvldgm_\Dim(f)$.
We observe that 
the non-equivalence 
is caused by 
the two central triangles which contain more than one critical value.
A subdivision of the two central triangles on the right
(so that no triangles contain more than one critical value)
renders $\subsp^\Dim_{(i,{i+1})}$
deformation retracting to $\cplx^\Dim_{(i,i+1)}$.
Based on the above observation, we formulate 
the following property,
which guarantees the equivalence of modules induced by $\lvldgm_\Dim(f)$
and $\clvldgm_\Dim(f)$:

\opt{DCG}{\medskip}
\begin{definition}\label{dfn:f-level-compat}
The complex $K$ is said to be
{\it compatible with the $\Dim$-th levelsets} of the PL function $f$
if for any simplex $\sG$ of $K$
and its convex hull $|\sG|$,
function values of
points in $|\sG|$ include
at most one $\Dim$-th critical value of $f$.
\end{definition}

\opt{DCG}{\medskip}
\begin{proposition}\label{prop:f-level-compat-valid}
If $K$ is compatible with the $\Dim$-th levelsets of $f$,
then $\subsp^\Dim_{(i,j)}$ deformation retracts to 
$\cplx^\Dim_{(i,j)}$ for any $i<j$,
which implies that $\Hm_\Dim(\lvldgm_\Dim(f))$
and $\Hm_\Dim(\clvldgm_\Dim(f))$ are isomorphic.
\end{proposition}
\begin{myproof}
See Appendix~\ref{apx:pf-prop-f-level-compat-valid}.
\end{myproof}

In this paper, we always work on a complex that is compatible 
with the $\Dim$-th levelsets of its PL function.
We consider this assumption reasonable because
 when the assumption is violated,
it becomes impossible to depict certain changes of homological features
on the discrete domain.
Notice that a complex can be refined to become compatible
if it is not already so.
 In practice, 
one may also choose to ignore some ``less significant'' critical values
so that the complex becomes compatible with the remaining critical values; see \Cref{sec:exp}
for details in our experiments.

\section{Computation}\label{sec:compute}

In this section, given a weak $(\Dim+1)$-pseudomanifold with $\Dim\geq 1$,
we present algorithms that compute
an optimal sequence of levelset persistent $\Dim$-cycles
for a $\Dim$-th interval.
Though the computation for all types of intervals is based on minimum cuts,
we address the algorithm for each type separately
in each subsection.
The reasons are as follows.
First,
one has to choose a subcomplex to work on in order to 
build a dual graph for the minimum cut computation. 
In the open-open case, the subcomplex is always a {\it$(\Dim+1)$-pseudomanifold}
without boundary (see Section~\ref{sec:lvlset-oo-min-cyc})
whose dual graph is obvious;
in the other cases, however,
we do not have such convenience 
and the dual graph construction is more involved.
Also,
the closed-open case has to deal with the so-called ``monkey saddles''
and the solution adopts a two-phase approach
(see Section~\ref{sec:lvlset-co-min-cyc}); 
in the open-open case, however,
no such issues occur and
the algorithm is much simpler.
We also notice that even for standard persistent cycles which have simpler definitions, 
the hardness results and the algorithms 
for the {\it finite} and {\it infinite} intervals are still different~\cite{dey2020computing}.
With all being said, we observe that the computation for 
the closed-closed case does exhibit resemblance to the closed-open
case and is only described briefly; see Section~\ref{sec:lvlset-cc-min-cyc}.

Other than the type of persistence interval,
all subsections make the same assumptions on input as the following:
\begin{itemize}
    \item $\Dim\geq 1$ is the dimension of interest.

\opt{DCG}{\medskip}
    \item $K$ is a finite weak $(\Dim+1)$-pseudomanifold 
    with a finite weight $w(\sG)\geq 0$
    for each $\Dim$-simplex $\sG$.

\opt{DCG}{\medskip}
    \item $f:|K|\to\Real$ is a generic PL function with $\Dim$-th critical values
    $\pcritval_0=-\infty<\pcritval_{1}<\cdots<\pcritval_{\pcritvcnt}<\pcritval_{\pcritvcnt+1}=\infty$ 
    and corresponding $\Dim$-th critical vertices $\pcritv_{1},\ldots,\pcritv_{\pcritvcnt}$.
    We also assume that $K$ is compatible with the $\Dim$-th levelsets of~$f$.

\opt{DCG}{\medskip}
    \item $\lvlfilt_\Dim(f): K_0\leftrightarrowsp{\fsimp{}{0}} K_1\leftrightarrowsp{\fsimp{}{1}}\cdots\leftrightarrowsp{\fsimp{}{\filtcnt-1}} K_\filtcnt$ 
    is a fixed simplex-wise levelset filtration.
    Each $K_i$, $K_{i+1}$ in $\lvlfilt_\Dim(f)$ differ by
    a simplex $\fsimp{}{i}$,
    and each linear map in $\Hm_\Dim(\lvlfilt_\Dim(f))$ is denoted as
    $\fmorph{\Dim}{i}:\Hm_\Dim(K_i)\leftrightarrow\Hm_\Dim(K_{i+1})$.
\end{itemize}

\subsection{Open-open case}
\label{sec:lvlset-oo-min-cyc}
\begin{figure}
  \centering
  \opt{arxiv}{\includegraphics[width=0.25\linewidth]{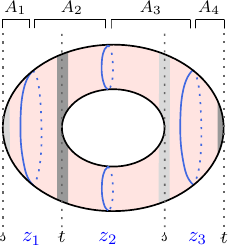}}
  \opt{DCG}{\includegraphics[width=0.3\linewidth]{fig/open-open-cyc}\medskip}
  \caption{A sequence of levelset persistent 1-cycles 
  for an open-open interval $\big(\critval^1_1,\critval^1_4\big)$;
  the complex (assuming the torus to be finely triangulated),
  the function,
  and the 1st critical values 
  are the same as in \Cref{fig:torus}.}
  \label{fig:open-open-cyc}
\end{figure}

Throughout this subsection,
assume that we aim to compute the optimal persistent $\Dim$-cycles 
for an {\it open-open} interval 
$\big(\pcritval_{\birth},\pcritval_{\death}\big)$ from $\Pers_\Dim(\lvldgm_\Dim(f))$,
which is produced by a simplex-wise interval $[K_\fbirth,K_\fdeath]$ 
from $\Pers_\Dim(\lvlfilt_\Dim(f))$.
\Cref{fig:open-open-cyc} illustrates
a sequence of persistent 1-cycles $z_1,z_2,z_3$ 
for an open-open interval $\big(\critval^1_1,\critval^1_4\big)$.

As seen from Section~\ref{sec:pers-cyc-dfn},
the following portion of $\lvlfilt_\Dim(f)$ is relevant to the definition 
(and hence the computation)
of levelset persistent $\Dim$-cycles for $\big(\pcritval_{\birth},\pcritval_{\death}\big)$:
\begin{align}
\begin{split}
\label{eqn:oo-intv-diag-cmplx-seq}
\cplx^\Dim_{(\birth-1,\birth+1)}
\bakincto\cdots\bakincto
K_{\fbirth-1}\bakinctosp{\fsimp{}{\fbirth-1}} K_{\fbirth}
\bakincto\cdots\bakincto
\cplx^\Dim_{(\birth,\birth+1)}
\incto
\cdots\\
\bakincto
\cplx^\Dim_{(\death-1,\death)}
\incto\cdots\incto
K_{\fdeath}
\inctosp{\fsimp{}{\fdeath}} K_{\fdeath+1}
\incto\cdots\incto
\cplx^\Dim_{(\death-1,\death+1).}
\end{split}
\end{align}
In the above sequence,
the simplices 
$\fsimp{}{\fbirth-1}$, $\fsimp{}{\fdeath}$
are the ones
creating and destroying the simplex-wise interval $[K_\fbirth,K_\fdeath]$,
which
are both $(\Dim+1)$-simplices~\cite{carlsson2009zigzag-realvalue}.
We 
restrict
the computation
to (a connected component of)
$\cplx^\Dim_{(\birth-1,\death+1)}$ because
each complex in Sequence~(\ref{eqn:oo-intv-diag-cmplx-seq})
is a subcomplex of $\cplx^\Dim_{(\birth-1,\death+1)}$.
However, instead of the usual one, we take
a special type of connected component which considers connectedness
in higher dimensions:

\opt{DCG}{\medskip}
\begin{definition}[$\diml$-connected~\cite{dey2020computing}]
\label{dfn:q-conn}
Let $\SG$ be a set of simplices,
and let $\sG$, $\sG'$ be two $\diml$-simplices of $\SG$
where $\diml\geq 1$.
A {\it $\diml$-path} from $\sG$ to $\sG'$ in $\SG$ 
is a sequence of $\diml$-simplices of $\SG$, $\tG_1,\ldots,\tG_\ell$,
such that $\tG_1=\sG$, $\tG_\ell=\sG'$, 
and each consecutive $\tG_i$, $\tG_{i+1}$ share a $(\diml-1)$-face in $\SG$. 
A maximal set of $\diml$-simplices of $\SG$, in which each pair 
is connected by a $\diml$-path,
constitutes a {\it $\diml$-connected component} of $\SG$.
We also say that $\SG$ is {\it $\diml$-connected} 
if it has only one $\diml$-connected component.
\end{definition}
\opt{DCG}{\medskip}

We now describe the algorithm.
Since the deletion of the $(\Dim+1)$-simplex $\fsimp{\Fcal}{\fbirth-1}$
gives birth to the interval $[K_\fbirth,K_\fdeath]$,
$\fsimp{\Fcal}{\fbirth-1}$ must be relevant to our computation.
So we let the complex that we work on, denoted $K'$,
be the closure of the $(\Dim+1)$-connected component 
of $K$ containing $\fsimp{\Fcal}{\fbirth-1}$.
(The \emph{closure} of a set of simplices
consists of all faces of the simplices in the set.)
We observe that $K'$ must be a
\emph{$(\Dim+1)$-pseudomanifold} without boundary,
i.e., each $\Dim$-simplex has exactly two $(\Dim+1)$-cofaces in $K'$; see Proposition~\ref{prop:oo-alg-cons-facts},
Claim~\ref{itm:K-prime-nobound}.
We then take  the dual graph $G$ of $K'$
and compute the optimal persistent $\Dim$-cycles
by computing a minimum cut on $(G,\gsrc,\gsink)$,
where $\gsrc$, $\gsink$ are some properly chosen sources  and sinks.
To set up $\gsrc$ and $\gsink$,
we first define the following set  of simplices:
\[\vsimpset{i}:=\cplx^\Dim_{(i-1,i+1)}\setminus
\big(\cplx^\Dim_{(i-1,i)}\union\cplx^\Dim_{(i,i+1)}\big).\]
Roughly speaking, $\vsimpset{i}$ consists of simplices
containing the critical value $\pcritval_i$
(e.g., the darker triangles in
\Cref{fig:levelset_compat} 
belong to $\vsimpset{i}$),
and also notice that $\vsimpset{i}$ may not be a simplicial complex.
We then alternately put vertices dual to the $(\Dim+1)$-simplices in
$\vsimpset{\birth},\ldots,\vsimpset{\death}$
into $\gsrc$ and $\gsink$.
For the example in \Cref{fig:open-open-cyc} where $K'$ is the entire torus,
the source $\gsrc$ contains vertices dual to 
2-simplices in $\vsimpset[1]{1}\union\vsimpset[1]{3}$,
and the sink $\gsink$ contains vertices dual to 
2-simplices in $\vsimpset[1]{2}\union\vsimpset[1]{4}$.
Notice that
$\vsimpset[1]{1},\ldots,\vsimpset[1]{4}$ are alternately shaded with light and dark gray
in \Cref{fig:open-open-cyc}.

The correctness of the above construction 
is based on the duality of the levelset persistent $\Dim$-cycles 
for $\big(\pcritval_{\birth},\pcritval_{\death}\big)$
and \sCommaT{}cuts on $(G,\gsrc,\gsink)$.
To see the duality, first consider
the sequence of persistent 1-cycles ${z_1,z_2,z_3}$ in \Cref{fig:open-open-cyc}.
By Definition~\ref{dfn:pers-cyc-oo}, there exist 2-chains 
\[A_1\subseteq \cplx^1_{(0,2)}\text{, }A_2\subseteq \cplx^1_{(1,3)}\text{, }A_3\subseteq \cplx^1_{(2,4)}\text{, and }A_4\subseteq \cplx^1_{(3,5)}\]
as shown in \Cref{fig:open-open-cyc}
such that
\[z_1=\partial(A_1)\text{, }z_1+z_2=\partial(A_2)\text{, }z_2+z_3=\partial(A_3)\text{, and }z_3=\partial(A_4).\]
Let $S$ contain the vertices dual to $A_1+A_3$
and $T$ contain the vertices dual to $A_2+A_4$.
Then, $(S,T)$ is a \sCommaT{}cut of $(G,\gsrc,\gsink)$.
Since edges in $E(S,T)$
are dual to 1-simplices in $z_1+z_2+z_3$,
we have that $w(S,T)=w(z_1)+w(z_2)+w(z_3)$.
So we have a cut $(S,T)$ dual to 
the given persistent 1-cycles ${z_1,z_2,z_3}$.
On the other hand, a \sCommaT{}cut of $(G,\gsrc,\gsink)$
produces a sequence of persistent $\Dim$-cycles for the given interval.
For the example in \Cref{fig:open-open-cyc}, let $(S,T)$ be a cut where 
$S$ contains the graph vertices dual to $A_1+A_3$
and $T$ contains the graph vertices dual to $A_2+A_4$,
as defined previously.
We then take the intersection of
the dual 1-simplices of $E(S,T)$
with $\cplx^1_{(1,2)},\cplx^1_{(2,3)},\cplx^1_{(3,4)}$.
The resulting 1-chains ${z_1,z_2,z_3}$
is a sequence of persistent 1-cycles for the interval 
$\big(\critval^1_1,\critval^1_4\big)$.
Hence,
by the duality, a minimum \sCommaT{}cut of $(G,\gsrc,\gsink)$
produces an optimal sequence of levelset persistent $\Dim$-cycles
for $\big(\pcritval_{\birth},\pcritval_{\death}\big)$.

We now present the details of our algorithm
as follows:

\opt{DCG}{\medskip}
\begin{algr}[Open-open case]
\label{alg:levelzz-open-open}
Given the input
as specified,
do the following:
\begin{enumerate}
\item
Let $K'$ be the closure of the $(\Dim+1)$-connected component 
of $K$ containing $\fsimp{\Fcal}{\fbirth-1}$.
Notice that $K'$ is a $(\Dim+1)$-pseudomanifold without boundary
(see Proposition~\ref{prop:oo-alg-cons-facts},
Claim~\ref{itm:K-prime-nobound}).

\opt{DCG}{\medskip}
\item
Build a weighted dual graph $G$ of $K'$,
where $V(G)$ corresponds to $(\Dim+1)$-simplices of $K'$
and $E(G)$ corresponds to $\Dim$-simplices of $K'$.
Let $\thG$ denote both the bijection from the $(\Dim+1)$-simplices to $V(G)$
and the bijection from the $\Dim$-simplices to $E(G)$.
For each edge $e$ of $G$,
if $\thG\inv(e)\in \cplx^\Dim_{(\pcycind,\pcycind+1)}$ for $\birth\leq\pcycind<\death$,
then set $w(e)$, the weight of $e$, as $w(\thG\inv(e))$;
otherwise, set $w(e)=\infty$.

\opt{DCG}{\medskip}
\item
For each $\pcycind$ s.t.\ $\birth\leq\pcycind\leq\death$,
let $\simpset_{\pcycind}$ denote the set of $(\Dim+1)$-simplices in 
$K'\intersect\vsimpset{\pcycind}$.
Also,
let $L_\mathsf{e}$ be the set of even integers in $\{0,1,\ldots,\death-\birth\}$
and $L_\mathsf{o}$ be the set of odd ones.
Then, let 
\[\gsrc=\thG\bigg(\bigunion_{i\in L_\mathsf{e}}\simpset_{\birth+i}\bigg)\text{, }
\gsink=\thG\bigg(\bigunion_{i\in L_\mathsf{o}}\simpset_{\birth+i}\bigg),\]
and compute the minimum \sCommaT{}cut $(S^*,T^*)$ of $(G,\gsrc,\gsink)$.

\opt{DCG}{\medskip}
\item
For each $\pcycind$ s.t.\ $\birth\leq\pcycind<\death$,
let $z^*_\pcycind=\cplx^\Dim_{(\pcycind,\pcycind+1)}\intersect\thG\inv(E(S^*,T^*))$.
Return $z^*_\birth,\ldots,z^*_{\death-1}$ 
as an optimal sequence of levelset persistent $\Dim$-cycles
for the interval $\big(\pcritval_{\birth},\pcritval_{\death}\big)$.

\end{enumerate}
\end{algr}

\subsubsection{Correctness of the algorithm}

To justify the correctness of Algorithm~\ref{alg:levelzz-open-open},
we first present
Proposition~\ref{prop:oo-alg-cons-facts} stating several facts about Algorithm~\ref{alg:levelzz-open-open}.
We then utilize Proposition~\ref{prop:oo-alg-cons-facts} to prove Propositions 
\ref{prop:oo-alg-pers-cyc-corr-cut} and~\ref{prop:oo-alg-cut-corr-pers-cyc},
which formally present the duality.
Then,
Propositions
\ref{prop:oo-alg-pers-cyc-corr-cut} and~\ref{prop:oo-alg-cut-corr-pers-cyc} 
lead to Theorem~\ref{thm:levelzz-oo-alg-correct},
which draws the conclusion.

\opt{DCG}{\medskip}
\begin{proposition}
\label{prop:oo-alg-cons-facts}
The following claims hold for Algorithm~\ref{alg:levelzz-open-open}:
\begin{enumerate}
    \item \label{itm:oo-intv-kill-simps-same-conn}
    The simplex $\fsimp{\Fcal}{\fdeath}$
    belongs to $K'$.

\opt{DCG}{\medskip}
    \item \label{itm:p+1chains-exist-sum-equal-all}
    Let $z_\birth,\ldots,z_{\death-1}$ be 
    any sequence of persistent $\Dim$-cycles for $\big(\pcritval_{\birth},\pcritval_{\death}\big)$;
    then, there exist $(\Dim+1)$-chains 
    $A_{\birth}\subseteq K_{\fbirth-1},
    A_{\birth+1}\subseteq \cplx^\Dim_{(\birth,\birth+2)},\ldots,
    A_{\death-1}\subseteq \cplx^\Dim_{(\death-2,\death)},\allowbreak
    A_{\death}\subseteq K_{\fdeath+1}$
    such that
    $\fsimp{\Fcal}{\fbirth-1}\in A_{\birth}$,
    $\fsimp{\Fcal}{\fdeath}\in A_{\death}$,
    $z_\birth=\partial(A_\birth)$,
    $z_{\death-1}=\partial(A_{\death})$,
    and $z_{\pcycind-1}+z_{\pcycind}=\partial(A_{\pcycind})$ 
    for each $\birth<\pcycind<\death$.
    Furthermore,
    let $z'_\pcycind=K'\intersect z_\pcycind$,
    $A'_\pcycind=K'\intersect A_\pcycind$ for each $\pcycind$;
    then,
    $\fsimp{\Fcal}{\fbirth-1}\in A'_{\birth}$,
    $\fsimp{\Fcal}{\fdeath}\in A'_{\death}$,
    $z'_\birth=\partial\big(A'_\birth\big)$,
    $z'_{\death-1}=\partial\big(A'_{\death}\big)$,
    and $z'_{\pcycind-1}+z'_{\pcycind}=\partial\big(A'_{\pcycind}\big)$ 
    for each $\birth<\pcycind<\death$.
    Finally, one has that $A'_\birth+\cdots+A'_{\death}$
    equals 
    the set of $(\Dim+1)$-simplices of $K'$
    and $A'_\birth,\ldots,A'_{\death}$ are pair-wise disjoint.

\opt{DCG}{\medskip}
    \item \label{itm:K-prime-nobound}
    The complex $K'$ is a $(\Dim+1)$-connected $(\Dim+1)$-pseudomanifold without boundary, i.e.,
    each $\Dim$-simplex has exactly two $(\Dim+1)$-cofaces in $K'$.
\end{enumerate}
\end{proposition}
\begin{myproof}
See Appendix~\ref{apx:pf-prop-oo-alg-cons-facts}.
\end{myproof}

\begin{proposition}
\label{prop:oo-alg-pers-cyc-corr-cut}
Let ${z_\birth,\ldots,z_{\death-1}}$ be 
any sequence of levelset persistent $\Dim$-cycles 
for $\big(\pcritval_{\birth},\pcritval_{\death}\big)$;
then, there exists a \sCommaT{}cut $(S,T)$ of $(G,\gsrc,\gsink)$
such that $w(S,T)\leq \sum_{\pcycind=\birth}^{\death-1} w(z_\pcycind)$.
\end{proposition}
\begin{myproof}
Let $A'_\birth,\ldots,A'_{\death}$ and
$z'_\birth,\ldots,z'_{\death-1}$
be as specified in 
Claim~\ref{itm:p+1chains-exist-sum-equal-all}
of Proposition~\ref{prop:oo-alg-cons-facts}
for the given 
${z_\birth,\ldots,z_{\death-1}}$,
and
let $S=\thG\big(\sum_{j\in L_\mathsf{e}}A'_{\birth+j}\big)$, 
$T=\thG\big(\sum_{j\in L_\mathsf{o}}A'_{\birth+j}\big)$.
We first show that for a $\simpset_\pcycind$ such that $\pcycind-\birth$ is even,
$\simpset_\pcycind$ does not intersect $\sum_{j\in L_\mathsf{o}}A'_{\birth+j}$.
For contradiction, suppose instead that there is a $\sG$ in both of them.
Then, since $\simpset_\pcycind\subseteq \vsimpset{\pcycind}\subseteq \cplx^\Dim_{(\pcycind-1,\pcycind+1)}$
and $A'_{\birth+j}\subseteq \cplx^\Dim_{(\birth+j-1,\birth+j+1)}$ 
for each $j\in L_\mathsf{o}$,
$\sG$ must be in $A'_{\pcycind-1}\subseteq \cplx^\Dim_{(\pcycind-2,\pcycind)}$
or $A'_{\pcycind+1}\subseteq \cplx^\Dim_{(\pcycind,\pcycind+2)}$
because other chains in $\Set{A'_{\birth+j}\given j\in L_\mathsf{o}}$ 
do not intersect $\cplx^\Dim_{(\pcycind-1,\pcycind+1)}$.
So we have that $\sG$ is in $\cplx^\Dim_{(\pcycind-2,\pcycind)}$ or $\cplx^\Dim_{(\pcycind,\pcycind+2)}$.
The fact that $\sG\in\simpset_\pcycind\subseteq \cplx^\Dim_{(\pcycind-1,\pcycind+1)}$
implies that $\sG$ is in $\cplx^\Dim_{(\pcycind-1,\pcycind)}$ or $\cplx^\Dim_{(\pcycind,\pcycind+1)}$,
a contradiction to $\sG\in\simpset_\pcycind\subseteq \vsimpset{\pcycind}
=\cplx^\Dim_{(\pcycind-1,\pcycind+1)}\setminus
\big(\cplx^\Dim_{(\pcycind-1,\pcycind)}\union\cplx^\Dim_{(\pcycind,\pcycind+1)}\big)$.
So $\simpset_\pcycind$ does not intersect $\sum_{j\in L_\mathsf{o}}A'_{\birth+j}$.
Then,
since $\sum_{j=\birth}^{\death} A'_j$ equals 
the set of $(\Dim+1)$-simplices of $K'$ 
by Claim~\ref{itm:p+1chains-exist-sum-equal-all} 
of Proposition~\ref{prop:oo-alg-cons-facts},
we have that $\simpset_\pcycind\subseteq\sum_{j\in L_\mathsf{e}}A'_{\birth+j}$,
i.e., $\thG(\simpset_\pcycind)\subseteq S$.
This means that $\gsrc\subseteq S$.
Similarly, we have $\gsink\subseteq T$.
Claim~\ref{itm:p+1chains-exist-sum-equal-all} 
of Proposition~\ref{prop:oo-alg-cons-facts} implies that
$S\union T=V(G)$ and $S\intersect T=\emptyset$,
and so $(S,T)$ is a \sCommaT{}cut of $(G,\gsrc,\gsink)$.
The fact that $\sum_{\pcycind=\birth}^{\death-1} z'_\pcycind=
\partial\big(\sum_{j\in L_\mathsf{e}}A'_{\birth+j}\big)
=\partial\big(\sum_{j\in L_\mathsf{o}}A'_{\birth+j}\big)$
implies that $\sum_{\pcycind=\birth}^{\death-1} z'_\pcycind=\thG\inv(E(S,T))$.
So we have $w(S,T)=\sum_{\pcycind=\birth}^{\death-1} w(z'_\pcycind)\leq \sum_{\pcycind=\birth}^{\death-1} w(z_\pcycind)$.
\end{myproof}

\begin{proposition}
\label{prop:oo-alg-cut-corr-pers-cyc}
For any \sCommaT{}cut $(S,T)$ of $(G,\gsrc,\gsink)$ with finite weight,
let $z_\pcycind=\cplx^\Dim_{(\pcycind,\pcycind+1)}\intersect\thG\inv(E(S,T))$ 
for each 
$\birth\leq\pcycind<\death$.
Then, ${z_\birth,\ldots,z_{\death-1}}$
is a sequence of levelset persistent $\Dim$-cycles
for $\big(\pcritval_{\birth},\pcritval_{\death}\big)$
with $\sum_{\pcycind=\birth}^{\death-1} w(z_\pcycind)=w(S,T)$.
\end{proposition}
\begin{myproof}
We first prove that,
for any $\pcycind$ s.t.\ $\birth<\pcycind<\death$ and $\pcycind-\birth$ is even,
$\partial\big(\thG\inv(S)\intsec \cplx^\Dim_{(\pcycind-1,\pcycind+1)}\big)
=z_{\pcycind-1}+z_\pcycind$.
To prove this, 
first consider any 
$\sG\in\partial\big(\thG\inv(S)\intsec \cplx^\Dim_{(\pcycind-1,\pcycind+1)}\big)$.
We have that $\sG$ is a face of only one $(\Dim+1)$-simplex $\tG_1$
in $\thG\inv(S)\intsec \cplx^\Dim_{(\pcycind-1,\pcycind+1)}$.
Note that $\tG_1\in\thG\inv(S)\subseteq K'$.
Since $K'$ is a $(\Dim+1)$-pseudomanifold without boundary
(Claim~\ref{itm:K-prime-nobound} of Proposition~\ref{prop:oo-alg-cons-facts}), 
$\sG$ has another $(\Dim+1)$-coface $\tG_2$ in $K'$.
Then, it must be true that $\tG_2\in\thG\inv(T)$.
To see this,
suppose instead that $\tG_2\in\thG\inv(S)$.
Note that $\tG_2\not\in \cplx^\Dim_{(\pcycind-1,\pcycind+1)}$
because otherwise 
$\tG_2$ would be in $\thG\inv(S)\intsec \cplx^\Dim_{(\pcycind-1,\pcycind+1)}$,
contradicting the fact that
$\sG$ has only one $(\Dim+1)$-coface
in $\thG\inv(S)\intsec \cplx^\Dim_{(\pcycind-1,\pcycind+1)}$.
Also note that $\tG_2$ is 
not in $\vsimpset{\pcycind-1}$ or $\vsimpset{\pcycind+1}$
because if $\tG_2$ is in one of them,
combining the fact that $\pcycind-1-\birth$ and $\pcycind+1-\birth$ are odd,
we would have that $\tG_2$ is in $\simpset_{\pcycind-1}$ or $\simpset_{\pcycind+1}$
and thus $\thG(\tG_2)\in \gsink \subseteq T$,
which is a contradiction.
Since $K'\subseteq \cplx^\Dim_{(\birth-1,\death+1)}$
and $\bigSet{\cplx^\Dim_{(\birth-1,\pcycind-1)},\vsimpset{\pcycind-1},\cplx^\Dim_{(\pcycind-1,\pcycind+1)},
\vsimpset{\pcycind+1},\cplx^\Dim_{(\pcycind+1,\death+1)}}$ covers $\cplx^\Dim_{(\birth-1,\death+1)}$,
we have that $\tG_2$ is in 
$\cplx^\Dim_{(\birth-1,\pcycind-1)}$ or $\cplx^\Dim_{(\pcycind+1,\death+1)}$.
This implies that $\sG\subseteq\tG_2$ is in 
$\cplx^\Dim_{(\birth-1,\pcycind-1)}$ or $\cplx^\Dim_{(\pcycind+1,\death+1)}$,
contradicting that $\sG\subseteq\tG_1\in \cplx^\Dim_{(\pcycind-1,\pcycind+1)}$.
It is now true that $\sG\in\thG\inv(E(S,T))$
because $\tG_1\in \thG\inv(S)$ and $\tG_2\in \thG\inv(T)$.
Since $(S,T)$ has finite weight, 
$\sG$ must come from a $\cplx^\Dim_{(j,j+1)}$ for $\birth\leq j<\death$
and thus must come from $\cplx^\Dim_{(\pcycind-1,\pcycind)}$ or $\cplx^\Dim_{(\pcycind,\pcycind+1)}$.
Then, $\sG$ is in $z_{\pcycind-1}$ or $z_{\pcycind}$.
Moreover, since $z_{\pcycind-1}$ and $z_{\pcycind}$ are disjoint,
we have $\sG\in z_{\pcycind-1}+z_{\pcycind}$.

On the other hand, 
for any $\sG\in z_{\pcycind-1}+z_{\pcycind}$,
first assume that $\sG\in z_{\pcycind-1}=\cplx^\Dim_{(\pcycind-1,\pcycind)}\intersect\thG\inv(E(S,T))$.
Since 
$\sG\in \thG\inv(E(S,T))$,
$\sG$ must be a face of a $(\Dim+1)$-simplex $\tG$ in $\thG\inv(S)$
and another $(\Dim+1)$-simplex in $\thG\inv(T)$.
We then show that $\tG\in \cplx^\Dim_{(\pcycind-1,\pcycind+1)}$.
Suppose instead that $\tG\not\in \cplx^\Dim_{(\pcycind-1,\pcycind+1)}$,
and let $v$ be the vertex belonging to $\tG$
but not $\sG$.
We have that $f(v)\not\in(\pcritval_{\pcycind-1},\pcritval_{\pcycind+1})$
because if $f(v)\in(\pcritval_{\pcycind-1},\pcritval_{\pcycind+1})$, 
the fact that $\sG\in \cplx^\Dim_{(\pcycind-1,\pcycind)}$ would imply that
$\tG$ is in $\cplx^\Dim_{(\pcycind-1,\pcycind+1)}$.
Note that $f(v)$ cannot be greater than or equal to $\pcritval_{\pcycind+1}$
because otherwise $K$ would not be compatible with the $\Dim$-th levelsets of $f$.
Therefore, $f(v)\leq \pcritval_{\pcycind-1}$,
and it must be true that $\tG\in \cplx^\Dim_{(\pcycind-2,\pcycind)}$.
This implies that $\tG\in\vsimpset{\pcycind-1}$.
We now have that $\tG\in\simpset_{\pcycind-1}$, where $\pcycind-1-\birth$ is odd.
Then, $\thG(\tG)\in \gsink\subseteq T$,
a contradiction to $\tG\in\thG\inv(S)$.
Combining the fact that $\tG\in \cplx^\Dim_{(\pcycind-1,\pcycind+1)}$
and $\tG$ is the only $(\Dim+1)$-coface of $\sG$ in $\thG\inv(S)$,
we have that $\tG$ is the only $(\Dim+1)$-coface of $\sG$ 
in $\thG\inv(S)\intsec \cplx^\Dim_{(\pcycind-1,\pcycind+1)}$.
If $\sG\in z_{\pcycind}$, we can have the same result.
Therefore, $\sG\in \partial\big(\thG\inv(S)\intsec \cplx^\Dim_{(\pcycind-1,\pcycind+1)}\big)$,
and we have proved that $\partial\big(\thG\inv(S)\intsec \cplx^\Dim_{(\pcycind-1,\pcycind+1)}\big)
=z_{\pcycind-1}+z_\pcycind$.

Similarly, we can prove that 
$\partial\big(\thG\inv(T)\intsec \cplx^\Dim_{(\pcycind-1,\pcycind+1)}\big)
=z_{\pcycind-1}+z_\pcycind$ 
for $\pcycind$ s.t.\ $\birth<\pcycind<\death$ 
and $\pcycind-\birth$ is odd,
$\partial\big(\thG\inv(S)\intsec K_{\fbirth-1})=z_\birth$,
and
$\partial\big(\thG\inv(S)\intsec K_{\fdeath+1})=z_{\death-1}$
or $\partial\big(\thG\inv(T)\intsec K_{\fdeath+1})=z_{\death-1}$
based on the parity of $\death-\birth$.
Since $\fsimp{\Fcal}{\fbirth-1}\in K_{\fbirth-1}\subseteq \cplx^\Dim_{[\birth,\birth+1)}$
and $\fsimp{\Fcal}{\fbirth-1}\not\in \cplx^\Dim_{(\birth,\birth+1)}$,
we have that $\fsimp{\Fcal}{\fbirth-1}\in\vsimpset{\birth}$,
which means that $\thG(\fsimp{\Fcal}{\fbirth-1})\in \gsrc\subseteq S$.
Therefore, $\fsimp{\Fcal}{\fbirth-1}\in\thG\inv(S)\intsec K_{\fbirth-1}$.
Since $\partial\big(\thG\inv(S)\intsec K_{\fbirth-1})=z_\birth$,
we have that $z_\birth\homolog\partial(\fsimp{\Fcal}{\fbirth-1})$ in $K_\fbirth$,
i.e., $[z_\birth]\in\Hm_\Dim(K_\fbirth)$ is the non-zero class 
in $\ker(\fmorph{\Dim}{\fbirth-1})$.
Analogously, $[z_{\death-1}]\in\Hm_\Dim(K_\fdeath)$ is the non-zero class
in $\ker(\fmorph{\Dim}{\fdeath})$.
The above facts imply that
$z_\birth,\ldots,z_{\death-1}$
is a sequence of levelset persistent $\Dim$-cycles
for $\big(\pcritval_{\birth},\pcritval_{\death}\big)$.
The equality of the weight follows from the disjointness of $z_\birth,\ldots,z_{\death-1}$
and the fact that $w(S,T)$ is finite.
\end{myproof}

\begin{theorem}
\label{thm:levelzz-oo-alg-correct}
Algorithm~\ref{alg:levelzz-open-open} computes an optimal sequence 
of levelset persistent \Dim-cycles
for a given {\rm open-open} interval.
\end{theorem}
\begin{proof}
First, by \Cref{prop:oo-alg-pers-cyc-corr-cut}, the min-cut $(S^*,T^*)$
in Algorithm~\ref{alg:levelzz-open-open} must have finite weight.
Then, by \Cref{prop:oo-alg-cut-corr-pers-cyc},
$z^*_\birth,\ldots,z^*_{\death-1}$ returned by the algorithm
is  a sequence of  persistent $\Dim$-cycles
for  $\big(\pcritval_{\birth},\pcritval_{\death}\big)$
with
$\sum_{\pcycind=\birth}^{\death-1} w(z^*_\pcycind)=w(S^*,T^*)$.
For contradiction, suppose instead that 
$z^*_\birth,\ldots,z^*_{\death-1}$ is not an optimal sequence of  persistent $\Dim$-cycles
for  $\big(\pcritval_{\birth},\pcritval_{\death}\big)$.
Let $z'_\birth,\ldots,z'_{\death-1}$ 
be an optimal sequence of  persistent $\Dim$-cycles
for  $\big(\pcritval_{\birth},\pcritval_{\death}\big)$.
We have $\sum_{\pcycind=\birth}^{\death-1} w(z'_\pcycind)<\sum_{\pcycind=\birth}^{\death-1} w(z^*_\pcycind)$.
By \Cref{prop:oo-alg-pers-cyc-corr-cut},
there exists a \sCommaT{}cut $(S',T')$ of $(G,\gsrc,\gsink)$
such that $w(S',T')\leq \sum_{\pcycind=\birth}^{\death-1} w(z'_\pcycind)
<\sum_{\pcycind=\birth}^{\death-1} w(z^*_\pcycind)=w(S^*,T^*)$,
contradicting that $(S^*,T^*)$ is a min-cut.
\end{proof}

\subsection{Closed-open case}
\label{sec:lvlset-co-min-cyc}

Throughout the subsection,
assume that we aim to compute the optimal persistent $\Dim$-cycles 
for a {\it closed-open} interval 
$\big[\pcritval_{\birth},\pcritval_{\death}\big)$ from $\Pers_\Dim(\lvldgm_\Dim(f))$,
which is produced by a simplex-wise interval $[K_\fbirth,K_\fdeath]$ 
from $\Pers_\Dim(\lvlfilt_\Dim(f))$. 
\Cref{fig:octopus,fig:broken_octopus} 
provide examples for $\Dim=1$,
where ${z'_1,z'_2,z'_3}$ 
and ${z''_1,z''_2,z''_3}$
are two sequences of levelset persistent 1-cycles 
for the interval $\big[\critval^1_2,\critval^1_4\big)$.

Similar to the previous case,
we have the following
portion of $\lvlfilt_\Dim(f)$ 
relevant to the definition and computation:
\begin{align}
\begin{split}
\label{eqn:co-intv-diag-cmplx-seq}
\cplx^\Dim_{(\birth-1,\birth)}
\incto\cdots\incto
K_{\fbirth-1}\inctosp{\fsimp{}{\fbirth-1}} K_{\fbirth}
\incto\cdots\incto
\cplx^\Dim_{(\birth-1,\birth+1)}
\bakincto\cdots\bakincto
\cplx^\Dim_{(\birth,\birth+1)}
\incto
\cdots \\
\bakincto
\cplx^\Dim_{(\death-1,\death)}
\incto\cdots\incto
K_{\fdeath}
\inctosp{\fsimp{}{\fdeath}} K_{\fdeath+1}
\incto\cdots\incto
\cplx^\Dim_{(\death-1,\death]}
\end{split}
\end{align}
The creator $\fsimp{}{\fbirth-1}$ of the simplex-wise interval $[K_\fbirth,K_\fdeath]$  
is a $\Dim$-simplex
and the destroyer $\fsimp{}{\fdeath}$
a $(\Dim+1)$-simplex~\cite{carlsson2009zigzag-realvalue}.
Notice that we end the sequence with $\cplx^\Dim_{(\death-1,\death]}$
instead of $\cplx^\Dim_{(\death-1,\death+1)}$ 
as in the case ``open death'' in Section~\ref{sec:pers-cyc-dfn}.
This is valid due to the following reasons: 
(i)~$\cplx^\Dim_{(\death-1,\death]}$
is derived from $\cplx^\Dim_{(\death-1,\death)}$
by adding the lower star of $\pcritv_\death$
and hence must appear in $\lvlfilt_\Dim(f)$ based on Definition~\ref{dfn:lvlfilt};
(ii) $K_{\fdeath+1}$ is a subcomplex of $\cplx^\Dim_{(\death-1,\death]}$
and the proof is similar to that of Proposition~\ref{prop:reg-cmplx-disjoint}.
Therefore, the computation can be restricted to $\cplx^\Dim_{(\birth-1,\death]}$
because each complex in Sequence~(\ref{eqn:co-intv-diag-cmplx-seq})
is a subcomplex of $\cplx^\Dim_{(\birth-1,\death]}$.

\subsubsection{Overview}\label{sec:lvlset-co-min-cyc-sum}
\begin{figure}
  \subfloat[]{
    \opt{arxiv}{\raisebox{0.01\linewidth}{
      \includegraphics[width=0.32\linewidth]{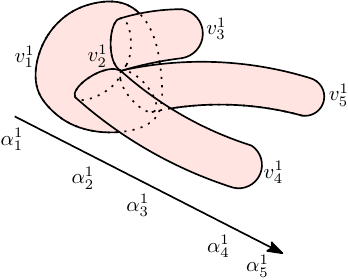}}}
    \opt{DCG}{\raisebox{0.01\linewidth}{
      \includegraphics[width=0.345\linewidth]{fig/octopus}}}
  \label{fig:octopus}}
  \opt{arxiv}{\hspace{2em}}
  \opt{DCG}{\hspace{0.5em}}
  \subfloat[]{
    \opt{arxiv}{
      \includegraphics[width=0.32\linewidth]{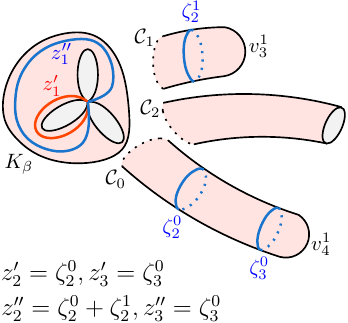}
    }
    \opt{DCG}{
      \includegraphics[width=0.345\linewidth]{fig/broken_octopus}
    }
    \label{fig:broken_octopus}}
  \opt{arxiv}{\hspace{2em}}
  \opt{DCG}{\hspace{0.5em}}
  \subfloat[]{
    \opt{arxiv}{
    \raisebox{0.12\linewidth}{
      \includegraphics[width=0.11\linewidth]{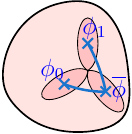}
    }}
    \opt{DCG}{
    \raisebox{0.12\linewidth}{
      \includegraphics[width=0.13\linewidth]{fig/K_b}
    }}
    \label{fig:K_b}}

  \opt{DCG}{\medskip}
  \caption{
  (a) A complex $K$ with all 1st critical vertices listed,
  in which $\critv^1_2$ is a monkey saddle;
  the direction of the height function is indicated by the arrow.
  (b) The relevant subcomplex $\cplx^\Dim_{(\birth-1,\death]}=\cplx^1_{(1,4]}$ 
  with $K_\fbirth$
  broken from the remaining parts
  for a better illustration.
  (c) The complex $K_\fbirth$ with boundaries filled by 
  2-dimensional ``cells'' drawn as darker regions.
  The blue edges are augmenting edges in the dual graph. 
  Notice that $K_\fbirth$ also contains boundary 1-simplices
  around the critical vertex $\critv^1_1$,
  which are not drawn.
  } 
  \label{fig:close-open-ex}
\end{figure}

For an overview of the idea of our algorithm,
we first use the example in \Cref{fig:close-open-ex}
to illustrate several important observations.
These observations provide insights into the solution
and introduce the key issue to solve.
We then discuss the key issue in detail.
Finally, we describe our solution 
in words, and postpone the formal pseudocode
to Section~\ref{sec:lvlset-co-min-cyc-code}.

Now consider the example in \Cref{fig:close-open-ex},
and let ${z_ 1,z_2,z_3}$ be
an arbitrary sequence of persistent 1-cycles for 
$\big[\critval^1_2,\critval^1_4\big)$.
By definition,
there exist 2-chains 
\[A_{2}\subseteq \cplx^1_{(1,3)}\text{, }
A_{3}\subseteq \cplx^1_{(2,4)}\text{, and }
A_{4}\subseteq \cplx^1_{(3,4]}\] 
such that
\[z_{1}+z_{2}=\partial(A_2)\text{, }
z_{2}+z_{3}=\partial(A_3)\text{, and }
z_{3}=\partial(A_{4}).\]
Assume that $\big[\critval^1_2,\critval^1_4\big)$
is produced by a simplex-wise interval which is still denoted $[K_\fbirth,K_\fdeath]$,
and let $A=A_2+A_3+A_4$.
We have $\partial(A)=z_1\subseteq K_\fbirth$.
One strategy we adopt 
for approaching the problem
is that we separate $K_\fbirth$ from
the remaining parts of $\cplx^\Dim_{(\birth-1,\death]}$ 
and tackle 
$K_\fbirth$ and $\cplx^\Dim_{(\birth-1,\death]}\setminus K_\fbirth$
individually.
So we separate
$A$ into the part that is in $K_\fbirth$
and the part that is not.
Since $\cplx^\Dim_{(\birth-1,\death]}=\cplx^1_{(1,4]}$ in our example,
the part of $A$ {\it not} in $K_\fbirth$ comes from
different 2-connected components of $\cplx^1_{(1,4]}\setminus K_\fbirth$,
which are $\comp_0$, $\comp_1$, and $\comp_2$
as shown in \Cref{fig:broken_octopus}.
We then observe the following: 
\begin{itemize}
    \item \emph{Any component of $\comp_0$, $\comp_1$, or $\comp_2$
that intersects $A$
must be completely included in $A$.}
\end{itemize}
This is because a 2-simplex of such a component (e.g., $\comp_1$) not in $A$
would cause $\partial(A)$ to contain 1-simplices not in $K_\fbirth$,
contradicting $z_1=\partial(A)\subseteq K_\fbirth$
(the formal justification is in \Cref{sec:lvlset-co-min-cyc-pf}).
For the same reason, we also observe:
\begin{itemize}
    \item \emph{Any component intersecting $A$
must have its boundary\footnote{The boundary   here means the boundary of the component
as a $(\Dim+1)$-chain.} contained in $K_\fbirth$.}
\end{itemize}
For example, in \Cref{fig:broken_octopus}, no 2-simplices in $\comp_2$ can fall in $A$
(because the boundary of $\comp_2$ is not contained in $K_\fbirth$),
while $\comp_1$ can either be totally in or disjoint with $A$.
The proof of Proposition~\ref{prop:levelzz-co-alg-cycles-cut}
formally justifies this observation.
We also notice that there is exactly one 2-connected component
of $\cplx^1_{(1,4]}\setminus K_\fbirth$ (i.e., $\comp_0$ in \Cref{fig:broken_octopus})
whose boundary resides in $K_\fbirth$ and contains $\fsimp{\Fcal}{\fbirth-1}$
(see Proposition~\ref{prop:only-one-comp-contain-creator}).
(While this is not drawn in \Cref{fig:close-open-ex},
we assume that $K$ is triangulated in a way that $\fsimp{}{\fbirth-1}$
is shared by the boundaries of $\comp_0$ and $\comp_2$.)
A fact about $\comp_0$ is
that it is always included in $A$
(see the proof of Proposition~\ref{prop:levelzz-co-alg-cycles-cut}).
For the other components
with boundaries contained in $K_\fbirth$ (e.g., $\comp_1$ in \Cref{fig:broken_octopus}),
in general, any subset of them can contribute to a certain $A$
and take part in forming the persistent cycles.
For example,
in \Cref{fig:broken_octopus}, 
only $\comp_0$ contributes to the persistent 1-cycles
${z'_1,z'_2,z'_3}$,
and both $\comp_0$, $\comp_1$ contribute to
${z''_1,z''_2,z''_3}$.

The crux of the algorithm, therefore, is to \emph{determine a subset
of the components} along with $\comp_0$ contributing to 
the optimal persistent cycles
(a complicated monkey saddle
with multiple forks may result in many such components),
because
we can compute the 
optimal persistent cycles 
under a \emph{fixed} choice of the subset.
To see this,
suppose that ${z''_1,z''_2,z''_3}$ in \Cref{fig:close-open-ex}
are the optimal persistent 1-cycles 
for $\big[\critval^1_2,\critval^1_4\big)$
under the choice of the subset $\bigSet{\comp_0,\comp_1}$,
i.e., ${z''_1,z''_2,z''_3}$ have the minimum sum of weight
among all persistent 1-cycles coming from  both
$\comp_0$ and $\comp_1$.
We first observe that
$z''_1$ must be the minimum 1-cycle 
homologous to $\partial(\comp_0)+\partial(\comp_1)$
in $K_\fbirth$.
Such a cycle $z''_1$ can be computed from a minimum \sCommaT{}cut
on a dual graph of $K_\fbirth$.
Also, the set of 1-cycles
$\bigSet{\zG_2^0\subseteq \cplx^1_{(2,3)}, \zG_3^0\subseteq \cplx^1_{(3,4)}}$
must be the ones in $\comp_0$ with the minimum sum of weight
such that 
\begin{equation}\label{eqn:c0-cyc-cond}
\zG_2^0\homolog \partial(\comp_0)\text{ in }\cplx^1_{(1,3)}\text{, }
\zG_2^0\homolog \zG_3^0\text{ in }\cplx^1_{(2,4)}\text{, and }
\zG_3^0\text{ null-homologous in }\cplx^1_{(3,4]}.\end{equation}
Additionally, $\zG_2^1\subseteq \cplx^1_{(2,3)}$
must be the minimum 1-cycle in $\comp_1$ 
such that 
\begin{equation}\label{eqn:c1-cyc-cond}
\zG_2^1\homolog \partial(\comp_1)\text{ in }\cplx^1_{(1,3)}
\text{ and }\zG_2^1\text{ is null-homologous in }\cplx^1_{(2,3]}.\end{equation}
See Step~\ref{alg-stp:co-conn-comp-close-at-right}
of Algorithm~\ref{alg:levelzz-closed-open} 
for a formal description.
To compute the minimum cycles $\bigSet{\zG_2^0, \zG_3^0}$,
$\bigSet{\zG_2^1}$,
we utilize an algorithm similar to Algorithm~\ref{alg:levelzz-open-open}.

Notice that a priori optimal selection of a subset of the components
is not obvious:
while introducing more components increases weights for cycles 
in the $\Dim$-th regular complexes 
(because the components are disjoint),
the cycle in $K_\fbirth$ corresponding to this choice may have a smaller weight
due to belonging to a different homology class
(e.g., $z''_1\homolog\partial(\comp_0)+\partial(\comp_1)$ may a have much smaller weight than
$z'_1\homolog\partial(\comp_0)$
in \Cref{fig:broken_octopus}).

Our solution is as follows:
generically, suppose that $\comp_0,\ldots,\comp_k$
are all the $(\Dim+1)$-connected components 
of $\cplx^\Dim_{(\birth-1,\death]}\setminus {K}_\fbirth$
with boundaries in $K_\fbirth$,
where $\comp_0$ is the one whose boundary contains $\fsimp{\Fcal}{\fbirth-1}$.
We do the following:
\begin{enumerate}
    \item For each $j=0,\ldots, k$, 
compute the minimum (possibly empty) $\Dim$-cycles
$\bigSet{\zG^j_\pcycind\given\birth\leq\pcycind<\death}$
in $\comp_j$
satisfying the conditions as in \Cref{eqn:c0-cyc-cond,eqn:c1-cyc-cond}
(see Step~\ref{alg-stp:co-conn-comp-close-at-right}
of Algorithm~\ref{alg:levelzz-closed-open} presented in Section~\ref{sec:lvlset-co-min-cyc-code}
for a formal description).
Notice that for $\comp_1$ in \Cref{fig:broken_octopus},
$\zG_3^1$ is empty,
which makes $\zG_2^1$ null-homologous in $\cplx^1_{(2,3]}$.

    \opt{DCG}{\medskip}
    \item
    Build a dual graph $\dgraphsec$ for $K_\fbirth$.
    Besides those vertices in $\dgraphsec$ corresponding to the $(\Dim+1)$-simplices,
we also add to $\dgraphsec$ \emph{dummy vertices} $\dummyV_0,\ldots,\dummyV_k$
corresponding to the boundaries 
$\partial(\comp_0),\ldots,\partial(\comp_k)$
and a single dummy vertex $\bar{\dummyV}$
corresponding to 
the remaining boundary portion of $K_\fbirth$.
Roughly speaking,
when a dummy vertex $\dummyV_j$ is said to ``correspond to''
$\partial(\comp_j)$,
one can imagine that a $(\Dim+1)$-dimensional ``cell'' 
with boundary $\partial(\comp_j)$
is added to $K_\fbirth$ 
and $\dummyV_j$ is the vertex dual to this cell.
In addition to the regular dual edges in $G$,
for each $\dummyV_j$,
we add to $\dgraphsec$ an {\it augmenting edge} connecting $\dummyV_j$ to $\bar{\dummyV}$
and let its weight be $\sum_{\pcycind=\birth}^{\death-1} w\big(\zG_\pcycind^j\big)$.
 Adding the augmenting edges helps us choose a subset of
$\comp_0,\ldots,\comp_k$ for forming the optimal persistence $\Dim$-cycles,
whose reason will be made clear later.
See also \Cref{fig:K_b} for an example of the dummy vertices 
and augmenting edges.

\opt{DCG}{\medskip}
    \item Compute the minimum \sCommaT{}cut $(S^*,T^*)$ 
of $\big(\dgraphsec,\dummyV_0,\bar{\dummyV}\big)$,
which produces an optimal sequence of levelset persistent $\Dim$-cycles 
for $\big[\pcritval_{\birth},\pcritval_{\death}\big)$.
\end{enumerate}

To see the correctness of the algorithm, consider an arbitrary \sCommaT{}cut $(S,T)$ 
of $\big(\dgraphsec,\dummyV_0,\bar{\dummyV}\big)$.
Whenever a $\dummyV_j$ is in $S$,
it means that the component $\comp_j$ is chosen to 
form the persistent cycles.
Since the augmenting edge $\bigSet{\dummyV_j,\bar{\dummyV}}$
is crossing the cut,
its weight $\sum_{\pcycind=\birth}^{\death-1} w\big(\zG_\pcycind^j\big)$ records the cost of introducing $\comp_j$
in forming the persistent cycles.
Moreover,
let $\dummyV_{\nu_0},\ldots,\dummyV_{\nu_\ell}$ 
be all the dummy vertices in $S$.
We then observe the following:

\opt{DCG}{\medskip}
\begin{observation}\label{obsv:non-aug-dual-cyc-homolog}
The non-augmenting edges in $E(S,T)$ 
produce a dual $\Dim$-cycle $z_{\birth-1}$ in
$K_\fbirth$ homologous to 
$\partial(\comp_{\nu_0})+\cdots+\partial(\comp_{\nu_\ell})$.
\end{observation}
\opt{DCG}{\medskip}

Then,
the $\Dim$-cycle $z_{\birth-1}$, along with 
all $\bigSet{\zG^{\nu_j}_\pcycind\given\birth\leq\pcycind<\death}$
from $\comp_{\nu_0},\ldots,\comp_{\nu_\ell}$,
form a sequence of persistent $\Dim$-cycles for $\big[\pcritval_{\birth},\pcritval_{\death}\big)$
whose sum of weight equals $w(S,T)$.
Section~\ref{sec:lvlset-co-min-cyc-pf} formally justifies our algorithm.
For a brief explanation  of \Cref{obsv:non-aug-dual-cyc-homolog},
recall that adding the dummy vertices to $G$
corresponds to adding the 
$(\Dim+1)$-dimensional ``cells'' to 
$K_\fbirth$,
making $K_\fbirth$  closed without boundary.
The cut $(S,T)$, with $S$  containing 
the dummy vertices $\dummyV_{\nu_0},\ldots,\dummyV_{\nu_\ell}$, thus becomes a separation of the 
boundaries
$\partial(\comp_{\nu_0})+\cdots+\partial(\comp_{\nu_\ell})$
with the remaining boundary portions in $K_\fbirth$.
Hence, the dual of the cut  $(S,T)$ must be homologous to 
$\partial(\comp_{\nu_0})+\cdots+\partial(\comp_{\nu_\ell})$.

\subsubsection{Pseudocode}\label{sec:lvlset-co-min-cyc-code}

We provide the full details of our algorithm in this subsection.
For the ease of exposition, 
so far we have let $\cplx^\Dim_{(\birth-1,\death]}$
be the complex on which we compute the optimal persistent cycles.
However, 
there is a problem with it,
which can be illustrated by
the example in \Cref{fig:close-open-ex}. Imagine that
$\critv^1_4$ and $\critv^1_5$ in the figure are pinched together,
so that $K$ is not a 2-manifold anymore (but still a weak 2-pseudomanifold).
The simplex-wise filtration $\lvlfilt_\Dim(f)$ can be constructed in a way that
the disc around $\critv^1_4$ is formed before the disc around $\critv^1_5$; 
such an $\lvlfilt_\Dim(f)$ is essentially the same as the one
before pinching. However, $\cplx^\Dim_{(\birth-1,\death]}$
now contains both 
$\critv^1_4$, $\critv^1_5$,
while the disc of $\critv^1_5$
should not be included in the computation\footnote{One problem with 
including $\critv^1_5$ is that there could be
another
2-connected component
($\comp_2$ in \Cref{fig:broken_octopus} 
with the right hole filled)
of $\cplx^\Dim_{(\birth-1,\death]}\setminus K_\fbirth$ 
whose boundary resides in $K_\fbirth$ and contains $\fsimp{\Fcal}{\fbirth-1}$,
breaking a critical fact our algorithm relies on.}.
Hence,
we make an adjustment to work on a complex $\wdtild{K}$ 
instead of $\cplx^\Dim_{(\birth-1,\death]}$;
see Step~\ref{alg-stp:co-conn-comp} of Algorithm~\ref{alg:levelzz-closed-open}
for the definition of $\wdtild{K}$.
It can be easily verified that each complex 
appearing in \Cref{dfn:pers-cyc-co}
is a subcomplex of $\wdtild{K}$.

Our exposition in Section~\ref{sec:lvlset-co-min-cyc-sum} also 
frequently deals with the complex $K_\fbirth$.
However, in the pseudocode (Algorithm~\ref{alg:levelzz-closed-open}),
$K_\fbirth$ takes a slightly different form:
we add to $K_\fbirth$ some missing $(\Dim+1)$-simplices
and denote the new complex as $\bar{K}_\fbirth$;
see Step~\ref{alg-stp:co-conn-comp} of the pseudocode
for definition.
Doing this
makes the description of the $(\Dim+1)$-connected components 
in Step~\ref{alg-stp:co-conn-comp-close-at-right} 
cleaner.

\opt{DCG}{\medskip}
\begin{algr}[Closed-open case]
\label{alg:levelzz-closed-open}
\hspace{1em}
\begin{enumerate}
    \item \label{alg-stp:co-conn-comp}
    Set the following:
    \begin{itemize}
    \item
    $\wdtild{K}=\cplx^\Dim_{(\birth-1,\death)}\union K_{\fdeath+1}$
    \item
    $\bar{K}_\fbirth=K_\fbirth\union\bigSet{(\Dim+1)
    \text{-simplices with all }\Dim\text{-faces in } K_\fbirth}$
    \end{itemize}

\opt{DCG}{\medskip}
    \item \label{alg-stp:co-conn-comp-close-at-right}
    Let $\comp_0,\ldots,\comp_k$
    be all the $(\Dim+1)$-connected components of $\wdtild{K}\setminus \bar{K}_\fbirth$
    such that $\partial(\comp_j)\subseteq\bar{K}_\fbirth$
    for each $j$,
    where $\comp_0$ is the unique one
    whose boundary contains $\fsimp{\Fcal}{\fbirth-1}$.
    {\rm(}Notice that the {\rm boundary} $\partial(\comp_j)$ here means 
    the boundary of the $(\Dim+1)$-chain $\comp_j$.{\rm)}

    For each $\comp_j$,
    let $M_j$ be the closure of $\comp_j$.
    Among all sets of $\Dim$-cycles of the form 
    \[\bigSet{z_\pcycind\subseteq M_j\intersect \cplx^\Dim_{(\pcycind,\pcycind+1)}\given\birth\leq\pcycind<\death}\]
    such that
    \begin{itemize}
        \item
    $z_\birth\homolog \partial(\comp_j)$ in $M_j\intersect \cplx^\Dim_{(\birth-1,\birth+1)}$,
        \item
    $z_{\pcycind-1}\homolog z_{\pcycind}$ in $M_j\intersect \cplx^\Dim_{(\pcycind-1,\pcycind+1)}$
    for each $\birth<\pcycind<\death$, and
        \item
    $z_{\death-1}$ is null-homologous in $M_j\intersect K_{\fdeath+1}$,
    \end{itemize}
    compute the set
    $\bigSet{\zG^j_\pcycind\given \birth\leq\pcycind<\death}$
    with the minimum sum of weight.

\opt{DCG}{\medskip}
    \item \label{alg-stp:co-dual-graph-K_b}
    Build a weighted dual graph $\dgraphsec$ from $\bar{K}_\fbirth$ as follows:

    Let each $(\Dim+1)$-simplex of $\bar{K}_\fbirth$
    correspond to a vertex in $\dgraphsec$,
    and add the dummy vertices $\bar{\dummyV},\dummyV_0,\ldots,\dummyV_k$ to $\dgraphsec$.
    Let $\thGvar$ denote the bijection from the $(\Dim+1)$-simplices
    to $V(\dgraphsec)\setminus\bigSet{\bar{\dummyV},\dummyV_0,\ldots,\dummyV_k}$.

    Let each $\Dim$-simplex $\sG$ of $\bar{K}_\fbirth$
    correspond to an edge $e$ in $\dgraphsec$,
    where the weight of $e$, $w(e)$, equals the weight of $\sG$.
    There are the following cases:
    \begin{itemize}
        \item {\rm $\sG$ has two $(\Dim+1)$-cofaces in $\bar{K}_\fbirth$:}
    $e$ is the usual one.

\opt{DCG}{\medskip}
        \item {\rm $\sG$ has one $(\Dim+1)$-coface $\tG$ in $\bar{K}_\fbirth$:}
    If $\sG\in\partial(\comp_j)$ for a $\comp_j$,
    let $e$ connect $\thGvar(\tG)$ and $\dummyV_j$ in $\dgraphsec$;
    otherwise, let $e$ connect $\thGvar(\tG)$ and $\bar{\dummyV}$.

\opt{DCG}{\medskip}
        \item {\rm $\sG$ has no $(\Dim+1)$-cofaces in $\bar{K}_\fbirth$:}
    If $\sG$ is in the boundaries of two components $\comp_i$ and $\comp_j$,
    let $e$ connect $\dummyV_i$ and $\dummyV_j$;
    if $\sG$ is in the boundary of only one component $\comp_j$,
    let $e$ connect $\dummyV_j$ and $\bar{\dummyV}$;
    otherwise, 
    let $e$ connect $\bar{\dummyV}$ on both ends.
    \end{itemize}
    In addition to the above edges,
    add the augmenting edges with weights as described.
    Let $\thGvar$ also denote the bijection from the $\Dim$-simplices
    to the non-augmenting edges
    and let $E'(S,T)$ denote the set of non-augmenting edges crossing 
    a cut $(S,T)$.

\opt{DCG}{\medskip}
    \item \label{alg-stp:co-final-cut}
    Compute the minimum \sCommaT{}cut $(S^*,T^*)$ of $\big(\dgraphsec,\dummyV_0,\bar{\dummyV}\big)$.
    Let $\dummyV_{\mu_0},\ldots,\dummyV_{\mu_l}$ be all the dummy vertices in $S^*$.
    Then, set 
        \[z^*_{\birth-1}=\thGvar\inv(E'(S^*,T^*))\text{ and }
        z^*_{\pcycind}=\sum_{j=0}^l \zG_\pcycind^{\mu_j}
        \text{ for each }\birth\leq\pcycind<\death.\]
    Return $z^*_{\birth-1},z^*_{\birth},\ldots,z^*_{\death-1}$
    as an optimal sequence of levelset persistent $\Dim$-cycles 
    for $\big[\pcritval_{\birth},\pcritval_{\death}\big)$.

\end{enumerate}
\end{algr}

As mentioned, 
the minimum cycles in Step~\ref{alg-stp:co-conn-comp-close-at-right}
can be computed using a similar approach of
Algorithm~\ref{alg:levelzz-open-open},
with a difference that Algorithm~\ref{alg:levelzz-open-open}
works on a complex ``closed on both ends''
while $M_j$ is ``closed only on the right''.
Therefore, we need to add a dummy vertex to the dual graph for the boundary,
which is put into the source.
Notice that we can build a single dual graph for all the $M_j$'s
and share the dummy vertex,
so that we only need to invoke one minimum cut computation.

\subsubsection{Correctness of the algorithm}\label{sec:lvlset-co-min-cyc-pf}
In this subsection, 
we prove the correctness of Algorithm~\ref{alg:levelzz-closed-open}.
We first state 
the following 
basic fact about $\fsimp{\Fcal}{\fbirth-1}$:

\opt{DCG}{\medskip}
\begin{proposition}
\label{prop:co-creator-no-coface}
The $\Dim$-simplex $\fsimp{\Fcal}{\fbirth-1}$ 
has no $(\Dim+1)$-cofaces in $\bar{K}_\fbirth$.
\end{proposition}
\begin{myproof}
Supposing instead that $\fsimp{\Fcal}{\fbirth-1}$ 
has a $(\Dim+1)$-coface $\tG$ in $\bar{K}_\fbirth$, 
then $\partial(\tG)\subseteq{K}_\fbirth$.
Since $\bar{K}_\fbirth\subseteq \cplx^\Dim_{(\birth-1,\birth]}$,
the $\Dim$-cycle $\partial(\tG)$ 
created by $\fsimp{\Fcal}{\fbirth-1}$
is a boundary in $\cplx^\Dim_{(\birth-1,\birth]}$.
Simulating a run of Algorithm~\ref{alg:zigzag-pers-abstr}
(presented in Appendix~\ref{sec:zigzag-pers-alg-abstr})
with input $\lvlfilt_\Dim(f)$,
at the $(\fbirth-1)$-th iteration,
we can let $\partial(\tG)$ be the representative $\Dim$-cycle at index
$\fbirth$ for the new interval $[\fbirth,\fbirth]$.
However,
since $\partial(\tG)$
is a boundary in $\cplx^\Dim_{(\birth-1,\birth]}$,
the interval starting with
$\fbirth$ must end with an index less than $\fdeath$,
which is a contradiction.
\end{myproof}

Proposition~\ref{prop:only-one-comp-contain-creator} justifies 
the operations in Step~\ref{alg-stp:co-conn-comp-close-at-right}:

\opt{DCG}{\medskip}
\begin{proposition}
\label{prop:only-one-comp-contain-creator}
Among all the $(\Dim+1)$-connected components of $\wdtild{K}\setminus \bar{K}_\fbirth$,
there is exactly one 
component
whose boundary resides in $\bar{K}_\fbirth$
and contains $\fsimp{\Fcal}{\fbirth-1}$.
\end{proposition}
\begin{myproof}
See Appendix~\ref{apx:pf-prop-only-one-comp-contain-creator}.
\end{myproof}

Finally, Propositions~\ref{prop:levelzz-co-alg-cut-cycles}
and~\ref{prop:levelzz-co-alg-cycles-cut}
lead to Theorem~\ref{thm:levelzz-co-alg-correct},
which is the conclusion.

\opt{DCG}{\medskip}
\begin{proposition}
\label{prop:levelzz-co-alg-cut-cycles}
For any \sCommaT{}cut $(S,T)$ of $\big(\dgraphsec,\dummyV_0,\bar{\dummyV}\big)$,
let $\dummyV_{\nu_0},\ldots,\dummyV_{\nu_\ell}$ be all the dummy vertices in $S$.
Furthermore,
let $z_{\birth-1}=\thGvar\inv(E'(S,T))$ 
and $z_{\pcycind}=\sum_{j=0}^\ell \zG_\pcycind^{\nu_j}$ 
for each $\birth\leq\pcycind<\death$.
Then, ${z_{\birth-1},z_\birth,\ldots,z_{\death-1}}$
is a sequence of levelset persistent $\Dim$-cycles 
for $\big[\pcritval_{\birth},\pcritval_{\death}\big)$
with $\sum_{\pcycind=\birth-1}^{\death-1} w(z_\pcycind)=w(S,T)$.
\end{proposition}
\begin{myproof}
Note that we can also consider $(S,T)$ as a \sCommaT{}cut of a graph 
derived by deleting the augmenting edges from $\dgraphsec$ 
where the sources are $\dummyV_{\nu_0},\ldots,\dummyV_{\nu_\ell}$
and the sinks are all the other dummy vertices.
This implies that $z_{\birth-1}=\thGvar\inv(E'(S,T))$ 
is homologous to $\partial(\comp_{\nu_0}+\cdots+\comp_{\nu_\ell})$
in $\bar{K}_\fbirth$.
Since $\dummyV_0$ is the source of $\dgraphsec$,
$\dummyV_0$ must be one of $\dummyV_{\nu_0},\ldots,\dummyV_{\nu_\ell}$.
Then, by Proposition~\ref{prop:only-one-comp-contain-creator},
$\partial(\comp_{\nu_0}+\cdots+\comp_{\nu_\ell})$ 
contains $\fsimp{\Fcal}{\fbirth-1}$.
So
$z_{\birth-1}$ must also contain $\fsimp{\Fcal}{\fbirth-1}$
because $z_{\birth-1}\homolog\partial(\comp_{\nu_0}+\cdots+\comp_{\nu_\ell})$ in $\bar{K}_\fbirth$
and
$\fsimp{\Fcal}{\fbirth-1}$ has no $(\Dim+1)$-coface 
in $\bar{K}_\fbirth$ (Proposition~\ref{prop:co-creator-no-coface}).
Furthermore, 
the properties of the cycles $\bigSet{\zG_\pcycind^j}$ computed in 
Step~\ref{alg-stp:co-conn-comp-close-at-right} 
of Algorithm~\ref{alg:levelzz-closed-open}
imply that $z_{\birth}=\zG_\birth^{\nu_0}+\cdots+\zG_\birth^{\nu_\ell}$
is homologous to $\partial(\comp_{\nu_0}+\cdots+\comp_{\nu_\ell})$ 
in $\cplx^\Dim_{(\birth-1,\birth+1)}$.
So $z_{\birth-1}\homolog z_{\birth}$ in $\cplx^\Dim_{(\birth-1,\birth+1)}$.

For 
${z_{\birth-1},z_\birth,\ldots,z_{\death-1}}$ 
to be
persistent $\Dim$-cycles for $\big[\pcritval_{\birth},\pcritval_{\death}\big)$,
we need to verify several other conditions in Definition~\ref{dfn:pers-cyc-co},
in which only one is non-trivial,
i.e., the condition that 
$[z_{\death-1}]\in\Hm_\Dim(K_{\fdeath})$ is the 
non-zero class in $\ker(\fmorph{\Dim}{\fdeath})$.
To see this, we first note that obviously $[z_{\death-1}]\in\ker(\fmorph{\Dim}{\fdeath})$.
To prove $[z_{\death-1}]\neq 0$, 
we use a similar approach in the proof of Proposition~\ref{prop:oo-alg-cons-facts},
i.e., simulate a run of Algorithm~\ref{alg:zigzag-pers-abstr}
for computing $\Pers_\Dim(\lvlfilt_\Dim(f))$
and show that 
$z_{\death-1}\subseteq K_\fdeath$ can be the representative cycle 
at index $\fdeath$ for the interval $[\fbirth,\fdeath]$.
The details are omitted.

For the weight, we have 
\begin{align*}
\begin{split}
w(S,T)
& =\sum_{e\in E'(S,T)}w(e)+\sum_{j=0}^\ell w\big(\bigSet{\dummyV_{\nu_j},\bar{\dummyV}}\big)
=w(z_{\birth-1})+\sum_{j=0}^\ell \sum_{\pcycind=\birth}^{\death-1} w\big(\zG_\pcycind^{\nu_j}\big)\\
& =w(z_{\birth-1})+\sum_{\pcycind=\birth}^{\death-1}\sum_{j=0}^\ell w\big(\zG_\pcycind^{\nu_j}\big)\allowbreak
=\sum_{\pcycind=\birth-1}^{\death-1} w(z_\pcycind)
\end{split}
\end{align*}
where $\bigSet{\dummyV_{\nu_j},\bar{\dummyV}}$ 
denotes the augmenting edge in $\dgraphsec$
connecting $\dummyV_{\nu_j}$ and $\bar{\dummyV}$.
\end{myproof}

\begin{proposition}
\label{prop:levelzz-co-alg-cycles-cut}
Let ${z_{\birth-1},z_\birth,\ldots,z_{\death-1}}$
be any sequence of levelset persistent $\Dim$-cycles 
for $\big[\pcritval_{\birth},\pcritval_{\death}\big)$;
then,
there exists a \sCommaT{}cut $(S,T)$ of $\big(\dgraphsec,\dummyV_0,\bar{\dummyV}\big)$
with $w(S,T)\leq\sum_{\pcycind=\birth-1}^{\death-1} w(z_\pcycind)$.
\end{proposition}
\begin{myproof}
By definition,
there exist $(\Dim+1)$-chains 
${A}_{\birth}\subseteq \cplx^\Dim_{(\birth-1,\birth+1)},\ldots,
{A}_{\death-1}\subseteq \cplx^\Dim_{(\death-2,\death)},
{A}_{\death}\subseteq K_{\fdeath+1}$ such that
${z}_{\birth-1}+{z}_{\birth}=\partial\big({A}_\birth\big),\ldots,
{z}_{\death-2}+{z}_{\death-1}=\partial\big({A}_{\death-1}\big),\allowbreak
{z}_{\death-1}=\partial\big({A}_{\death}\big)$.
Let $A=\sum_{\pcycind=\birth}^{\death}A_\pcycind$;
then, $\partial(A)=z_{\birth-1}$.
Let $\comp_{\nu_0},\ldots,\comp_{\nu_\ell}$
be all the components 
defined in Step~\ref{alg-stp:co-conn-comp-close-at-right} of Algorithm~\ref{alg:levelzz-closed-open} 
which intersect $A$.
We claim that each $\comp_{\nu_j}\subseteq A$.
For contradiction,
suppose instead that there is a $\sG\in\comp_{\nu_j}$
not in $A$.
Let $\sG'$
be a simplex in $A\intsec \comp_{\nu_j}$.
Since $\sG$, $\sG'$ are both in $\comp_{\nu_j}$,
there must be 
a $(\Dim+1)$-path $\tG_1,\ldots,\tG_q$ 
from $\sG$ to $\sG'$ in $\wdtild{K}\setminus \bar{K}_\fbirth$.
Note that $\sG\not\in A$ and $\sG'\in A$,
and so there is an $\iG$ 
such that $\tG_{\iG}\not\in A$ and $\tG_{\iG+1}\in A$.
Let $\tG^\Dim$ be a $\Dim$-face shared by $\tG_\iG$ and $\tG_{\iG+1}$ 
in $\wdtild{K}\setminus \bar{K}_\fbirth$;
then, $\tG^\Dim\in\partial(A)$ and $\tG^\Dim\not\in\bar{K}_\fbirth$.
This contradicts $\partial(A)=z_{\birth-1}\subseteq\bar{K}_\fbirth$.
So $\comp_{\nu_j}\subseteq A$.
We also note that
$\comp_{\nu_0},\ldots,\comp_{\nu_\ell}$ are all
the $(\Dim+1)$-connected components of $\wdtild{K}\setminus \bar{K}_\fbirth$
intersecting $A$.
The reason is that, 
if $\wdhat{\comp}$ is a component 
intersecting $A$
whose boundary 
is not completely in $\bar{K}_\fbirth$,
then we also have $\wdhat{\comp}\subseteq A$
and the justification is similar as above.
Let $\sG$ be a simplex in $\partial\big(\wdhat{\comp}\big)$ 
but not $\bar{K}_\fbirth$;
then, $\sG\in\partial(A)$.
To see this,
suppose instead that $\sG\not\in\partial(A)$.
Then $\sG$ has a $(\Dim+1)$-coface $\tG_1\in\wdhat{\comp}\subseteq A$
and a $(\Dim+1)$-coface $\tG_2\in A\setminus\wdhat{\comp}$.
We have $\tG_2\in\bar{K}_\fbirth$ because if not,
combining the fact that $\sG,\tG_1,\tG_2\in \wdtild{K}\setminus\bar{K}_\fbirth$
and $\tG_1\in\wdhat{\comp}$,
$\tG_2$ would be in $\wdhat{\comp}$.
As a face of $\tG_2$, $\sG$ must also be in $\bar{K}_\fbirth$,
which is a contradiction.
So we have $\sG\in\partial(A)$. Note that $\sG\not\in\bar{K}_\fbirth$,
which contradicts $\partial(A)\subseteq\bar{K}_\fbirth$,
and hence such a $\wdhat{\comp}$ cannot exist.
We then have
$\partial\big(A\setminus\bigunion_{j=0}^\ell \comp_{\nu_j}\big)
=\partial\big(A+\comp_{\nu_0}+\cdots+\comp_{\nu_\ell}\big)
=z_{\birth-1}+\partial\big(\comp_{\nu_0}\big)+\cdots+\partial\big(\comp_{\nu_\ell}\big)$,
where $A\setminus\bigunion_{j=0}^\ell \comp_{\nu_j}\subseteq \bar{K}_\fbirth$.
Now
$\partial\big(\comp_{\nu_0}\big)+\cdots+\partial\big(\comp_{\nu_\ell}\big)$
is homologous to $z_{\birth-1}$ 
in $\bar{K}_\fbirth$,
which means that
it must contain $\fsimp{\Fcal}{\fbirth-1}$
because $z_{\birth-1}$ contains $\fsimp{\Fcal}{\fbirth-1}$
and $\fsimp{\Fcal}{\fbirth-1}$ has no $(\Dim+1)$-coface 
in $\bar{K}_\fbirth$ (Proposition~\ref{prop:co-creator-no-coface}).
This implies that $\bigSet{\comp_{\nu_0},\ldots,\comp_{\nu_\ell}}$
contains $\comp_0$ by Proposition~\ref{prop:only-one-comp-contain-creator}.
Let $S=\thGvar\big(A\setminus\bigunion_{j=0}^\ell\comp_{\nu_j}\big)
\union\bigSet{\dummyV_{\nu_0},\ldots,\dummyV_{\nu_\ell}}$
and $T=V(\dgraphsec)\setminus S$.
It can be verified that  
$(S,T)$ is a \sCommaT{}cut of $\big(\dgraphsec,\dummyV_0,\bar{\dummyV}\big)$
and $z_{\birth-1}=\thGvar\inv(E'(S,T))$.

We then prove that $w(S,T)\leq\sum_{\pcycind=\birth-1}^{\death-1} w(z_\pcycind)$.
Let $A_\pcycind^{\nu_j}=M_{\nu_j}\intersect A_\pcycind$,
$z_{\pcycind}^{\nu_j}=M_{\nu_j}\intersect z_{\pcycind}$
for each $\pcycind$ and $j$.
For any $j$,
we claim the following
\begin{equation}\label{eqn:sum-A-nu-j-eq-z}
\partial\Bigg(\sum_{\pcycind=\birth+1}^{\death}A_\pcycind^{\nu_j}\Bigg)
=z_{\birth}^{\nu_j}
\end{equation}

To prove Equation~(\ref{eqn:sum-A-nu-j-eq-z}),
we first note the following
\begin{align*}
\begin{split}
\partial\Bigg(\sum_{\pcycind=\birth+1}^{\death}A_\pcycind^{\nu_j}\Bigg)=
\partial\Bigg(M_{\nu_j}\intersect\sum_{\pcycind=\birth+1}^{\death}A_\pcycind\Bigg)\text{, }
z_{\birth}^{\nu_j}
=M_{\nu_j}\intersect z_{\birth}
=M_{\nu_j}\intersect\partial\Bigg(\sum_{\pcycind=\birth+1}^{\death}A_\pcycind\Bigg)
\end{split}
\end{align*}

So we only need to show that 
$\partial\big(M_{\nu_j}\intersect\sum_{\pcycind=\birth+1}^{\death}A_\pcycind\big)
=M_{\nu_j}\intersect\partial\big(\sum_{\pcycind=\birth+1}^{\death}A_\pcycind\big)
$.
Letting $B=\sum_{\pcycind=\birth+1}^{\death}A_\pcycind$,
what we need to prove now becomes 
$\partial(M_{\nu_j}\intersect B)
=M_{\nu_j}\intersect\partial(B)$.
Consider an arbitrary 
$\sG\in\partial(M_{\nu_j}\intersect B)$.
We have that $\sG$ is a face of only one 
$(\Dim+1)$-simplex $\tG\in M_{\nu_j}\intersect B$.
Note that $\tG\in B$,
and we show that $\tG$ is the only $(\Dim+1)$-coface
of $\sG$ in $B$.
Suppose instead that 
$\sG$ has another 
$(\Dim+1)$-coface $\tG'$ in $B$.
Then, $\tG'\not\in M_{\nu_j}$ because $\tG'\not\in M_{\nu_j}\intersect B$.
Note that $B\subseteq \cplx^\Dim_{(\birth,\death]}$,
which means that $B$ is disjoint with 
$\bar{K}_\fbirth\subseteq \cplx^\Dim_{(\birth-1,\birth]}$.
So $\tG'\in B\subseteq \wdtild{K}\setminus\bar{K}_\fbirth$.
It is then true that $\sG\in \bar{K}_\fbirth$ because if not,
i.e., $\sG\in \wdtild{K}\setminus \bar{K}_\fbirth$,
then $\tG'$ would reside in $\comp_{\nu_j}\subseteq M_{\nu_j}$
(following from $\tG\in \comp_{\nu_j}$).
We now have $\tG\in B\subseteq \cplx^\Dim_{(\birth,\death]}$
and $\sG\in \bar{K}_\fbirth\subseteq \cplx^\Dim_{(\birth-1,\birth]}$,
which implies that $\sG\intersect\tG=\emptyset$,
contradicting $\sG\subseteq\tG$.
Therefore, $\sG\in\partial(B)$.
Since $\tG\in M_{\nu_j}$, we have $\sG\in M_{\nu_j}$,
and so $\sG\in M_{\nu_j}\intersect\partial(B)$.
On the other hand,
let $\sG$ be any $\Dim$-simplex in $M_{\nu_j}\intersect\partial(B)$.
Since $\sG\in\partial(B)$,
$\sG$ is a face of only one $(\Dim+1)$-simplex $\tG$ in $B$.
We then prove that $\tG\in M_{\nu_j}$.
Suppose instead that $\tG\not\in M_{\nu_j}$.
Then, since $\sG\in M_{\nu_j}$,
$\sG$ must be a face of $(\Dim+1)$-simplex $\tG'\in M_{\nu_j}$.
It follows that $\sG\in\bar{K}_\fbirth$, because if not,
$\tG$ and $\tG'$ would both be in $M_{\nu_j}$.
We then reach the contradiction that 
$\sG\intersect\tG=\emptyset$
because $\tG\in B\subseteq \cplx^\Dim_{(\birth,\death]}$
and $\sG\in \bar{K}_\fbirth\subseteq \cplx^\Dim_{(\birth-1,\birth]}$.
Therefore, $\sG$ is a face of 
only one $(\Dim+1)$-simplex $\tG$ in $M_{\nu_j}\intersect B$,
which means that $\sG\in\partial(M_{\nu_j}\intersect B)$.

Note that $\sum_{\pcycind=\birth}^{\death}A_\pcycind^{\nu_j}
=M_{\nu_j}\intersect A=\comp_{\nu_j}$
because $\comp_{\nu_j}\subseteq A$.
Hence,
by Equation~(\ref{eqn:sum-A-nu-j-eq-z})
\[
z_{\birth}^{\nu_j}=
\partial\Bigg(\sum_{\pcycind=\birth+1}^{\death}A_\pcycind^{\nu_j}\Bigg)=
\partial\Bigg(\sum_{\pcycind=\birth}^{\death}A_\pcycind^{\nu_j}\Bigg)+
\partial\big(A_\birth^{\nu_j}\big)=
\partial\big(\comp_{\nu_j}\big)+
\partial\big(A_\birth^{\nu_j}\big)
\]
Now we have
$z_{\birth}^{\nu_j}+\partial\big(\comp_{\nu_j}\big)=
\partial\big(A_\birth^{\nu_j}\big)$,
i.e.,
$z_{\birth}^{\nu_j}\homolog\partial\big(\comp_{\nu_j}\big)$
in $M_{\nu_j}\intersect \cplx^\Dim_{(\birth-1,\birth+1)}$.
Similar to Equation~(\ref{eqn:sum-A-nu-j-eq-z}), 
for each $\pcycind$ s.t.\ $\birth<\pcycind<\death$,
we have 
$\partial\big(\sum_{\eta=\pcycind}^{\death}A_\eta^{\nu_j}\big)=
z_{\pcycind-1}^{\nu_j}$
and
$\partial\big(\sum_{\eta=\pcycind+1}^{\death}A_\eta^{\nu_j}\big)=
z_{\pcycind}^{\nu_j}$.
Therefore,
$\partial\big(A_\pcycind^{\nu_j}\big)=
z_{\pcycind-1}^{\nu_j}+z_{\pcycind}^{\nu_j}$,
i.e.,
$z_{\pcycind-1}^{\nu_j}\homolog z_{\pcycind}^{\nu_j}$
in $M_{\nu_j}\intersect \cplx^\Dim_{(\pcycind-1,\pcycind+1)}$.
We also have that 
$\partial\big(\sum_{\eta=\death}^{\death}A_\eta^{\nu_j}\big)=
z_{\death-1}^{\nu_j}$,
i.e.,
$z_{\death-1}^{\nu_j}$ is null homologous in 
$M_{\nu_j}\intersect K_{\fdeath+1}$.
So $\bigSet{z_\pcycind^{\nu_j}\given \birth\leq\pcycind<\death}$ 
is a set of $\Dim$-cycles satisfying 
the condition specified in Step~\ref{alg-stp:co-conn-comp-close-at-right}
of Algorithm~\ref{alg:levelzz-closed-open},
which means that
$\sum_{\pcycind=\birth}^{\death-1} w(\zG^{\nu_j}_\pcycind)\leq \sum_{\pcycind=\birth}^{\death-1} w(z^{\nu_j}_\pcycind)$.

Finally, 
we have
\begin{align*}
w(S,T) 
& =\sum_{e\in E'(S,T)}w(e)+\sum_{j=0}^\ell w\big(\bigSet{\dummyV_{\nu_j},\bar{\dummyV}}\big)
=w(z_{\birth-1})+\sum_{j=0}^\ell \sum_{\pcycind=\birth}^{\death-1} w\big(\zG_\pcycind^{\nu_j}\big)\\
& \leq w(z_{\birth-1})+\sum_{\pcycind=\birth}^{\death-1}\sum_{j=0}^\ell w\big(z_\pcycind^{\nu_j}\big)
=\sum_{\pcycind=\birth-1}^{\death-1} w(z_\pcycind)
\end{align*}
where $\bigSet{\dummyV_{\nu_j},\bar{\dummyV}}$ 
denotes the augmenting edge in $\dgraphsec$
connecting $\dummyV_{\nu_j}$ and $\bar{\dummyV}$.
\end{myproof}

\begin{theorem}
\label{thm:levelzz-co-alg-correct}
Algorithm~\ref{alg:levelzz-closed-open} computes an optimal sequence of level persistent \Dim-cycles
for a given {\rm closed-open} interval.
\end{theorem}

\subsection{Closed-closed case}
\label{sec:lvlset-cc-min-cyc}

\begin{figure}
  \centering
  \subfloat[]{
       \opt{arxiv}{\includegraphics[width=0.43\linewidth]{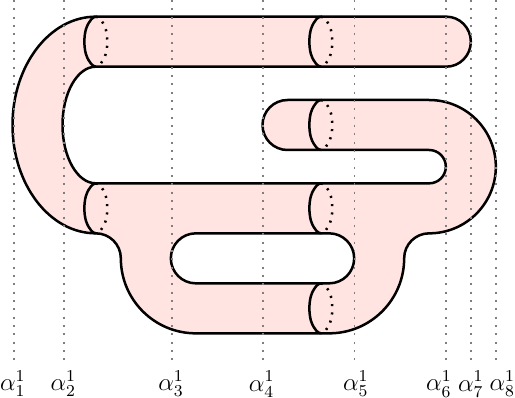}}
       \opt{DCG}{\includegraphics[width=0.45\linewidth]{fig/tube}}
  \label{fig:tube}}
  \opt{arxiv}{\hspace{1.5em}}
  \opt{DCG}{\hspace{1.4em}}
  \subfloat[]{
       \opt{arxiv}{\includegraphics[width=0.43\linewidth]{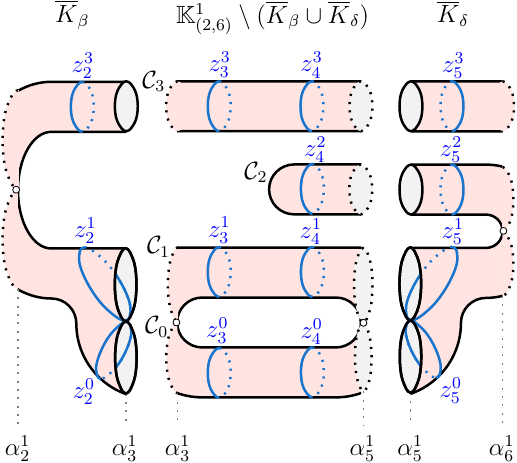}}
       \opt{DCG}{\includegraphics[width=0.45\linewidth]{fig/broken_tube}}
    \label{fig:broken_tube}}

\opt{DCG}{\medskip}
  \caption{(a) A complex $K$ with  height function $f$ taken over the horizontal line
  and  1st critical values listed
  at the bottom.
  (b) The relevant subcomplex $\cplx^\Dim_{(\birth-1,\death+1)}=\cplx^1_{(2,6)}$ 
  for the interval $\big[\critval^1_3,\critval^1_5\big]$,
  where $\bar{K}_\fbirth$ and $\bar{K}_\fdeath$ 
  are broken from the remaining parts
  for a better illustration.
  An empty dot indicates that the point is not included in the space.} 
  \label{fig:closed-ex}
\end{figure}

In the subsection,
we describe the computation of
the optimal persistent $\Dim$-cycles 
for a {\it closed-closed} interval 
$\big[\pcritval_{\birth},\pcritval_{\death}\big]$ from $\Pers_\Dim(\lvldgm_\Dim(f))$,
which is produced by a simplex-wise interval $[K_\fbirth,K_\fdeath]$ 
from $\Pers_\Dim(\lvlfilt_\Dim(f))$. 
Due to the similarity to the closed-open case,
we only describe the algorithm briefly.
\Cref{fig:closed-ex}
provides an example for $\Dim=1$,
in which different sequences of persistent 1-cycles are formed for
the interval $\big[\critval^1_3,\critval^1_5\big]$,
and two of them are
${z_2^1+z_2^3,z_3^1+z_3^3,z_4^1+z_4^3,z_5^1+z_5^3}$
and ${z_2^0,z_3^0,z_4^0+z_4^2,z_5^0+z_5^2}$.

Similar to the previous cases,
we have the following
relevant portion of $\lvlfilt_\Dim(f)$:
\begin{align}
\begin{split}
\label{eqn:cc-intv-diag-cmplx-seq}
\cplx^\Dim_{(\birth-1,\birth)}
\incto\cdots\incto
K_{\fbirth-1}\inctosp{\fsimp{}{\fbirth-1}} K_{\fbirth}
\incto\cdots\incto
\cplx^\Dim_{(\birth-1,\birth+1)}
\bakincto\cdots\bakincto
\cplx^\Dim_{(\birth,\birth+1)}
\incto
\cdots\\
\bakincto
\cplx^\Dim_{(\death-1,\death)}
\incto\cdots\incto
\cplx^\Dim_{(\death-1,\death+1)}
\bakincto\cdots\bakincto
K_{\fdeath}
\bakinctosp{\fsimp{}{\fdeath}} K_{\fdeath+1}
\bakincto\cdots\bakincto
\cplx^\Dim_{(\death,\death+1)}.
\end{split}
\end{align}
The creator $\fsimp{}{\fbirth-1}$
and the destroyer $\fsimp{}{\fdeath}$
of the simplex-wise interval $[K_\fbirth,K_\fdeath]$ 
are both $\Dim$-simplices~\cite{carlsson2009zigzag-realvalue},
and the computation can be restricted to the subcomplex $\cplx^\Dim_{(\birth-1,\death+1)}$.
Roughly speaking,
the algorithm for the closed-closed case
resembles the algorithm for the closed-open case
in that it now performs similar operations on {\it both} $K_\fbirth$ and $K_\fdeath$
as Algorithm~\ref{alg:levelzz-closed-open} does on $K_\fbirth$.
The idea is as follows:
\begin{enumerate}
    \item
    First, instead of directly working on $K_\fbirth$ and $K_\fdeath$,
    we work on $\bar{K}_\fbirth$ and $\bar{K}_\fdeath$,
    which include some missing $(\Dim+1)$-simplices. 
    Formally,
    $\bar{K}_\fbirth= K_\fbirth\union\bigSet{(\Dim+1)
    \text{-simplices with all }\Dim\text{-faces in } K_\fbirth}$,
    and $\bar{K}_\fdeath$ is defined similarly.

\opt{DCG}{\medskip}
    \item
    Let $\comp_0,\ldots,\comp_k$
    be all the $(\Dim+1)$-connected components of 
    $\cplx^\Dim_{(\birth-1,\death+1)}\setminus \big(\bar{K}_\fbirth\union \bar{K}_\fdeath\big)$
    with boundaries in $\bar{K}_\fbirth\union \bar{K}_\fdeath$.
    Then, only $\comp_0,\ldots,\comp_k$
    can be used to form the persistent $\Dim$-cycles in the $\Dim$-th regular complexes.
    Re-index these components 
    such that $\comp_0,\ldots,\comp_h$ ($h\leq k$) are all the ones 
    in $\comp_0,\ldots,\comp_k$
    whose boundaries contain {\it both}
    $\fsimp{\Fcal}{\fbirth-1}$ and $\fsimp{\Fcal}{\fdeath}$.
    We have that $h=0\text{ or }1$.
    If $h=0$, then $\comp_0$ must take part in forming a sequence of persistent 
    cycles for $\big[\pcritval_{\birth},\pcritval_{\death}\big]$.
    If $h=1$, then either $\comp_0$
    or $\comp_1$ but not both must take part in forming persistent cycles for the interval.

\opt{DCG}{\medskip}
    \item
    Compute minimum $\Dim$-cycles in the $\Dim$-th regular complexes 
    similarly as in Step~\ref{alg-stp:co-conn-comp-close-at-right}
    of Algorithm~\ref{alg:levelzz-closed-open}.
    For a $\comp_j$, let $M_j$ be its closure.
    If the boundary of $\comp_j$ lies completely in $\bar{K}_\fbirth$,
    the computed $\Dim$-cycles $\bigSet{\zG^j_\pcycind\subseteq 
    M_j\intsec\cplx^\Dim_{(\pcycind,\pcycind+1)} \given \birth\leq\pcycind<\death}$ 
    is the set 
    with the minimum sum of weight satisfying
    the conditions as in Step~\ref{alg-stp:co-conn-comp-close-at-right} 
    of Algorithm~\ref{alg:levelzz-closed-open}.
    If the boundary of $\comp_j$ lies completely in $\bar{K}_\fdeath$,
    the computed minimum $\Dim$-cycles
    satisfy symmetric conditions.
    If the boundary of $\comp_j$ intersects both $\bar{K}_\fbirth$ and $\bar{K}_\fdeath$,
    the computed minimum $\Dim$-cycles satisfy:
    $\zG^j_{\birth}\homolog\partial(\comp_j)\intersect \bar{K}_\fbirth$ 
    in $\cplx^\Dim_{(\birth-1,\birth+1)}$,
    $\zG^j_{\death-1}\homolog\partial(\comp_j)\intersect \bar{K}_\fdeath$ 
    in $\cplx^\Dim_{(\death-1,\death+1)}$,
    and the consecutive cycles are homologous.

\opt{DCG}{\medskip}
    \item To compute the optimal persistent $\Dim$-cycles, we build
    a dual graph $G$ for $\bar{K}_\fbirth\union\bar{K}_\fdeath$, in which
    the boundary of each $\comp_j$ corresponds to a dummy vertex $\dummyV_j$,
    and the remaining 
    boundary portion corresponds to a dummy vertex $\bar{\dummyV}$.
    We also add the augmenting edges to $G$
    and set their weights similarly to Algorithm~\ref{alg:levelzz-closed-open}.
    For each $i$ s.t.\ $0\leq i\leq h$,
    we 
    compute the minimum cut 
    on $G$
    with source being $\bigSet{\dummyV_i}$
    and sink being $\bigSet{\bar{\dummyV},\dummyV_0,\ldots,\dummyV_h}
    \setminus\bigSet{\dummyV_i}$.
    The minimum of the min-cuts for all $i$
    produces an optimal sequence of persistent $\Dim$-cycles for $\big[\pcritval_{\birth},\pcritval_{\death}\big]$.    
\end{enumerate}

We can look at \Cref{fig:closed-ex} for intuitions
of the above algorithm.
In \Cref{fig:broken_tube}, there are four 2-connected components 
of $\cplx^1_{(2,6)}\setminus\big(\bar{K}_\fbirth\union \bar{K}_\fdeath\big)$
with boundaries in $\bar{K}_\fbirth\union \bar{K}_\fdeath$,
which are $\comp_0$, $\comp_1$, $\comp_2$, and $\comp_3$.
Among them, $\comp_0$, $\comp_1$ are the ones whose boundaries
contain both $\fsimp{\Fcal}{\fbirth-1}$ and $\fsimp{\Fcal}{\fdeath}$.
The persistent 1-cycles 
${z_2^1+z_2^3,z_3^1+z_3^3,z_4^1+z_4^3,z_5^1+z_5^3}$
come from the components $\comp_1$ and $\comp_3$,
in which the starting cycle $z_2^1+z_2^3$ is homologous
to $\partial(\comp_1)\intersect\bar{K}_\fbirth
+\partial(\comp_3)\intersect\bar{K}_\fbirth$,
and the ending cycle $z_5^1+z_5^3$ is homologous
to $\partial(\comp_1)\intersect\bar{K}_\fdeath
+\partial(\comp_3)\intersect\bar{K}_\fdeath$.
Another sequence
${z_2^0,z_3^0,z_4^0+z_4^2,z_5^0+z_5^2}$
comes from $\comp_0$ and $\comp_2$,
in which the starting cycle $z_2^0$ is homologous
to $\partial(\comp_0)\intersect\bar{K}_\fbirth$,
and the ending cycle $z_5^0+z_5^2$ is homologous
to $\partial(\comp_0)\intersect\bar{K}_\fdeath
+\partial(\comp_2)$.
To compute the optimal sequence of persistent 1-cycles,
one first computes the minimum 1-cycles (e.g., $\bigSet{\zG_3^3,\zG_4^3}$) 
in each component of $\comp_0,\ldots,\comp_3$.
Then, to determine the optimal combination of the components 
and the persistent $\Dim$-cycles in $K_\fbirth$ and $K_\fdeath$,
one leverages the dual graph of $\bar{K}_\fbirth\union \bar{K}_\fdeath$
and the augmenting edges.

We finally notice that for the degenerate case of $\birth=\death$,
since there are no $\Dim$-th regular complexes between
$K_\fbirth$ and $K_\fdeath$,
the algorithm needs an adjustment:
one simply does not add augmenting edges at all.

\paragraph{Complexity.}
Let $n$ be the size of $K$.
Then, for the three algorithms in this section,
operations other than the minimum cut computation 
can be done in $O(n\log n)$ time.
Using the max-flow algorithm by Orlin~\cite{orlin2013max},
the time complexity of all three algorithms is $O(n^2)$.
Notice that we assume persistence intervals to be given
so that the time used for computing the levelset zigzag barcode 
is not included.

\subsection{Experiments}\label{sec:exp}
\begin{figure}
  \centering
  \opt{arxiv}{\includegraphics[width=0.4\linewidth]{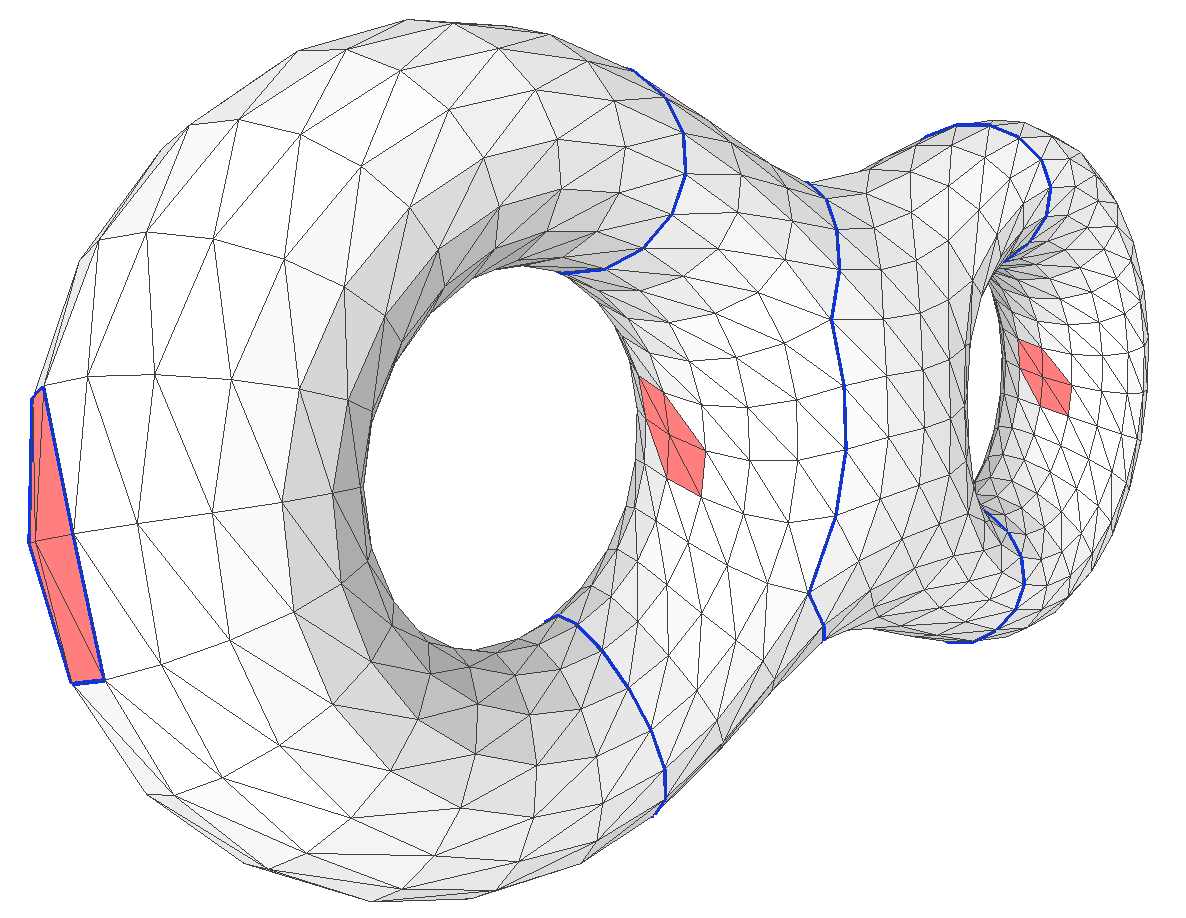}\medskip}
  \opt{DCG}{\includegraphics[width=0.45\linewidth]{fig/dtorus1}\medskip}
  \caption{Optimal levelset persistent 1-cycles (blue) computed by our software for an open-open interval
  for a double torus. Discs of critical vertices are colored red. Parts of the cycles and meshes
  hidden from the view are symmetric to what are shown.} 
  \label{fig:dtorus}
\end{figure}

\begin{figure}
  \centering
  \opt{arxiv}{\includegraphics[width=0.55\linewidth]{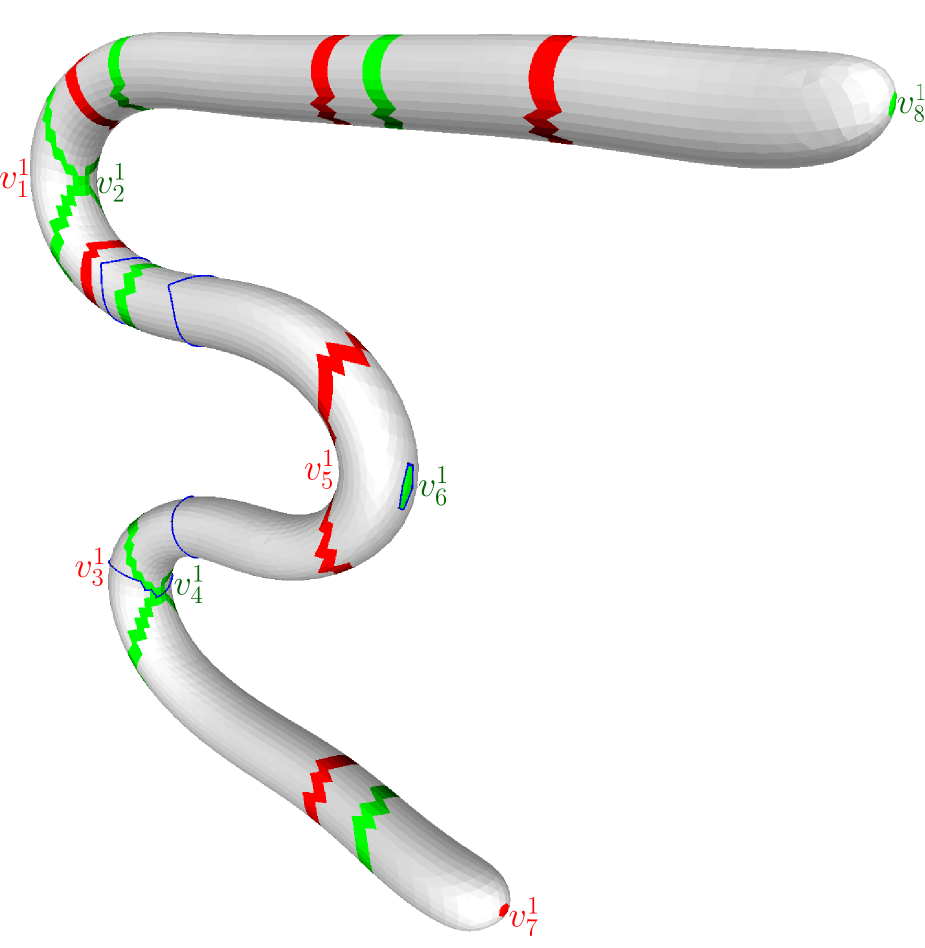}\medskip}
  \opt{DCG}{\includegraphics[width=0.65\linewidth]{fig/snake4}\medskip}
  \caption{An optimal sequence of three persistent 1-cycles (blue) for a closed-open interval 
  $\big[\critval^1_{4},\critval^1_{6}\big)$ on a snake-shaped model
  where triangles containing 1st critical values are alternatively colored
  as red and green.
  Those (red) triangles containing $\critval^1_1$ are completely hidden.
  Notice that the first cycle in the sequence (between $v^1_3$, $v^1_4$
  and touching $v^1_4$)
  contains two separate components.} 
  \label{fig:snake}
\end{figure}

We implemented our algorithms for the open-open and closed-open 
intervals for $p=1$ and performed experiments
on some triangular meshes
with height functions taken.
See \Cref{fig:dtorus,fig:snake}
for the computed optimal levelset persistent 1-cycles.
The experiments demonstrate that our algorithms produce optimal cycles
with nice quality which also capture variations of the
topological features within the persistence intervals.

\section{Equivalence of \texorpdfstring{$\Dim$}{p}-th and classical levelset filtrations}
\label{sec:equiv-p-levelzz}

In this section, we prove that the $\Dim$-th levelset filtration
defined in Section~\ref{sec:p-levelzz} and the classical definition
by Carlsson et al.~\cite{carlsson2009zigzag-realvalue}
produce equivalent
$\Dim$-th persistence intervals.
We first recall the classical definition
in Section~\ref{sec:levelzz}
and provide the proof in Section~\ref{sec:equiv-p-levelzz-equiv}.

\subsection{Classical levelset zigzag}\label{sec:levelzz}

Throughout this section,
let $K$ be a finite simplicial complex with underlying space $X=|K|$
and $f:X\to\Real$ be a generic PL function with critical values
$\critval_0=-\infty<\critval_1<\cdots<\critval_\critvcnt<\critval_{\critvcnt+1}=\infty$.
The original construction~\cite{carlsson2009zigzag-realvalue} 
of levelset zigzag persistence
picks regular values $\regval_0,\regval_1,\ldots,\regval_{\critvcnt}$ 
such that $\critval_i<\regval_i<\critval_{i+1}$ for each $i$.
Then, the {\it levelset filtration}
of $f$, 
denoted $\clvldgm(f)$, is defined as
\begin{align}
\begin{split}
\label{eqn:old-lvldgm}
\clvldgm(f):f\inv(\regval_0) \incto
f\inv[\regval_0,\regval_1]\bakincto 
f\inv(\regval_1) \incto 
f\inv[\regval_1,\regval_2] \bakincto
\cdots 
\incto f\inv[\regval_{\critvcnt-1},\regval_{\critvcnt}]
\bakincto f\inv(\regval_{\critvcnt}).
\end{split}
\end{align}

In order to align with our constructions in Section~\ref{sec:p-levelzz},
we adopt an alternative but equivalent definition of $\clvldgm(f)$
as follows, where we denote $f\inv(\critval_i,\critval_j)$ as $\subsp_{(i,j)}$:
\begin{align}
\begin{split}
\label{eqn:new-lvldgm}
\clvldgm(f): \subsp_{(0,1)} \incto 
\subsp_{({0},{2})}\bakincto 
\subsp_{(1,2)} \incto 
\subsp_{(1,3)} \bakincto 
\cdots 
\incto \subsp_{({\critvcnt-1},{\critvcnt+1})}
\bakincto \subsp_{({\critvcnt},{\critvcnt+1})} 
\end{split}
\end{align}
Notice that each $\subsp_{(i,{i+1})}$ deformation retracts to $f\inv({\regval_i})$
and each $\subsp_{({i-1},{i+1})}$ deformation retracts to $f\inv{[\regval_{i-1},\regval_i]}$,
so that zigzag modules induced by the two filtrations in~(\ref{eqn:old-lvldgm})
and~(\ref{eqn:new-lvldgm}) are isomorphic.

The barcode $\Pers_\Dim(\clvldgm(f))$ is then
the classical version of $\Dim$-th levelset barcode defined in~\cite{carlsson2009zigzag-realvalue}.
Intervals in $\Pers_\Dim(\clvldgm(f))$
can also be mapped to
real-value intervals in which the homological features persist:

\opt{DCG}{\bigskip}
\begin{center}
\begin{tabular}{ l l c l }
 \toprule
 {closed-open:}   & $\big[\subsp_{(\birth-1,\birth+1)},\subsp_{(\death-1,\death)}\big]$ & 
 $\Leftrightarrow$ & $[\critval_\birth,\critval_\death)$ \\

 {open-closed:}   & $\big[\subsp_{(\birth,\birth+1)},\subsp_{(\death-1,\death+1)}\big]$ & 
 $\Leftrightarrow$ & $(\critval_\birth,\critval_\death]$ \\

 {closed-closed:} & $\big[\subsp_{(\birth-1,\birth+1)},\subsp_{(\death-1,\death+1)}\big]$ & 
 $\Leftrightarrow$ & $[\critval_\birth,\critval_\death]$ \\

 {open-open:}     & $\big[\subsp_{(\birth,\birth+1)},\subsp_{(\death-1,\death)}\big]$ & 
 $\Leftrightarrow$ & $(\critval_\birth,\critval_\death)$ \\[4pt] 
 \bottomrule
\end{tabular}
\end{center}

\subsection{Equivalence}\label{sec:equiv-p-levelzz-equiv}

The following theorem is the major conclusion of this section 
(recall that $\clvldgm_\Dim(f)$ is the continuous version of 
$\Dim$-th levelset filtration
of $f$ as in Definition~\ref{dfn:p-lvlpers}):

\opt{DCG}{\medskip}
\begin{theorem}\label{thm:p-lvldgm-valid}
For an arbitrary PL function $f$ as defined above, the real-value intervals in
$\Pers_\Dim(\clvldgm(f))$ and $\Pers_\Dim(\clvldgm_\Dim(f))$ are the same.
\end{theorem}
\opt{DCG}{\medskip}

To prove Theorem~\ref{thm:p-lvldgm-valid}, we first provide the following proposition:

\opt{DCG}{\medskip}
\begin{proposition}\label{prop:cross-non-p-critval-iso}
Let $\critval_\ell\leq\critval_i<\critval_j\leq\critval_k$
be critical values of $f$.
If for each $h$ such that $\ell<h\leq i$ or $j\leq h<k$,
$\critval_h$ is not a $\Dim$-th homologically critical value,
then the map $\Hm_\Dim(\subsp_{(i,j)})\to\Hm_\Dim(\subsp_{(\ell,k)})$
induced by inclusion
is an isomorphism.
\end{proposition}
\begin{myproof}

\begin{figure}
    \centering\normalsize 
\begin{tikzpicture}%
\draw (0,0) node(00) {$\emptyset$} ;
\draw (2,0) node(20) {$\subsp_{(i,i+1)}$} ;
\draw (4,0) node(40) {$\subsp_{(i+1,i+2)}$} ;
\draw (6,0) node(60) {$\subsp_{(i+2,j)}$} ;
\draw (8,0) node(80) {$\subsp_{(i+3,i+4)}$} ;
\draw (10,0) node(100) {$\subsp_{(i+4,k)}$} ;

\draw (1,1) node(11) {$\subsp_{(i,i+1)}$} ;
\draw (3,1) node(31) {$\subsp_{(i,i+2)}$} ;
\draw (5,1) node(51) {$\subsp_{(i+1,j)}$} ;
\draw (7,1) node(71) {$\subsp_{(i+2,i+4)}$} ;
\draw (9,1) node(91) {$\subsp_{(i+3,k)}$} ;

\draw (2,2) node(22) {$\subsp_{(i,i+2)}$} ;
\draw (4,2) node(42) {$\subsp_{(i,j)}$} ;
\draw (6,2) node(62) {$\subsp_{(i+1,i+4)}$} ;
\draw (8,2) node(82) {$\subsp_{(i+2,k)}$} ;

\draw (3,3) node(33) {$\subsp_{(i,j)}$} ;
\draw (5,3) node(53) {$\subsp_{(i,i+4)}$} ;
\draw (7,3) node(73) {$\subsp_{(i+1,k)}$} ;

\draw (4,4) node(44) {$\subsp_{(i,i+4)}$} ;
\draw (6,4) node(64) {$\subsp_{(i,k)}$} ;

\draw (5,5) node(55) {$\subsp_{(i,k)}$} ;

\draw[->] (00) -- (11);
\draw[->] (20) -- (11); 
\draw[->] (20) -- (31);
\draw[->] (40) -- (31);
\draw[->] (40) -- (51);
\draw[->] (60) -- (51);
\draw[->] (60) -- node[above,sloped,yshift=-2pt]{${\scriptstyle\approx}$} (71);
\draw[->] (80) -- node[above,sloped,yshift=-2pt]{${\scriptstyle\approx}$} (71);
\draw[->] (80) -- node[above,sloped,yshift=-2pt]{${\scriptstyle\approx}$} (91);
\draw[->] (100) -- node[above,sloped,yshift=-2pt]{${\scriptstyle\approx}$} (91);

\draw[->] (11) -- (22);
\draw[densely dashed,->] (31) -- (22); 
\draw[densely dashed,->] (31) -- (42);
\draw[densely dashed,->] (51) -- (42); 
\draw[densely dashed,->] (51) -- (62);
\draw[densely dashed,->] (71) -- (62);
\draw[densely dashed,->] (71) -- (82);
\draw[densely dashed,->] (91) -- (82);

\draw[->] (22) -- (33);
\draw[densely dashed,->] (42) -- (33);
\draw[densely dashed,->] (42) -- (53);
\draw[densely dashed,->] (62) -- (53);
\draw[densely dashed,->] (62) -- (73);
\draw[densely dashed,->] (82) -- (73);

\draw[->] (33) -- (44);
\draw[densely dashed,->] (53) -- (44); 
\draw[densely dashed,->] (53) -- (64);
\draw[densely dashed,->] (73) -- (64);  

\draw[->] (44) -- (55);
\draw[densely dashed,->] (64) -- (55);

\draw [decorate,decoration={brace,amplitude=10pt,mirror},xshift=-4pt,yshift=0pt]
(0,-0.35) -- 
node [midway,below,yshift=-10pt] {$\Dcal_2$}
(10.8,-0.35);

\draw [decorate,decoration={brace,amplitude=10pt},xshift=-4pt,yshift=0pt]
(0-0.4,0+0.4) --
node [midway,above,sloped,yshift=10pt] {$\Dcal_1$}
 (5-0.4,5+0.4);
\end{tikzpicture}
    \caption{Mayer-Vietoris pyramid for $j=i+3$, $k=i+5$.}
    \label{fig:pyramid}
\end{figure}

We first prove that the inclusion-induced map 
$\Hm_\Dim(\subsp_{(i,j)})\to\Hm_\Dim(\subsp_{(i,k)})$
is an isomorphism. 
For this, we build a Mayer-Vietoris pyramid
similar to the one 
in~\cite{carlsson2009zigzag-realvalue} for proving the Pyramid Theorem.
Moreover, in the pyramid,
let $\Dcal_1$ be the filtration along the northeastbound diagonal
and $\Dcal_2$ be the filtration along the bottom.
An example is shown in \Cref{fig:pyramid} for $j=i+3$, $k=i+5$,
where inclusion arrows in $\Dcal_1$, $\Dcal_2$ are solid
and the remaining arrows are dashed.
Since all diamonds in the pyramid are 
Mayer-Vietoris diamonds~\cite{carlsson2009zigzag-realvalue},
each interval
$[\subsp_{(i,i+\birth)},\subsp_{(i,i+\death)}]$ in $\Pers_\Dim(\Dcal_1)$
corresponds to the following interval in $\Pers_\Dim(\Dcal_2)$:
\[
\begin{cases}
\big[\subsp_{(i,i+1)},\subsp_{(i+\death-1,i+\death)}\big]
& \text{if }\birth=1 \\
\big[\subsp_{(i+\birth-2,i+\birth)},\subsp_{(i+\death-1,i+\death)}\big]
& \text{otherwise}
\end{cases}
\]
The fact that $\critval_h$ is not a $\Dim$-th critical value
for $j\leq h<k$ implies that linear maps in
$\Hm_\Dim(\Dcal_2)$ induced by arrows between
$\subsp_{(j-1,j)}$ and $\subsp_{(k-1,k)}$ (i.e., those arrows marked with
`$\approx$' in the example) are isomorphisms.
This means that no interval in $\Pers_\Dim(\Dcal_2)$
starts with $\subsp_{(h-1,h+1)}$ or ends with $\subsp_{(h-1,h)}$ for $j\leq h<k$.
So we have that no interval in $\Pers_\Dim(\Dcal_1)$
starts with $\subsp_{(i,h+1)}$ or ends with $\subsp_{(i,h)}$ for $j\leq h<k$.
This in turn means that each 
$\Hm_\Dim(\subsp_{(i,h)})\to\Hm_\Dim(\subsp_{(i,h+1)})$
in $\Hm_\Dim(\Dcal_1)$ is an isomorphism for $j\leq h<k$,
which implies that their composition 
$\Hm_\Dim(\subsp_{(i,j)})\to\Hm_\Dim(\subsp_{(i,k)})$
is an isomorphism.

Symmetrically, we have that $\Hm_\Dim(\subsp_{(i,k)})\to\Hm_\Dim(\subsp_{(\ell,k)})$
is an isomorphism, which implies that $\Hm_\Dim(\subsp_{(i,j)})\to\Hm_\Dim(\subsp_{(\ell,k)})$
is an isomorphism.
\end{myproof}

\begin{myproof}[Proof of Theorem~\ref{thm:p-lvldgm-valid}]
Let 
$\pcritval_0=-\infty<\pcritval_{1}<\cdots<\pcritval_{\pcritvcnt}<\pcritval_{\pcritvcnt+1}=\infty$ 
be all the 
$\Dim$-th homologically critical values of $f$,
and let $\pcritval_i=\critval_{\pcritind_i}$ for each~$i$.
Note that $\subsp^\Dim_{(i,j)}=\subsp_{(\pcritind_i,\pcritind_j)}$
for $i<j$.
We first show that the two zigzag modules as defined in \Cref{fig:twozzmods} are isomorphic,
where the upper module is $\Hm_\Dim(\clvldgm(f))$,
and the lower module is a version of $\Hm_\Dim(\clvldgm_\Dim(f))$
elongated
by making several copies of $\Dim$-th homology groups
of the regular subspaces and connecting them
by identity maps. 
The commutativity of the diagram is easily seen
because all maps are induced by inclusion. 
The vertical maps
are isomorphisms by Proposition~\ref{prop:cross-non-p-critval-iso}.
Hence, the two modules in \Cref{fig:twozzmods} are isomorphic.
This means that persistence intervals of the two modules
bijectively map to each other, 
and we also have that
their corresponding real-value
intervals are the same. For example, an interval 
$[\subsp_{(\pcritind_\birth-1,\pcritind_\birth+1)},\subsp_{(\pcritind_\death-1,\pcritind_\death)}]$
from $\Hm_\Dim(\clvldgm(f))$ corresponds to an interval 
$[\subsp^\Dim_{({\birth-1},{\birth+1})},\subsp^\Dim_{({\death-1},\death)}]$
from $\Hm_\Dim(\clvldgm_\Dim(f))$,
and they both produce the real-value interval
$[\critval_{\pcritind_\birth},\critval_{\pcritind_\death})$.
\end{myproof}

\begin{figure}
    \centering\normalsize 
\begin{tikzcd}[column sep=0.85em]
\cdots \arrow[r] &
  \Hm_\Dim\big(\subsp_{(\pcritind_{i}-1,\pcritind_{i}+1)}\big) 
    \arrow[r,leftarrow] \arrow[d,"\rotatebox{270}{$\approx$}"] &
  \Hm_\Dim\big(\subsp_{(\pcritind_{i},\pcritind_{i}+1)}\big) 
    \arrow[r] \arrow[d,"\rotatebox{270}{$\approx$}"] &
  \Hm_\Dim\big(\subsp_{(\pcritind_{i},\pcritind_{i}+2)}\big) 
    \arrow[r,leftarrow] \arrow[d,"\rotatebox{270}{$\approx$}"] &
  \cdots \arrow[r,leftarrow] &
  \Hm_\Dim\big(\subsp_{(\pcritind_{i+1}-1,\pcritind_{i+1})}\big) 
    \arrow[r] \arrow[d,"\rotatebox{270}{$\approx$}"] &
  \cdots\\
\cdots \arrow[r] & 
  \Hm_\Dim\big(\subsp^\Dim_{({i-1},{i+1})}\big) \arrow[r,leftarrow] &
  \Hm_\Dim\big(\subsp^\Dim_{({i},{i+1})}\big) \arrow[r,equal] &
  \Hm_\Dim\big(\subsp^\Dim_{({i},{i+1})}\big) \arrow[r,equal] &
  \cdots \arrow[r,equal] &
  \Hm_\Dim\big(\subsp^\Dim_{({i},{i+1})}\big) \arrow[r] &
  \cdots 
\end{tikzcd}
    \caption{Two isomorphic zigzag modules where the upper module is $\Hm_\Dim(\clvldgm(f))$
and the lower module is an elongated version of $\Hm_\Dim(\clvldgm_\Dim(f))$.}
    \label{fig:twozzmods}
\end{figure}

\section{Connection to interval decomposition}\label{sec:conn-to-interv-decomp}

In this section, we connect our levelset persistent cycles 
to the interval decomposition of zigzag modules.
Specifically, for a generic PL function $f$,
we show that
levelset persistent $\Dim$-cycles induce the entire 
interval decomposition for $\Hm_\Dim(\lvldgm_\Dim(f))$ (Theorem~\ref{thm:pers-cyc-ind-decomp}),
and part of
an interval decomposition for $\Hm_\Dim(\lvlfilt_\Dim(f))$
with the rest being from the trivial intervals
(Theorem~\ref{thm:per-cyc-ind-rep-cyc}).

To reach the conclusions,
we first define the general zigzag representatives~\cite{maria2014zigzag,DBLP:conf/soda/Dey0M25}
as mentioned in \Cref{sec:pers-cyc-dfn},
which generate an interval submodule
in a straightforward way,
i.e., picking a cycle for a homology class at
each position.

\opt{DCG}{\medskip}
\begin{definition}%
\label{dfn:rep-cyc}
Let 
$\Dim\geq 0$,
$\lfilt: \lcplx_0 \leftrightarrow \cdots \leftrightarrow \lcplx_\lfiltcnt$
be a simplex-wise zigzag filtration,
and $[\birth,\death]$ be an interval in $\Pers_{\Dim}({\lfilt})$.
Denote each linear map in $\Hm_\Dim(\lfilt)$ as 
$\morph{\lfilt}{j}:\Hm_\Dim(\lcplx_j)\leftrightarrow\Hm_\Dim(\lcplx_{j+1})$.
The {\it representative $\Dim$-cycles} for $[\birth,\death]$
is a sequence of $\Dim$-cycles
$\Set{{z}_i\subseteq \lcplx_i\given \birth\leq i\leq\death}$
such that:
\begin{enumerate}
    \item \label{itm:dfn-rep-cyc-birth-cond}
    For $\birth>0$,
    $[z_\birth]$ is not in $\img(\morph{\lfilt}{\birth-1})$ 
    if 
    $\lcplx_{\birth-1}\incto \lcplx_{\birth}$ is forward,
    or $[z_\birth]$ is the non-zero class in $\ker(\morph{\lfilt}{\birth-1})$ otherwise.

    \opt{DCG}{\medskip}
    \item \label{itm:dfn-rep-cyc-death-cond}
    For $\death<\lfiltcnt$,
    $[z_\death]$ is not in $\img(\morph{\lfilt}{\death})$ 
    if $\lcplx_{\death}\bakincto \lcplx_{\death+1}$ is backward,
    or $[z_\death]$ is the non-zero class in $\ker(\morph{\lfilt}{\death})$ otherwise.

    \opt{DCG}{\medskip}
    \item \label{itm:dfn-rep-cyc-bi-mapsto}
    For each $i\in[\birth,\death-1]$, 
    $[z_i]\leftrightarrow[z_{i+1}]$ by $\morph{\lfilt}{i}$,
    i.e., $[z_i]\mapsto[z_{i+1}]$ 
    or $[z_i]\mapsfrom[z_{i+1}]$.
\end{enumerate}
The interval submodule $\Ical$ of $\Hm_\Dim(\lfilt)$
{\it induced by} the representative $\Dim$-cycles 
is a module
such that $\Ical(i)$ equals the 1-dimensional vector space generated 
by $[{z}_i]$ for $i\in[\birth,\death]$ 
and equals $0$ otherwise,
where $\Ical(i)$ is the $i$-th vector space in $\Ical$.
\end{definition}
\opt{DCG}{\medskip}

The following proposition connects
representative cycles
to the interval decomposition:

\opt{DCG}{\medskip}
\begin{proposition}
\label{prop:rep-ind-mod-decomp}
Let 
$\Dim\geq 0$,
$\lfilt: \lcplx_0 \leftrightarrow \cdots \leftrightarrow \lcplx_\lfiltcnt$
be a simplex-wise zigzag filtration with $\Hm_\Dim(\lcplx_0)=0$,
and
$\Pers_\Dim(\lfilt)=\Set{[\birth_\aG,\death_\aG]\given \aG\in\Acal}$ 
be indexed by a set $\Acal$.
One has that 
$\Hm_\Dim(\lfilt)$ is 
\textbf{equal to}
a direct sum of interval submodules
$\bigoplus_{\aG\in \Acal}\Ical^{[\birth_\aG,\death_\aG]}$
if and only if for each $\aG$, 
$\Ical^{[\birth_\aG,\death_\aG]}$ is induced by 
a sequence of representative $\Dim$-cycles for $[\birth_\aG,\death_\aG]$.
\end{proposition}
\begin{myproof}
Suppose that $\Hm_\Dim(\lfilt)=\bigoplus_{\aG\in\Acal}\Ical^{[\birth_\aG,\death_\aG]}$ 
is an interval decomposition.
For each $\aG$, define a sequence of representative $\Dim$-cycles 
$\Set{{\cyc}_i^\aG\given \birth_\aG\leq i\leq\death_\aG}$
for $[\birth_\aG,\death_\aG]$
by letting ${\cyc}_i^\aG$ be an arbitrary cycle in the non-zero class
of the $i$-th vector space of 
$\Ical^{[\birth_\aG,\death_\aG]}$.
It can be verified that $\Set{{\cyc}_i^\aG\given \birth_\aG\leq i\leq\death_\aG}$
are valid representative $\Dim$-cycles for $[\birth_\aG,\death_\aG]$
inducing $\Ical^{[\birth_\aG,\death_\aG]}$.
This finishes the ``only if'' part of the proof.
The ``if'' part follows from the proof of Proposition~9 in~\cite{dey2021computing}.
\end{myproof}

Now consider
a generic PL function $f:|K|\to\Real$ on a finite simplicial complex $K$
and a non-trivial interval $[K_\fbirth,K_\fdeath]$ 
of $\Pers_\Dim(\lvlfilt_\Dim(f))$ for $\Dim\geq 1$.
A sequence of levelset persistent $\Dim$-cycles $\Set{z_i}$ 
for $[K_\fbirth,K_\fdeath]$
{\it induces} a sequence of representative $\Dim$-cycles 
$\Set{\zG_j\given \fbirth\leq j\leq\fdeath}$ for this interval
as follows:
for any $K_j\in[K_\fbirth,K_\fdeath]$, we can always find a $z_i$
satisfying $z_i\subseteq K_j$, i.e., the complex 
that $z_i$ originally belongs to (as in \Cref{dfn:pers-cyc-oo,dfn:pers-cyc-co,dfn:pers-cyc-cc})
is included in $K_j$;
then, set $\zG_j=z_i$.
It can be verified that the induced representative $\Dim$-cycles
are valid so that levelset persistent cycles also induce interval submodules. 
We then have the following fact:

\opt{DCG}{\medskip}
\begin{theorem}\label{thm:per-cyc-ind-rep-cyc}
For any non-trivial interval $J$ of $\Pers_\Dim(\lvlfilt_\Dim(f))$,
a sequence of levelset persistent $\Dim$-cycles for $J$ induces an interval submodule
of $\Hm_\Dim(\lvlfilt_\Dim(f))$ over $J$.
These induced interval submodules constitute part of an interval decomposition
for $\Hm_\Dim(\lvlfilt_\Dim(f))$,
where the remaining parts are from the trivial intervals.
\end{theorem}
\begin{myproof}
This follows from Proposition~\ref{prop:rep-ind-mod-decomp}.
Note that in order to apply Proposition~\ref{prop:rep-ind-mod-decomp},
$\Hm_\Dim(\cplx^\Dim_{(0,1)})$ has to be trivial,
where $\cplx^\Dim_{(0,1)}$ is the starting complex of $\lvlfilt_\Dim(f)$.
If the minimum value of $f$ is $\Dim$-th critical,
then $\cplx^\Dim_{(0,1)}=\cplx_{(0,1)}=\emptyset$,
and so $\Hm_\Dim(\cplx^\Dim_{(0,1)})$ is trivial.
Otherwise, since
$\Hm_\Dim(\cplx^\Dim_{(0,1)})=\Hm_\Dim(\cplx_{(0,2)})$ 
(Proposition~\ref{prop:cross-non-p-critval-iso})
and $\cplx_{(0,2)}$ deformation retracts to a point,
we have that $\Hm_\Dim(\cplx^\Dim_{(0,1)})$ is trivial.
\end{myproof}

Similarly as for $\Hm_\Dim(\lvlfilt_\Dim(f))$,
levelset persistent $\Dim$-cycles 
can also induce 
interval submodules for $\Hm_\Dim(\lvldgm_\Dim(f))$,
the details of which are omitted.
The following fact follows:

\opt{DCG}{\medskip}
\begin{theorem}\label{thm:pers-cyc-ind-decomp}
Let $\Pers_\Dim(\lvldgm_\Dim(f))=\Set{J_k\given k\in\LG}$
where  $\LG$ is an index set. 
For any interval $J_k$ of $\Pers_\Dim(\lvldgm_\Dim(f))$,
a sequence of levelset persistent $\Dim$-cycles 
for $J_k$ induces an interval submodule
$\Ical_k$ of $\Hm_\Dim(\lvldgm_\Dim(f))$ over $J_k$.
Combining all the modules, 
one has an interval decomposition
$\Hm_\Dim(\lvldgm_\Dim(f))=\bigoplus_{k\in\LG}\Ical_k$.
\end{theorem}
\begin{myproof}
This follows from Theorem~\ref{thm:per-cyc-ind-rep-cyc}.
Note that $\Hm_\Dim(\lvldgm_\Dim(f))$ can be viewed as being ``contracted''
from $\Hm_\Dim(\lvlfilt_\Dim(f))$. 
While in Theorem~\ref{thm:per-cyc-ind-rep-cyc},
the induced interval submodules
form only part of the interval decomposition of $\Hm_\Dim(\lvlfilt_\Dim(f))$,
the remaining submodules from the trivial intervals disappear in the interval decomposition 
of $\Hm_\Dim(\lvldgm_\Dim(f))$.
\end{myproof}

\section*{Acknowledgment}
T.\ K.\ Dey was supported by NSF grants CCF 1839252 and 2049010.
T.\ Hou was supported by NSF grants CCF  1839252, 2049010, and 2439255.
A.\ Pulavarthy was supported by NSF grant CCF 2439255 and the graduate research assistant fund of the School of Computing 
at DePaul University.

\bibliographystyle{plain}
\bibliography{refs}

\appendix

\section{Missing proofs}

\subsection{Proof of Proposition~\ref{prop:reg-cmplx-disjoint}}
\label{sec:pf-prop-reg-cmplx-disjoint}

We only prove that $K_\fbirth\subseteq\cplx^\Dim_{(\birth-1,\birth]}$
because the proof for $K_\fdeath\subseteq\cplx^\Dim_{[\death,\death+1)}$
is similar.
For contradiction, assume instead that 
$K_\fbirth\nsubseteq\cplx^\Dim_{(\birth-1,\birth]}$.
Note that from $\cplx^\Dim_{(\birth-1,\birth]}$ 
to $\cplx^\Dim_{(\birth-1,\birth+1)}$,
we are not crossing any 
$\Dim$-th critical values,
and so the linear map 
$\Hm_\Dim(\cplx^\Dim_{(\birth-1,\birth]})\to\Hm_\Dim(\cplx^\Dim_{(\birth-1,\birth+1)})$
is an isomorphism (see Proposition~\ref{prop:cross-non-p-critval-iso}).
Since $\cplx^\Dim_{(\birth-1,\birth]}$ 
appears between $\cplx^\Dim_{(\birth-1,\birth)}$ and 
$\cplx^\Dim_{(\birth-1,\birth+1)}$ in $\lvlfilt_\Dim(f)$
(see Definition~\ref{dfn:lvlfilt}),
we have the following subsequence in $\lvlfilt_\Dim(f)$:
\[\cplx^\Dim_{(\birth-1,\birth)}\incto\cdots\incto 
\cplx^\Dim_{(\birth-1,\birth]}\incto\cdots\incto 
K_\fbirth\incto\cdots\incto
\cplx^\Dim_{(\birth-1,\birth+1)}\incto\cdots\incto
K_\fdeath\]
The fact that $[K_\fbirth,K_\fdeath]$ forms an interval in $\Pers_\Dim(\lvlfilt_\Dim(f))$
indicates that 
a $\Dim$-th homology class is born (and persists) 
when $\cplx^\Dim_{(\birth-1,\birth]}$
is included into $\cplx^\Dim_{(\birth-1,\birth+1)}$,
contradicting the fact that 
$\Hm_\Dim(\cplx^\Dim_{(\birth-1,\birth]})\to\Hm_\Dim(\cplx^\Dim_{(\birth-1,\birth+1)})$
is an isomorphism.

\subsection{Proof of Proposition~\ref{prop:f-level-compat-valid}}
\label{apx:pf-prop-f-level-compat-valid}
Let $S$ consist of simplices of $K$ not in $\cplx^\Dim_{(i,j)}$ 
whose interiors intersect $\subsp^\Dim_{(i,j)}$.
Then, let $\sG$ be a simplex of $S$ with no proper cofaces in $S$.
We have that there exists a $u\in\sG$ with $f(u)\in(\pcritval_i,\pcritval_j)$
and a $w\in\sG$ with $f(w)\not\in(\pcritval_i,\pcritval_j)$.
If $f(w)\leq \pcritval_i$, then all vertices in $\sG$
must have the function values falling in 
$(\pcritval_{i-1},\pcritval_{i+1})$
because $K$ is compatible with the $\Dim$-th levelsets of $f$.
We then have that 
$|\sG|\intsec \subsp^\Dim_{(i,j)}$
deformation retracts to 
$\bd(|\sG|)\intsec \subsp^\Dim_{(i,j)}$,
where $\bd(|\sG|)$ denotes the boundary of the 
topological disc $|\sG|$.
This implies that $\subsp^\Dim_{(i,j)}$
deformation retracts to $\subsp^\Dim_{(i,j)}\setminus\interior(\sG)$,
where $\interior(\sG)$ denotes the interior of $|\sG|$.
If $f(w)\geq \pcritval_j$,
the result is similar.
After doing the above for the all such $\sG$ in $S$,
we have that $\subsp^\Dim_{(i,j)}$
deformation retracts to 
$\subsp^\Dim_{(i,j)}\setminus\bigunion_{\sG\in S}\interior(\sG)$.
Note that 
$\subsp^\Dim_{(i,j)}\setminus\bigunion_{\sG\in S}\interior(\sG)=\big|\cplx^\Dim_{(i,j)}\big|$,
and so the proof is done.

\subsection{Proof of Proposition~\ref{prop:oo-alg-cons-facts}}
\label{apx:pf-prop-oo-alg-cons-facts}
For the proof, we first observe the following fact which
follows immediately
from Proposition~\ref{prop:rep-ind-mod-decomp}:

\opt{DCG}{\medskip}
\begin{proposition}
\label{prop:rep-cyc-non-zero}
Let $\Dim\geq 0$,
$\lfilt: \lcplx_0 \leftrightarrow \cdots \leftrightarrow \lcplx_\lfiltcnt$
be a simplex-wise filtration with $\Hm_\Dim(\lcplx_0)=0$,
$[\bG',\dG']$ be an interval in $\Pers_\Dim(\lfilt)$,
and ${\zG_{\bG'},\ldots,\zG_{\dG'}}$ 
be a sequence of representative $\Dim$-cycles 
for $[\bG',\dG']$.
One has that $\zG_i$ is not a boundary in $\lcplx_i$ for each $\bG'\leq i\leq\dG'$.
\end{proposition}
\opt{DCG}{\medskip}

The following fact is also helpful for our proof:

\opt{DCG}{\medskip}
\begin{proposition}
\label{prop:conn-comp-bound-indp}
Let $X$ be a simplicial complex,  
$A$ be a $\diml$-chain of $X$ where $\diml\geq 1$,
and $X'$ be the closure of a $\diml$-connected component of $X$;
one has that $X'\intersect \partial(A)=\partial(X'\intersect A)$.
\end{proposition}
\begin{myproof}
First, let $B$ be an arbitrary $\diml$-chain of $X$
and $\sG^{\diml-1}$ be an arbitrary $(\diml-1)$-simplex in $X$.
Define $\cof_\diml(B,\sG^{\diml-1})$ as the set of $\diml$-simplices in $B$
having $\sG^{\diml-1}$ as a face.
It can be verified that 
$\cof_\diml(B,\sG^{\diml-1})=\cof_\diml(X'\intersect B,\sG^{\diml-1})$
if $\sG^{\diml-1}\in X'$.

To prove the proposition, 
let $\sG^{\diml-1}$ be an arbitrary $(\diml-1)$-simplex in $X'\intersect \partial(A)$.
Since $\sG^{\diml-1}\in \partial(A)$,
we have that $\bigCard{\cof_\diml(A,\sG^{\diml-1})}$ is an odd number.
Since $\sG^{\diml-1}\in X'$,
the fact in the previous paragraph implies that
$\bigCard{\cof_\diml(X'\intersect A,\sG^{\diml-1})}=\bigCard{\cof_\diml(A,\sG^{\diml-1})}$ is also an odd number.
Therefore, $\sG^{\diml-1}\in\partial(X'\intersect A)$.
On the other hand, 
let $\sG^{\diml-1}$ be an arbitrary $(\diml-1)$-simplex in $\partial(X'\intersect A)$.
Then, $\bigCard{\cof_\diml(X'\intersect A,\sG^{\diml-1})}$ is an odd number.
Since $\sG^{\diml-1}$ is a face of a $\diml$-simplex in $X'$,
we have that $\sG^{\diml-1}\in X'$. 
Therefore,
$\bigCard{\cof_\diml(A,\sG^{\diml-1})}
=\bigCard{\cof_\diml(X'\intersect A,\sG^{\diml-1})}$ is an odd number.
So we have that $\sG^{\diml-1}\in \partial(A)$
and then $\sG^{\diml-1}\in X'\intersect\partial(A)$.
\end{myproof}

Now we prove Proposition~\ref{prop:oo-alg-cons-facts}.
Let $z_\birth,\ldots,z_{\death-1}$ be 
a sequence of persistent $\Dim$-cycles 
for $\big(\pcritval_{\birth},\pcritval_{\death}\big)$ as claimed.
Note that $[\partial(\fsimp{\Fcal}{\fbirth-1})]$ 
is the non-zero class in $\ker(\fmorph{\Dim}{\fbirth-1})$.
Therefore, by Definition~\ref{dfn:pers-cyc-oo},
$\partial(\fsimp{\Fcal}{\fbirth-1})\homolog z_\birth$ in $K_\fbirth$.
This means that
there exists a $(\Dim+1)$-chain $C\subseteq K_\fbirth$ such that 
$z_\birth+\partial(\fsimp{\Fcal}{\fbirth-1})=\partial(C)$.
Let $A_\birth=C+\fsimp{\Fcal}{\fbirth-1}$;
then, $z_\birth=\partial(A_\birth)$ where $A_\birth$ is a $(\Dim+1)$-chain 
in $K_{\fbirth-1}$ containing $\fsimp{\Fcal}{\fbirth-1}$.
Similarly, we have that
$z_{\death-1}=\partial(A_{\death})$ for a $(\Dim+1)$-chain 
$A_{\death}\subseteq K_{\fdeath+1}$ containing $\fsimp{\Fcal}{\fdeath}$.
By Definition~\ref{dfn:pers-cyc-oo},
there exists a $(\Dim+1)$-chain $A_\pcycind\subseteq \cplx^\Dim_{(\pcycind-1,\pcycind+1)}$ 
for each $\birth<\pcycind<\death$ such that 
$z_{\pcycind-1}+z_{\pcycind}=\partial(A_\pcycind)$.
Thus, $A_\birth,\ldots,A_{\death}$ are the $(\Dim+1)$-chains
satisfying the condition in Claim~\ref{itm:p+1chains-exist-sum-equal-all}.
Let $z'_\pcycind=K'\intersect z_\pcycind$
and $A'_\pcycind=K'\intersect A_\pcycind$ for each $\pcycind$.
By Proposition~\ref{prop:conn-comp-bound-indp}, $z'_\birth=\partial(A'_\birth)$.
Since $A'_\birth$ contains $\fsimp{\Fcal}{\fbirth-1}$,
it follows that $z'_\birth+\partial(\fsimp{\Fcal}{\fbirth-1})
=\partial\big(A'_\birth\setminus \Set{\fsimp{\Fcal}{\fbirth-1}}\big)$,
where $A'_\birth\setminus \Set{\fsimp{\Fcal}{\fbirth-1}}\subseteq K_\fbirth$.
It is then true that 
$z'_\birth\homolog\partial(\fsimp{\Fcal}{\fbirth-1})$ in $K_{\fbirth}$.
Now we simulate a run of Algorithm~\ref{alg:zigzag-pers-abstr}
for computing $\Pers_\Dim(\lvlfilt_\Dim(f))$.
Then,
at the $(\fbirth-1)$-th iteration of the run,
we can let $z'_\birth\subseteq K_{\fbirth}$ 
be the representative cycle at index $\fbirth$
for the new interval $[\fbirth,\fbirth]$.

Let $\lG$ be the index of the complex $\cplx^\Dim_{(\birth,\birth+2)}$
in $\lvlfilt_\Dim(f)$, i.e., $K_\lG=\cplx^\Dim_{(\birth,\birth+2)}$. 
In the run of Algorithm~\ref{alg:zigzag-pers-abstr},
the interval starting with $\fbirth$ must persist to $\lG$
because this interval ends with $\fdeath$.
At any $j$-th iteration for $\fbirth\leq j\leq\lG-2$,
other than the case that 
$\fmorph{\Dim}{j}$ is backward with a non-trivial cokernel,
the setting of representative cycles for 
all intervals persisting through
follows the trivial setting rule.
For the case that $\fmorph{\Dim}{j}$ is backward with a non-trivial cokernel,
since $z'_\birth\subseteq K_{j+1}$,
the setting of the representative cycles for the interval $[\fbirth,j+1]$
must also follow the trivial setting rule.
Hence, at the end of the $(\lG-2)$-th iteration,
$z'_\birth\subseteq K_{\lG-1}$ can be the representative cycle at index $\lG-1$
for the interval $[\fbirth,\lG-1]$.
Meanwhile,
it is true that 
$K'\intersect(z_\birth+z_{\birth+1})=K'\intersect z_\birth+K'\intersect z_{\birth+1}$.
So $z'_\birth+z'_{\birth+1}=K'\intersect\partial(A_{\birth+1})
=\partial(K'\intersect A_{\birth+1})=\partial\big(A'_{\birth+1}\big)$,
which means that $z'_\birth\homolog z'_{\birth+1}$ in $\cplx^\Dim_{(\birth,\birth+2)}=K_\lG$.
Therefore, $[z'_\birth]\mapsto[z'_{\birth+1}]$ by $\fmorph{\Dim}{\lG-1}$,
which means that $z'_{\birth+1}\subseteq K_{\lG}$ can be the representative cycle 
at index $\lG$ for the interval $[\fbirth,\lG]$.
By repeating the above arguments on each $z'_\pcycind$ that follows,
we have that 
$z'_{\death-1}\subseteq K_\fdeath$ can be the representative cycle 
at index $\fdeath$ for the interval $[\fbirth,\fdeath]$.
Finally, for contradiction, assume instead that
$\fsimp{\Fcal}{\fdeath}\not\in K'$.
This means that $\fsimp{\Fcal}{\fdeath}\not\in A'_{\death}$,
and hence $A'_{\death}\subseteq K_\fdeath$.
Since $z'_{\death-1}=\partial\big(A'_{\death}\big)$,
we then have that $z'_{\death-1}$ is a boundary in $K_\fdeath$.
However, 
by Proposition~\ref{prop:rep-cyc-non-zero},
$z'_{\death-1}$ cannot be a boundary in $K_\fdeath$,
which is a contradiction.
Therefore, Claim~\ref{itm:oo-intv-kill-simps-same-conn} is proved.
Furthermore, we have that $z'_\birth,\ldots,z'_{\death-1}$ and $A'_\birth,\ldots,A'_{\death}$
satisfy the condition in Claim~\ref{itm:p+1chains-exist-sum-equal-all}.

To prove the last statement of Claim~\ref{itm:p+1chains-exist-sum-equal-all},
first note that $\partial\big(\sum_{\pcycind=\birth}^{\death} A'_\pcycind\big)=0$.
Let $A'=\sum_{\pcycind=\birth}^{\death} A'_\pcycind$.
Since $\fsimp{\Fcal}{\fbirth-1}\in \cplx^\Dim_{(\birth-1,\birth+1)}$
and $\fsimp{\Fcal}{\fbirth-1}\not\in \cplx^\Dim_{(\birth,\birth+1)}$,
there must be a vertex in $\fsimp{\Fcal}{\fbirth-1}$
with function value in $\big(\pcritval_{\birth-1},\pcritval_\birth\big]$.
So $\fsimp{\Fcal}{\fbirth-1}\not\in \cplx^\Dim_{(\birth,\death+1)}$,
which means that $\fsimp{\Fcal}{\fbirth-1}\not\in A'_\pcycind$
for any $b<\pcycind\leq\death$.
We also have that $\fsimp{\Fcal}{\fbirth-1}\in A'_\birth$,
and hence $\fsimp{\Fcal}{\fbirth-1}\in A'$.
We then show that $A'$ equals the set of $(\Dim+1)$-simplices of $K'$.
First note that $A'\subseteq K'$.
Then, for contradiction,
suppose that there is a $(\Dim+1)$-simplex $\sG\in K'$ not in $A'$.
Since $\sG\in K'$,
there is a $(\Dim+1)$-path $\tG_1,\ldots,\tG_\ell$
from $\sG$ to $\fsimp{\Fcal}{\fbirth-1}$ in $K'$.
Since $\sG\not\in A'$ and $\fsimp{\Fcal}{\fbirth-1}\in A'$,
there must be a $j$ such that $\tG_j\not\in A'$ and $\tG_{j+1}\in A'$.
Let $\tG_j$ and $\tG_{j+1}$ share a $\Dim$-face $\tG^\Dim$;
then, $\tG^\Dim\in\partial(A')$,
contradicting the fact that $\partial(A')=0$.
For the disjointness of $A'_\birth,\ldots,A'_{\death}$,
suppose instead that there is a $\sG$ residing in more than one
of $A'_\birth,\ldots,A'_{\death}$.
Then,
$\sG$ can only reside in two consecutive chains $A'_\pcycind$ and $A'_{\pcycind+1}$,
because pairs of chains of other kinds are disjoint.
This implies that $\sG\not\in A'$, contradicting the fact that $A'$ 
contains all $(\Dim+1)$-simplices of $K'$.
Thus, Claim~\ref{itm:p+1chains-exist-sum-equal-all} is proved.

Combining the fact that $\partial(A')=0$,
$K'$ is a pure weak $(\Dim+1)$-pseudomanifold,
and Claim~\ref{itm:p+1chains-exist-sum-equal-all},
we can reach Claim~\ref{itm:K-prime-nobound}.

\subsection{Proof of Proposition~\ref{prop:only-one-comp-contain-creator}}
\label{apx:pf-prop-only-one-comp-contain-creator}
We first show that there is at least 
one such component.
Let ${z_{\birth-1},z_\birth,\ldots,z_{\death-1}}$
be a sequence of persistent $\Dim$-cycles 
for $\big[\pcritval_{\birth},\pcritval_{\death}\big)$.
Then, by definition,
there exist $(\Dim+1)$-chains 
$A_{\birth}\subseteq \cplx^\Dim_{(\birth-1,\birth+1)},\ldots,
A_{\death-1}\subseteq \cplx^\Dim_{(\death-2,\death)},
A_{\death}\subseteq K_{\fdeath+1}$ such that
$z_{\birth-1}+z_{\birth}=\partial(A_\birth),\ldots,
z_{\death-2}+z_{\death-1}=\partial(A_{\death-1}),
z_{\death-1}=\partial(A_{\death})$.
Let $A=\sum_{\pcycind=\birth}^{\death}A_\pcycind$;
then, $\partial(A)=z_{\birth-1}\subseteq 
\bar{K}_\fbirth$.
Note that 
$\fsimp{\Fcal}{\fbirth-1}\in z_{\birth-1}$ by definition,
which implies that $\fsimp{\Fcal}{\fbirth-1}$ is a face of 
only one $(\Dim+1)$-simplex $\tG\in A$.
Note that $\tG\not\in\bar{K}_\fbirth$ 
by Proposition~\ref{prop:co-creator-no-coface},
which means that $\tG\in \wdtild{K}\setminus \bar{K}_\fbirth$.
Let $\comp$ be the $(\Dim+1)$-connected component 
of $\wdtild{K}\setminus \bar{K}_\fbirth$ containing $\tG$.
We show that $\comp\subseteq A$.
For contradiction,
suppose instead that there is a $\tG'\in\comp$ which is not in $A$.
Since $\tG,\tG'\in\comp$,
there is a $(\Dim+1)$-path $\tG_1,\ldots,\tG_\ell$ 
from $\tG$ to $\tG'$ in $\wdtild{K}\setminus \bar{K}_\fbirth$.
Also since $\tG_1\in A$ and $\tG_\ell\not\in A$,
there must be an $\iG$ such that $\tG_{\iG}\in A$ and $\tG_{\iG+1}\not\in A$.
Let $\tG^\Dim$ be a $\Dim$-face shared by $\tG_\iG$ and $\tG_{\iG+1}$ 
in $\wdtild{K}\setminus \bar{K}_\fbirth$;
then, $\tG^\Dim\in\partial(A)$ and $\tG^\Dim\not\in\bar{K}_\fbirth$.
This contradicts $\partial(A)\subseteq\bar{K}_\fbirth$.
Since $\comp\subseteq A$,  
we have that
$\tG$ is the only $(\Dim+1)$-coface of 
$\fsimp{\Fcal}{\fbirth-1}$ in $\comp$,
which means that
$\fsimp{\Fcal}{\fbirth-1}\in\partial(\comp)$.
We then show that $\partial(\comp)\subseteq \bar{K}_\fbirth$.
For contradiction, suppose instead that there is a $\sG\in\partial(\comp)$ 
which is not in $\bar{K}_\fbirth$,
and let $\tG'$ be the only $(\Dim+1)$-coface of $\sG$ in $\comp$.
If $\sG$ has only one $(\Dim+1)$-coface in $\wdtild{K}$,
the fact that $\comp\subseteq A$
implies that $\tG'$ is the only $(\Dim+1)$-coface of $\sG$ in $A$.
Hence, $\sG\in\partial(A)$,
contradicting $\partial(A)\subseteq\bar{K}_\fbirth$.
If $\sG$ has another $(\Dim+1)$-coface $\tG''$ in $\wdtild{K}$,
then $\tG''$ must not be in $\bar{K}_\fbirth$ because
the $\Dim$-face $\sG$ of $\tG''$ is not in $\bar{K}_\fbirth$.
So $\tG''\in \wdtild{K}\setminus\bar{K}_\fbirth$.
Then, $\tG''\in\comp$
because it shares a $\Dim$-face $\sG\in\wdtild{K}\setminus\bar{K}_\fbirth$ 
with $\tG'\in\comp$,
contradicting the fact that 
$\tG'$ is the only $(\Dim+1)$-coface of $\sG$ in $\comp$.
Now we have constructed a $(\Dim+1)$-connected component $\comp$
of $\wdtild{K}\setminus \bar{K}_\fbirth$
whose boundary resides in $\bar{K}_\fbirth$
and contains $\fsimp{\Fcal}{\fbirth-1}$.

We then prove that there is only one such component.
For contradiction, suppose that there are two
components 
$\comp_l$, $\comp_j$ among
$\comp_0,\ldots,\comp_k$
whose boundaries contain $\fsimp{\Fcal}{\fbirth-1}$.
Then, at least one of $\comp_l$, $\comp_j$
does not contain $\fsimp{\Fcal}{\fdeath}$.
Let $\comp_j$ be the one {\it not} containing $\fsimp{\Fcal}{\fdeath}$.
Note that the set $\bigSet{\zG^j_\pcycind\given\birth\leq\pcycind<\death}$
computed in Step~\ref{alg-stp:co-conn-comp-close-at-right}
of Algorithm~\ref{alg:levelzz-closed-open}
satisfies that $\zG^j_{\death-1}$ is null-homologous
in $M_j\intsec K_{\fdeath+1}$.
The fact that $\fsimp{\Fcal}{\fdeath}\not\in M_j$
implies that $\zG^j_{\death-1}$ is also null-homologous
in $K_{\fdeath}$.
Then,
similar to the proof for Claim~\ref{itm:oo-intv-kill-simps-same-conn} 
of Proposition~\ref{prop:oo-alg-cons-facts},
we can derive a representative cycle $\zG^j_{\death-1}$
for the interval $[\fbirth,\fdeath]$
at index $\fdeath$ which is a boundary,
and thus a contradiction.

\section{The algorithm used in
the proof of \Cref{prop:oo-alg-cons-facts,prop:co-creator-no-coface,prop:levelzz-co-alg-cut-cycles}}
\label{sec:zigzag-pers-alg-abstr}

We describe an algorithm for computing zigzag persistence 
that helps us 
prove some results in this paper.
 This algorithm is a rephrasing (for the purpose of proofs) of
the algorithm proposed in~\cite{DBLP:conf/soda/Dey0M25}.
Given $\Dim\geq 0$
and a simplex-wise zigzag filtration $\lfilt: \emptyset=\lcplx_0 \leftrightarrow 
\cdots \leftrightarrow \lcplx_\lfiltcnt$ starting with an empty complex,
the algorithm computes
the $\Dim$-th zigzag persistence intervals
and their representative $\Dim$-cycles for $\lfilt$.
We denote each linear map in $\Hm_\Dim(\lfilt)$  as 
$\morph{\lfilt}{i}:\Hm_\Dim(\lcplx_i)\leftrightarrow\Hm_\Dim(\lcplx_{i+1})$.
Also, 
for any $i$ s.t.\ $0\leq i\leq\lfiltcnt$,
let $\lfilt^i$ denote the filtration
$\lcplx_0\leftrightarrow\lcplx_1\leftrightarrow\cdots\leftrightarrow \lcplx_i$,
which is a {\it prefix} of $\lfilt$.
The idea of the algorithm~\cite{DBLP:conf/soda/Dey0M25} is
to directly compute an interval decomposition
by maintaining representative cycles
for all intervals:

\opt{DCG}{\medskip}
\begin{algr}[Zigzag persistence algorithm]
\label{alg:zigzag-pers-abstr}
First
set $\Pers_\Dim(\lfilt^{0})=\emptyset$.
The algorithm then iterates for $i\leftarrow0,\ldots,\lfiltcnt-1$.
At the beginning of the $i$-th iteration,
the intervals and their representative cycles 
for $\Hm_\Dim(\lfilt^{i})$ have already been computed.
The aim of the $i$-th iteration is to compute these for $\Hm_\Dim(\lfilt^{i+1})$.
For describing the $i$-th iteration,
let $\Pers_\Dim(\lfilt^{i})=
\Set{[\birth_\aG,\death_\aG]\given \aG\in\Acal^i}$
be indexed by a set $\Acal^i$,
and let 
$\Set{\cyc^{\aG}_{k}\subseteq \lcplx_k\given \birth_\aG\leq k\leq\death_\aG}$
be a sequence of representative $\Dim$-cycles 
for each $[\birth_\aG,\death_\aG]$.
For ease of presentation,
we also let $\cyc^\aG_k=0$
for each $\aG\in\Acal^i$
and each $k\in[0,i]\setminus[\birth_\aG,\death_\aG]$.
We call
intervals of $\Pers_\Dim(\lfilt^{i})$ ending with $i$ 
{\rm surviving intervals}
at index $i$.
Each non-surviving interval of $\Pers_\Dim(\lfilt^{i})$
is directly included in $\Pers_\Dim(\lfilt^{i+1})$
and its representative cycles stay the same.
For surviving intervals of $\Pers_\Dim(\lfilt^{i})$,
the $i$-th iteration proceeds with the following cases:
\begin{itemize}
    \item \emph{$\morph{\Dim}{i}$ is an isomorphism:} 
    In this case, no intervals are created or cease to persist.
    For each surviving interval $[\birth_\aG,\death_\aG]$ in $\Pers_\Dim(\lfilt^{i})$,
    $[\birth_\aG,\death_\aG]=[\birth_\aG,i]$ now corresponds to an interval 
    $[\birth_\aG,i+1]$ in $\Pers_\Dim(\lfilt^{i+1})$.
    The representative cycles for $[\birth_\aG,i+1]$ are set by the following rule:

    \opt{DCG}{\medskip}
    \vspace{\itemsep}
    \noindent
    {\rm Trivial setting rule of representative cycles:}
    For each $j$ with $\birth_\aG\leq j\leq i$, 
    the representative cycle for $[\birth_\aG,i+1]$
    at index $j$ stays the same.
    The representative cycle for $[\birth_\aG,i+1]$
    at $i+1$ is set to a $\cyc^{\aG}_{i+1}\subseteq \lcplx_{i+1}$
    such that $[\cyc^{\aG}_{i}]\leftrightarrow [\cyc^{\aG}_{i+1}]$ by $\morph{\lfilt}{i}$
    (i.e., $[\cyc^{\aG}_i]\mapsto[\cyc^{\aG}_{i+1}]$ 
    or $[\cyc^{\aG}_i]\mapsfrom[\cyc^{\aG}_{i+1}]$).

    \opt{DCG}{\medskip}
    \item \emph{$\morph{\Dim}{i}$ is forward with non-trivial cokernel:}
    A new interval $[i+1,i+1]$ is added to $\Pers_\Dim(\lfilt^{i+1})$
    and its representative cycle at $i+1$ is set to
    a $\Dim$-cycle in $\lcplx_{i+1}$ containing $\fsimp{\lfilt}{i}$ ($\fsimp{\lfilt}{i}$ is a $\Dim$-simplex).
    All surviving intervals of $\Pers_\Dim(\lfilt^{i})$ persist to index $i+1$
    and are automatically added to $\Pers_\Dim(\lfilt^{i+1})$;
    their representative cycles are set by the trivial setting rule.

    \opt{DCG}{\medskip}
    \item \emph{$\morph{\Dim}{i}$ is backward with non-trivial kernel:}
    A new interval $[i+1,i+1]$ is added to $\Pers_\Dim(\lfilt^{i+1})$
    and its representative cycle at $i+1$ is set to 
    a $\Dim$-cycle homologous to $\partial(\fsimp{\lfilt}{i})$ in $\lcplx_{i+1}$
    ($\fsimp{\lfilt}{i}$ is a $(\Dim+1)$-simplex).
    All surviving intervals of $\Pers_\Dim(\lfilt^{i})$ persist to index $i+1$
    and their representative cycles are set by the trivial setting rule.

    \opt{DCG}{\medskip}
    \item \emph{$\morph{\Dim}{i}$ is forward with non-trivial kernel:}
    A surviving interval of $\Pers_\Dim(\lfilt^{i})$ does not persist to $i+1$.
    Let $\Bcal^i\subseteq\Acal^i$ consist of indices of all surviving intervals.
    We have that
    $\Set{[\cyc^{\aG}_{i}]\given \aG\in\Bcal^{i}}$
    forms a basis of $\Hm_\Dim(\lcplx_i)$.
    Suppose that $\morph{\lfilt}{i}\big([\cyc^{\aG_1}_{i}]+\cdots+[\cyc^{\aG_h}_{i}]\big)=0$,
    where $\aG_1,\ldots,\aG_h\in \Bcal^{i}$.
    We can rearrange the indices such that
    $\birth_{\aG_1}<\birth_{\aG_2}<\cdots<\birth_{\aG_h}$
    and $\aG_1<\aG_2<\cdots<\aG_h$.
    Let $\lG$ be $\aG_1$ if 
    $\morph{\Dim}{\birth_\aG-1}$
    is backward for every $\aG\in\Set{\aG_1,\ldots,\aG_h}$
    and otherwise be the largest $\aG\in\Set{\aG_1,\ldots,\aG_h}$ 
    such that 
    $\morph{\Dim}{\birth_\aG-1}$
    is forward.
    Then, $[\birth_{\lG},i]$ forms an interval of $\Pers_\Dim(\lfilt^{i+1})$.
    For each $k\in[\birth_{\lG},i]$,
    let $\cyc'_k=
    \cyc^{\aG_1}_k+\cdots+\cyc^{\aG_h}_k$;
    then, $\Set{\cyc'_k\given \birth_{\lG}\leq k\leq i}$
    is a sequence of representative cycles for $[\birth_{\lG},i]$.
    All the other surviving intervals of $\Pers_\Dim(\lfilt^{i})$ persist to $i+1$
    and their representative cycles are set by the trivial setting rule.

    \opt{DCG}{\medskip}
    \item \emph{$\morph{\Dim}{i}$ is backward with non-trivial cokernel:}
    A surviving interval of $\Pers_\Dim(\lfilt^{i})$ does not persist to $i+1$.
    Let $\Bcal^i\subseteq\Acal^i$ consist of indices of all surviving intervals,
    and let $\cyc^{\aG_1}_{i},\ldots,\cyc^{\aG_h}_{i}$
    be the cycles in $\Set{\cyc^{\aG}_{i}\given\aG\in\Bcal^i}$
    containing $\fsimp{\lfilt}{i}$
    {\rm(}$\fsimp{\lfilt}{i}$ is a $\Dim$-simplex{\rm)}.
    We can rearrange the indices such that
    $\birth_{\aG_1}<\birth_{\aG_2}<\cdots<\birth_{\aG_h}$
    and $\aG_1<\aG_2<\cdots<\aG_h$.
    Let $\lG$ be $\aG_1$ if $\morph{\Dim}{\birth_\aG-1}$
    is forward for every $\aG\in\Set{\aG_1,\ldots,\aG_h}$
    and otherwise be the largest $\aG\in\Set{\aG_1,\ldots,\aG_h}$ 
    such that $\morph{\Dim}{\birth_\aG-1}$
    is backward.
    Then, $[\birth_{\lG},i]$ forms an interval of $\Pers_\Dim(\lfilt^{i+1})$
    and the representative cycles for $[\birth_{\lG},i]$ stay the same.
    For each $\aG\in\Set{\aG_1,\ldots,\aG_h}\setminus\Set{\lG}$,
    let $\cyc'_k=\cyc^{\aG}_k+\cyc^{\lG}_k$ for each $k$ s.t.\ $\birth_{\aG}\leq k\leq i$,
    and let $\cyc'_{i+1}=\cyc'_i$; 
    then, $\Set{\cyc'_k\given \birth_{\aG}\leq k\leq i+1}$
    is a sequence of representative cycles for $[\birth_{\aG},i+1]$.
    For the other surviving intervals,
    the setting of representative cycles follows the trivial setting rule.
\end{itemize}

\end{algr}

See~\cite{DBLP:conf/soda/Dey0M25} for the correctness of \Cref{alg:zigzag-pers-abstr}.

\end{document}